\pdfoutput=1
\documentclass[runningheads]{llncs}
\usepackage[T1]{fontenc}
\usepackage{graphicx}
\usepackage{amsmath}
\usepackage{pgfplots}
\pgfplotsset{compat=1.18}
\usepackage{pgfplotstable}
\usepackage{booktabs}
\usepackage{makecell}
\usepackage{float}
\usepackage{dsfont}
\usepackage[numbers,sort&compress]{natbib}
\usepackage{xcolor}
\definecolor{linkblue}{RGB}{0,70,160}
\usepackage[colorlinks=true,linkcolor=linkblue,citecolor=linkblue,urlcolor=linkblue]{hyperref}
\usepackage{cleveref}
\usepackage{ifthen}

\provideboolean{showcredits}
\newif\ifshort \shorttrue

\usepackage{todonotes}
\renewcommand{\todo}[2][]{}
\AtBeginDocument{%
  \definecolor{blue}{rgb}{0,0,0}%
  \definecolor{red}{rgb}{0,0,0}%
}

\newcommand{\paraheader}[1]{\smallskip\noindent{\bfseries #1}}
\title{Not All LPs Are Equal: The Active-Passive Gap in Automated Market Maker Liquidity Provision}
\titlerunning{The Active-Passive Gap in AMM Liquidity Provision}
\author{Agathe Sadeghi\inst{1} \and
Dingyue Liu\inst{1} \and
Ciamac Moallemi\inst{2,1} \and
Xin Wan\inst{1} \and
Brian Zhu\inst{2}}
\authorrunning{A. Sadeghi et al.}
\institute{Uniswap Labs
\and
Columbia University}

\begin{document}
\maketitle

\begin{abstract}
Liquidity provision in automated market makers is typically analyzed at the pool level, implicitly assuming LP homogeneity.
This aggregate view can hide how liquidity provision outcomes differ between LP strategies, particularly as concentrated liquidity AMM designs operating on high-performance blockchains allow liquidity to be actively repositioned around trades.
We develop a markout-based framework to decompose Uniswap LP profitability into active and passive components using two complementary methods: a LIFO subtraction method that matches short-lived mint-burn positions and attributes swap-level markouts by liquidity share; and an infinitesimal LP benchmark that estimates the performance of a fully passive, always-in-range marginal LP directly from the AMM price path.

We apply these methods to Uniswap v2, v3, and v4 pools on Ethereum, Arbitrum, and Base chains, and find passive profitability can materially differ from aggregate pool profitability. 
In Uniswap v2, with liquidity distributed evenly and active LP behavior nearly absent, the overall and passive markouts are almost the same. 
In contrast, concentrated-liquidity pools have a systematic active-passive gap: passive LPs tend to underperform aggregate pool-level measures. 
The gap is wider on Ethereum than on L2s, consistent with active liquidity provision being more useful when block times and ordering conditions allow LPs to react to incoming flow.  
In general, passive LPs perform better on higher fee pools.
The LIFO and infinitesimal estimates are generally consistent directionally across most pools, providing evidence of the robustness of the decomposition. 
The results suggest that adverse selection in AMMs is not evenly distributed among LPs, with important implications for LP strategy, fee-tier design and measurement of DEX market quality.
\end{abstract}

%%% Local Variables:
%%% mode: LaTeX
%%% TeX-master: "../main"
%%% End:

\keywords{automated market makers \and decentralized exchanges \and liquidity provision \and adverse selection \and markout}

\section{Introduction}\label{sec:intro}
% \begin{itemize}
%     \item Motivation: Economic significance of DEX liquidity provision, the scale of capital at risk, the growth of on-chain markets, and why LP profitability is a first-order question
%     \item Problem Statement: Passive LPs systematically underperform active/JIT LPs, but this gap is poorly measured and poorly understood. Existing work focuses on trader-side toxicity; the LP-side decomposition is underexplored
%     \item Research Gap: No prior work has applied markout methodology from the LP's perspective to decompose active vs passive LP P\&L at scale, across chains, with multiple estimation methods
%     \item Contributions (methodology + empirics + L1/L2 chains)
%     \item Walking the reader through the paper structure
% \end{itemize}

Automated market makers (AMMs) have become a central pillar of on-chain financial infrastructure.
Uniswap alone has facilitated over \$4.4 trillion in cumulative trading volume, with billions of
dollars in liquidity deposited across thousands of pools at any given time.  The economic
viability of these markets depends on liquidity providers (LPs), who supply capital to automated
AMM pools and earn trading fees in return.  Understanding whether liquidity provision is
profitable --- and for whom --- is therefore a first-order question in decentralized finance.

\ifshort
Classical market microstructure theory establishes that liquidity providers earn the spread from uninformed traders but lose to better-informed ones
\citep{glosten1985bid, kyle1985continuous}; in AMMs this adverse selection takes a specific form, as arbitrageurs trade against deterministically stale on-chain prices whenever the external market moves, a cost formalized as loss-versus-rebalancing (LVR) \citep{milionis2022automated}. 
Concentrated liquidity in Uniswap v3 opened a spectrum of strategies around this cost: at one extreme, just-in-time (JIT) providers mint a position immediately before a swap and burn it immediately after, capturing fees while minimizing exposure to adverse price movements; at the other, passive LPs supply liquidity continuously and absorb informed and uninformed flow indiscriminately.
\else

Classical market microstructure theory establishes that liquidity providers face a fundamental tension: they earn the bid-ask spread from uninformed traders but lose to those with superior information \citep{glosten1985bid, kyle1985continuous}.
In the AMM setting, this adverse selection cost takes a specific form.
Because AMM pricing curves are deterministic and publicly observable, arbitrageurs trade against stale on-chain prices whenever the external market moves, extracting value from LPs.
\citet{milionis2022automated} formalize this cost as loss-versus-rebalancing (LVR), providing a theoretical benchmark for the aggregate losses borne by liquidity providers.

However, the aggregate view obscures a critical distinction.
The introduction of concentrated liquidity in Uniswap v3 enabled a spectrum of liquidity provision strategies, ranging from passive positions that remain deployed over long horizons to highly active strategies that dynamically manage liquidity placement around individual trades.
At the extreme end, Just-in-Time (JIT) liquidity providers mint concentrated positions immediately before a swap and burn them immediately after, capturing trading fees while minimizing exposure to adverse price movements.
Active LPs thus selectively participate in fee-generating volume while avoiding periods of arbitrage risk, whereas passive LPs supply liquidity continuously and absorb both uninformed and informed flow indiscriminately.
\fi

This strategic heterogeneity creates a profitability gap between active and passive LPs that is
poorly measured and poorly understood.  Existing empirical work on DEX markets has largely focused
on trader-side flow toxicity, identifying which swaps are driven by arbitrage and measuring the
resulting markout losses at the pool level \citep{heimbach2022risks}.  The LP-side decomposition
--- how these losses are distributed across different types of liquidity providers --- remains
underexplored.  Without this decomposition, aggregate LP profitability statistics conflate the
returns of sophisticated active strategies with those of passive capital, potentially masking the
true economic costs faced by the majority of liquidity providers.

This paper addresses this gap.
We develop and apply a markout-based methodology from the LP's perspective to decompose pool-level profit and loss into active and passive components.
Our analysis spans three chains (Ethereum, Arbitrum, and Base) and multiple pools, allowing us to examine how the active-passive gap varies across execution environments with different block times, gas costs, and participant compositions.

\ifshort
{\color{blue}
Throughout, profitability means fee-inclusive, short-horizon LP profit and loss measured through markout: the value transferred through swap execution relative to an external reference price observed after the trade. 
Total LP P\&L combines a market-risk (beta-like) component from price movements of the underlying
assets with a microstructural (alpha-like) component reflecting fees net of adverse selection;
markout isolates the latter. Markout also excludes gas, liquidity-management, and capital costs and inventory revaluation outside the horizon---second-order for passive positions, though material for the realized profit of active strategies. 
A positive markout means the pool received more value than it delivered when both legs are valued at subsequent reference prices, and is favorable to the LP; a negative markout is a short-horizon adverse-selection loss.
}

\else
{\color{blue}
Throughout the paper, profitability refers to fee-inclusive, short-horizon LP profit and loss measured through markout. 
This measure captures the value transferred through swap execution relative to an external reference price observed after the trade. 
It does not represent the complete return of an LP position, since it excludes gas costs, liquidity-management expenses, capital costs, and inventory revaluation outside the markout horizon. 
Total LP P\&L combines a market-risk (beta-like) component from price movements of the underlying assets with a microstructural (alpha-like) component reflecting fees net of adverse selection; markout is designed to isolate the latter. 
Because passive positions transact infrequently, the excluded gas and rebalancing costs are second-order for the passive measure, though they are material for the realized profit of active strategies. 
We use profitability in this markout-based sense throughout the remainder of the paper.

Markouts are stated from the LP perspective.
A positive markout indicates that, when both sides of the swap are valued at the subsequent external reference prices, the pool received more value than it delivered to the trader and is therefore favorable to the LP. 
A negative markout indicates that the trader received more subsequent benchmark value than the pool received and is interpreted as a short-horizon adverse-selection loss for the LP.}
\fi

% \item \paraheader{An LP-side markout framework with two complementary estimators.} Prior empirical
%   work on AMMs measures adverse selection at the pool or trader level; we instead develop a
%   markout-based decomposition from the \emph{passive liquidity provider's} perspective. 
%   We propose two
%   complementary estimators of passive LP profitability: a \emph{LIFO subtraction method} that
%   identifies active LP positions via mint-burn matching and attributes swap-level markouts by
%   liquidity share, and an \emph{infinitesimal LP benchmark} that prices the markout of a marginal,
%   always-in-range passive LP directly from the realized AMM price path. 
%   The infinitesimal benchmark makes no assumptions about active LP behavior, while the LIFO method preserves the dollar scale of passive P\&L and additionally profiles active LP behavior.

We make two contributions:
\begin{enumerate}
\item \emph{An LP-side markout framework with two complementary estimators.} Prior empirical work on AMMs measures adverse selection at the pool or trader level; we instead develop a markout-based decomposition from the passive liquidity provider's perspective. 
We propose two complementary estimators of passive LP profitability: a LIFO subtraction method that identifies active LP positions via mint-burn matching and attributes swap-level markouts by liquidity share, and an infinitesimal LP benchmark that prices the markout of a marginal, always-in-range passive LP directly from the realized AMM price path. 
The infinitesimal benchmark makes no assumptions about active LP behavior, while the LIFO method preserves the dollar scale of passive P\&L and additionally profiles active LP behavior.
\item \emph{An empirical map of the active-passive gap across Uniswap versions, chains, and fee tiers.} We apply the framework to Uniswap v2, v3, and v4 pools on Ethereum, Arbitrum, and Base. 
Under our LIFO identification, no short-lived mint-burn positions are identified in the selected v2 pools, so aggregate and passive markouts coincide there. 
In concentrated-liquidity v3 and v4 pools, passive profitability can diverge materially from aggregate pool profitability. 
The gap is wider on Ethereum than on Arbitrum and Base\footnote{We classify Arbitrum and Base as L2 execution environments for the main cross-chain analysis because both provide substantially faster and less costly execution than Ethereum mainnet. 
This grouping is intended to capture broad L1-versus-L2 differences rather than to treat the two chains as technologically identical.}, consistent with shorter L2 block times reducing the adverse selection rents active LPs can capture, and is most pronounced in low-fee tiers, where concentrated liquidity can be targeted most efficiently around incoming flow.
\end{enumerate}

{\color{blue}
We validate the decomposition extensively: the two estimators are constructed very differently---one removes identified active liquidity ex post, the other simulates an idealized passive LP along the realized price path---yet they agree directionally and lie within each other's confidence intervals across the great majority of pools, and the estimates are insensitive to replacing LIFO with FIFO, widening the matching horizon, or varying the markout horizon (Appendices~\ref{app:matchrobust} and~\ref{app:horizonrobust}). 
The gap is therefore a feature of the data rather than an artifact of a single identification strategy.}

% We validate the decomposition extensively. 
% The LIFO and infinitesimal estimates are constructed very differently---one identifies and removes active liquidity ex post, the other simulates an idealized passive LP along the realized price path---yet they agree directionally and lie within
% each other's confidence intervals across the great majority of pools. 
% The decomposition is also insensitive to implementation choices: replacing LIFO with FIFO matching, widening the mint-burn matching horizon, or varying the markout horizon leaves the estimates essentially unchanged (Appendices~\ref{app:matchrobust} and~\ref{app:horizonrobust}). 
% This consistency indicates that the active-passive gap is a feature of the data rather than an artifact of any single identification strategy.}

% The remainder of the paper is organized as follows.
% \Cref{sec:lit_rev} reviews the literature most closely related to our paper. 
% \Cref{sec:data} provides background on the Uniswap protocol mechanics and describes our data sources.
% \Cref{sec:methodology} presents the empirical methodology, including the markout framework, LP segmentation criteria, and the two estimation approaches for passive LP P\&L.
% \Cref{sec:result} reports the main results.
% \Cref{sec:conclusion} discusses the economic implications and concludes.

The remainder of the paper is organized as follows.
\Cref{sec:lit_rev} reviews related literature.
\Cref{sec:back_data} covers protocol background and data. 
\Cref{sec:methodology} develops the methodology.
\Cref{sec:result} reports the results, and \Cref{sec:conclusion} concludes.

%%% Local Variables:
%%% mode: LaTeX
%%% TeX-master: "../main"
%%% End:

\section{Literature Review}\label{sec:lit_rev}

% The profitability of liquidity provision in decentralized exchanges (DEXs) lies at the intersection of classical market microstructure and the design of automated market makers (AMMs).
% While prior work has characterized the aggregate economics of automated liquidity provision, relatively little empirical research distinguishes between different liquidity provision strategies within the same pool.
% This paper studies the profitability gap between passive and active liquidity providers (LPs) on Uniswap and investigates how this gap arises from adverse selection, arbitrage, and execution latency.
% We review the strands of literature that inform our analysis.

The profitability of DEX liquidity provision lies at the intersection of
classical market microstructure and AMM design. 
We review the strands most relevant to our analysis. An extended review can be found in Appendix~\ref{app:related-empirical}.

\paraheader{Market microstructure and adverse selection.}
\ifshort
Foundational models show that liquidity providers systematically lose to traders with superior information, with the spread compensating for adverse selection \citep{glosten1985bid}. \cite{kyle1985continuous} show that toxic flow, e.g., order flow that consistently predicts price movements, is a primary driver of market-maker losses. 
In modern electronic markets this cost is measured with markout P\&L, the difference between a trade's execution price and a subsequent reference mid-price: the empirical analogue of a realized spread from the LP perspective. 
Because markout is attributable at the swap level, it can be divided between the active and passive liquidity participating in each trade, which is precisely the granularity our decomposition requires. In the AMM setting, arbitrageurs play the role of informed traders.
\else
The economics of liquidity provision on decentralized exchanges closely mirrors the classical dealer problem studied in market microstructure.
Foundational models by \cite{glosten1985bid} and \cite{kyle1985continuous} show that liquidity providers systematically lose to traders with superior information.
In equilibrium, the bid–ask spread compensates market makers for this adverse selection risk, ensuring that expected profits from uninformed trades offset losses to informed traders. 

In modern electronic markets, adverse selection is commonly measured using markout profit-and-loss (P\&L), i.e.,  the difference between a trade's execution price and a subsequent reference mid-price. 
{\color{blue}Markout is particularly suitable for our analysis because it isolates the profitability associated with a specific swap. It asks whether the execution remains favorable to the liquidity supplying side after allowing time for the information contained in the trade to be incorporated into an external reference market. In this sense, markout is the empirical analogue of a realized spread from the LP perspective. Unlike position-level return measures, it can be attributed at the swap level and consequently divided between the active and passive liquidity participating in the trade.}
\citet{easley2012flow} show that order flow which consistently predicts price movements, often referred to as toxic flow, is a primary driver of market maker losses.

Our work applies this framework to decentralized exchanges.
Because AMMs cannot condition quotes on trader identity or information, arbitrageurs play a role analogous to informed traders in traditional markets.
We therefore use markout analysis to quantify how much value transfers from LPs to overall flows.
\fi

\paraheader{AMM design and loss-versus-rebalancing (LVR).}
AMMs replace discrete limit orders with deterministic pricing curves whose pricing and arbitrage properties were characterized by early theoretical work \citep{angeris2020improved}. 
Uniswap v3 introduced concentrated liquidity \citep{adams2021uniswapv3} and v4 adds programmable hooks. 
Deterministic pricing gives rise to arbitrage against stale prices, which \citet{milionis2022automated} formalize as LVR: the difference between the value of an AMM position and a portfolio that continuously rebalances at market prices. 
\citet{milionis2023flair} argue that LVR captures the cost of toxic flow but not competition among LPs, and introduce FLAIR to measure how fee opportunities are distributed across heterogeneous providers. 
Our markout measure is closely related to LVR: markouts evaluate trades at fixed post-trade
offsets, while LVR also does so but at periodic rehedging intervals.

\paraheader{Active liquidity provision and just-in-time strategies.}
Concentrated liquidity created a spectrum of strategies from long-horizon passive positions to active management \citep{fan2023strategic}, with JIT liquidity the extreme case. \citet{wan2022jit} first document the JIT mechanism on Uniswap v3; \citet{xiong2023demystifying} identify tens of thousands of JIT events and several thousand ETH of extracted profit over a 20-month window; \citet{capponi2024paradox} show game-theoretically that JIT reduces trader slippage but can redistribute fee revenue away from passive LPs and paradoxically reduce equilibrium liquidity; and \citet{trotti2025strategic} formalize the JIT provider's optimization, showing strategic deployment can erode per-trade passive LP profits by up to 44\%. Our active category is broader than same-block JIT (Section~\ref{sec:lp-seg}), and our swap-level markout decomposition is distinct from realized JIT strategy profit, which additionally reflects fee revenue, gas, rebalancing, and retained inventory \citep{xiong2023demystifying}.

\paraheader{Empirical measurement of LP profitability and on-chain flow.}
\ifshort
A growing empirical literature measures LP performance and on-chain flow composition, from cross-sectional LP returns \citep{heimbach2022risks} and AMM-versus-LOB comparisons \citep{lehar2025decentralized} to structural models of concentrated liquidity \citep{hasbrouck2025economic} and arbitrage-loss-versus-fee accounting \citep{fritsch2024measuring}, alongside work quantifying extractable value and using markout to identify informed flow \citep{qin2022quantifying,zhu2024drives}. 
Most of this work, however, analyzes LP performance at the pool level, implicitly treating LPs as homogeneous. 
Our paper addresses this gap by decomposing pool-level P\&L into active and passive components using markout-based metrics across multiple chains; Appendix~\ref{app:related-empirical} reviews this literature in detail.
\else
% Empirical-measurement related work. Inline under Literature Review in the
% AFT/LIPIcs build; relocated to the appendix in the WINE build.
Empirical studies of decentralized exchanges have progressed from simple return calculations to more detailed analyses of LP performance and flow composition.
Early empirical work such as \citet{heimbach2022risks} evaluates overall LP returns across Uniswap v3 pools, documenting wide cross-sectional dispersion in realized profitability.
\citet{lehar2025decentralized} use the full Uniswap transaction history to compare automated market making to limit-order-book equilibria, characterizing when AMM liquidity dominates and showing the absence of long-lived arbitrage.
\citet{hasbrouck2025economic} develop a structural economic model of concentrated liquidity provision in which LP supply is pinned down by a no-arbitrage risk-adjusted return condition, providing a benchmark against which realized LP performance can be evaluated.
Most directly related to our setting, \citet{fritsch2024measuring} measure arbitrage losses against fee income across the largest Uniswap pools and find that arbitrage losses exceed fees in many pools, with v2 positions often more profitable than their v3 counterparts in the passive limit.

A parallel literature analyzes the composition of on-chain order flow.
\citet{qin2022quantifying} provide a large-scale quantification of blockchain extractable value across DEX arbitrage, sandwich attacks, and liquidations, establishing the empirical scale of value transferred from passive participants to searchers.
Markout-based methodologies have also been adopted in both academic and practitioner analyses to identify informed or arbitrage-driven flow in Uniswap markets; for example, \citet{zhu2024drives} use markout as an explanatory variable for liquidity supply and document its predictive power for future market depth.
These approaches use post-trade price movements to estimate the extent to which LPs trade against price-moving order flow.

However, most existing empirical work analyzes LP performance at the pool level, implicitly treating liquidity providers as homogeneous.
In practice, concentrated liquidity enables a wide range of strategies with different exposure to arbitrage and fee income.

Our paper addresses this gap by decomposing pool-level P\&L into active and passive liquidity provision using markout-based metrics across multiple chains.
This allows us to quantify how arbitrage and trading flow are distributed across different types of LPs and to measure the resulting profitability gap between passive and strategically managed liquidity.
\fi

\section{Background and Data}
\label{sec:back_data}

This section provides background on the Uniswap protocol and describes the dataset used in our empirical analysis. 
We first review the mechanics of concentrated liquidity, then introduce a conceptual taxonomy of liquidity provider (LP) strategies, and finally describe the on-chain and market data used in the study.

\ifshort
% WINE build: protocol-mechanics background relocated to Appendix~\ref{app:protocol-mechanics}.
Uniswap v3 introduced concentrated liquidity: liquidity providers allocate capital within user-specified price ranges $[P_a,P_b]$ (discretized into ticks) rather than across the whole constant-product curve, earning fees pro rata only while the pool price is in range. 
V4 preserves this mechanism while adding a singleton architecture and hooks, without altering the provisioning mechanics relevant here. We reconstruct LP activity from three on-chain events---\textbf{Mint}, \textbf{Swap}, and \textbf{Burn}; Appendix~\ref{app:mechanics} reviews these mechanics in more detail.
\else
\subsection{Uniswap Protocol Mechanics}
% Uniswap protocol mechanics (background). Inline under Background and Data in the
% AFT/LIPIcs build; relocated to the appendix in the WINE build.
{\color{blue} Uniswap v2 implements a constant-product automated market maker in which token reserves ($x$) and ($y$) satisfy $xy=k$. 
Liquidity in v2 is supplied across the full price range, so all liquidity in the pool participates in a swap at the prevailing price. 
LPs may add or remove liquidity over time, but they cannot choose a narrower price range over which their capital is active.}

Uniswap v3 introduced concentrated liquidity, extending the constant-product automated market maker framework.
While the underlying invariant remains based on the constant-product relationship ($xy = k$), liquidity providers can allocate capital within user-specified price ranges rather than across the entire price curve.

Prices in Uniswap are discretized into ticks, where the price at tick $i$ is defined as $P(i) = 1.0001^i$. 
Each liquidity position specifies a price interval $[P_a, P_b]$ over which the provided liquidity is active.
When the pool price $P$ lies within this interval, the position participates in trades and earns a pro-rata share of trading fees.
If the price moves outside the specified range, the position becomes fully composed of a single asset and no longer earns fees until the price re-enters the range.

Each LP position is represented as a non-fungible position defined by its price bounds and liquidity amount.
This design allows liquidity providers to adopt heterogeneous strategies ranging from wide passive ranges to highly concentrated liquidity that closely tracks the market price. 

% Uniswap v4 preserves the core concentrated liquidity mechanism while introducing architectural modifications such as a singleton contract and extensible hooks.
% These changes allow additional logic to be executed during pool operations but do not fundamentally alter the liquidity provisioning mechanics considered in this study.

% To reconstruct liquidity provider activity, we track three primary on-chain events:

% \begin{itemize}
% \item \textbf{Mint}: creation of a liquidity position and deposit of assets
% \item \textbf{Swap}: execution of a trade that moves the pool price and generates trading fees
% \item \textbf{Burn}: withdrawal of liquidity and realization of accrued fees
% \end{itemize}

% These design allows us to reconstruct both trading activity and liquidity provisioning behavior over time.

{\color{blue}Uniswap v4 retains the concentrated-liquidity framework of v3 while introducing a singleton architecture and programmable hooks that allow additional pool-specific logic. The core distinction relevant for our analysis remains the same: liquidity can be concentrated over specific price ranges and actively repositioned over time. The version-specific event data used to reconstruct swaps and liquidity changes are described in Section~\ref{sec:data}.}
\fi

\subsection{Liquidity Provider Strategies}

Concentrated liquidity enables a wide range of liquidity provision strategies. 
In practice, liquidity providers differ substantially in how actively they manage their positions. 
Rather than forming discrete categories, LP behavior can be viewed along a spectrum ranging from passive to highly active liquidity provision. 
At one end of the spectrum are passive positions that remain largely unchanged over long horizons, while at the other end are highly dynamic strategies that adjust liquidity in response to individual trades.

\paraheader{Passive LPs.}
Passive LPs maintain liquidity positions over relatively long horizons with wide price ranges. 
These positions are typically adjusted infrequently and prioritize simplicity and low operational overhead rather than precise tracking of the current market price. 
As a result, passive LPs often provide a stable base layer of liquidity available to traders but may experience greater exposure to adverse selection when prices move.

\paraheader{Active LPs.}
Active LPs dynamically manage their liquidity positions in response to price movements. 
By periodically adjusting their liquidity ranges to remain close to the current trading price, these participants aim to maximize fee income per unit of capital while reducing exposure to price movements outside their active range. 
At the extreme end of this spectrum is Just-in-Time (JIT) liquidity, where liquidity is supplied immediately before a specific trade and removed shortly afterward. 
% In this strategy, an LP provides highly concentrated liquidity around a certain price, captures a large share of the fees generated by the trade, and withdraws liquidity immediately after execution. 
% Because liquidity is only present during the execution of the swap, JIT strategies can substantially reduce exposure to price risk while capturing significant fee revenue.

% Throughout the paper, we use the terms passive and active as convenient descriptors of behavior along this spectrum. 
% The empirical methodology in Section~\ref{sec:methodology} formalizes this distinction by identifying liquidity provisioning patterns associated with different levels of activity.

{\color{blue}Throughout the paper, active and passive describe the economic distinction between liquidity that is strategically managed over short horizons and liquidity that remains deployed more continuously. 
Section~\ref{sec:methodology} formalizes active liquidity through short-lived additions and removals observed within a matching horizon. 
Positions not attributed to such short-lived liquidity form the passive residual under the LIFO method. 
The infinitesimal method provides a complementary passive benchmark without classifying observed LP positions.}

\subsection{Data Sources}\label{sec:data}

Our empirical analysis combines on-chain event data with external market price data.

\paraheader{On-chain data.}
For the pools we analyze, we collect version specific {\color{blue}Uniswap event logs from Allium. We do not operate independent nodes on any of the chains. For each event, the data contain the chain, pool or contract identifier, block number, block timestamp, transaction hash, transaction index, log index, and the relevant event fields. We order events lexicographically by block number, transaction index, and log index. This ordering preserves the execution sequence of swaps and liquidity events within a block even when several transactions share the same block timestamp.}

{\color{blue}The swap timestamp used in the markout calculation is the canonical block timestamp as provided by the data source. It is not a transaction receipt timestamp. All swaps included in the same block may therefore share the same timestamp, e.g., on Ethereum, although their execution order remains observable through the transaction and log indices.}

On the LP side, for Uniswap v2 and v3 we use \texttt{Swap}, \texttt{Mint} and \texttt{Burn} events to reconstruct the execution of trades and the liquidity provisioning activity. 
\texttt{Swap} is the execution of a trade that moves the pool price and generates trading fees.
\texttt{Mint} is the creation of a liquidity position and deposit of assets. 
\texttt{Burn} is the withdrawal of liquidity and realization of accrued fees. 
In Uniswap v4, liquidity updates are stored in the singleton \texttt{PoolManager} as \texttt{ModifyLiquidity} events. 
We interpret positive liquidity deltas, i.e., $\texttt{LiquidityDelta} > 0$, as additions of liquidity, and negative liquidity deltas as removals of liquidity. 
We use such signed updates as the v4 analogue of mint and burn events in the LIFO matcher. 

Our dataset covers Ethereum Mainnet, Arbitrum, and Base over the period from February 1\textsuperscript{st} to November 30\textsuperscript{th}, 2025. 
These chains differ in block times and execution environments, providing variation in latency and block production that may influence liquidity provision strategies such as Just-in-Time liquidity.

\paraheader{Pool selection.}
We focus on high-volume pools (across chains and fee tiers) to have enough trading activity, liquidity provision activity, and arbitrage participation. To keep the cross-chain comparison consistent, we choose pools that have the same set of main benchmark assets (WETH, USDC, USDT, WBTC and cbBTC), which are the main liquid crypto and stablecoin markets on Uniswap. For each protocol version and chain we include the available pools for these pairs across the relevant fee tiers, mainly 5 bps and 30 bps, and 1 bps where available on Ethereum. We exclude pools with total sample-period volume below \$10 million, since these pools generate noisy markout estimates and little active LP activity.

\paraheader{External price data.}
All markout calculations use external reference prices derived from Binance spot order-book data. 
For each token $i$, we construct a second-level benchmark price in USDT from the mid-price at the top of the book as
$p_{i,t}=\frac{\text{bid}_{i,t}+\text{ask}_{i,t}}{2}$.
{\color{blue}For each swap occurring at timestamp $t$ and markout horizon $h$, the target benchmark time is $t+h$. We assign the benchmark using the last quote at or before $t+h$. We remove the swaps for which no admissible quote is available.}

The price of the input and output tokens are in USDT at the benchmark time. We then calculate the benchmark exchange rate between the two tokens by the ratio of their USDT prices:
\ifshort
$p_{b,h}
=
\frac{p_{\mathrm{in},t+h}}
     {p_{\mathrm{out},t+h}}.$
\else
$$
p_{b,h}
=
\frac{p_{\mathrm{in},t+h}}
     {p_{\mathrm{out},t+h}}.
$$
\fi
The construction of markout measures and the empirical methodology used to estimate LP profitability are described in Section~\ref{sec:methodology}.
\section{Empirical Methodology}
\label{sec:methodology}

% \begin{itemize}
%     \item Markout Framework: Formal definition of markout P\&L for a swap; Price reference construction from Binance; Markout windows: choices and justification; From swap markout to LP markout: sign convention, aggregation

%     \item LP Segmentation: Active (JIT) Identification: Formal definition: mint–burn pair within N blocks bracketing a swap; N-block threshold: L1 vs L2 differences and rationale; Edge cases: partial burns, overlapping ranges, multi-position JIT

%     \item Passive LP P\&L: Subtraction Method: LIFO matching algorithm (Method B2); Exact matching algorithm with confidence weighting (Method B1); How active LP contribution is removed to isolate passive P\&L; Limitations and assumptions

%     \item Passive LP P\&L: Infinitesimal Method: Theoretical derivation of instantaneous LP P\&L from the AMM curve; Connection to LVR literature; Implementation from on-chain state; Comparison to subtraction method: what each captures
% \end{itemize}

\subsection{Markout Framework}\label{sec:markout}

The \textit{markout} of a swap is a metric used to evaluate the execution quality and degree of adverse selection of that swap by comparing the execution price of a swap with a benchmark price at some point in the future. For a swap $s$ between tokens X and Y in which the user puts $q_{s,in}$ tokens X into the AMM and receives $q_{s,out}$ tokens Y from the AMM, we define the execution price as
% \ifshort
% $p_e = \frac{q_{s,out}}{q_{s,in}}.$
% \else
\begin{gather*}
    p_e = \frac{q_{s,out}}{q_{s,in}}.
\end{gather*}
% \fi
Let $p_{s,in,h}$ and $p_{s,out,h}$ denote the reference prices, in the same numeraire (USDT), of the tokens traded in and out, respectively, on a centralized exchange (CEX), in our case, Binance, at a time horizon $h$ after the swap occurred.\footnote{This reference price is relative to a numeraire used on the reference platform, typically either a fiat currency or stablecoin. In this paper, we use Binance for reference prices, which uses Tether stablecoin (USDT) as the numeraire.} We define benchmark price ratio (at horizon $h$) of token X relative to token Y as
% \ifshort
% $p_{b,h} = \frac{p_{in,h}}{p_{out,h}}.$
% \else
\begin{gather*}
    p_{b,h} = \frac{p_{in,h}}{p_{out,h}}.
\end{gather*}
% \fi
We define the \textit{dollar markout} of swap $s$ from the liquidity pool's perspective, denoted by $M_s$, as the value of tokens deposited into the pool less the value of tokens received from the pool for that swap, where the value is evaluated at horizon $h$ using reference prices from the CEX:
\begin{align*}
        M_s &= q_{s,in}p_{in,h}-q_{s,out}p_{out,h} &= q_{s,in}p_{in,h}\left(1-\frac{q_{s,out}p_{out,h}}{q_{s,in}p_{in,h}}\right) = q_{s,in}p_{in,h}\cdot\frac{p_{b,h}-p_e}{p_{b,h}}.
\end{align*}
Our definition of dollar markout has a natural interpretation as the P\&L of the pool --- the
counterparty to the swap --- and can be cleanly decomposed into two components: (i) the dollar volume of the amount traded in, i.e.\ $q_{s,in}p_{s,in,h}$, and (ii) the price improvement of the benchmark price relative to the execution price, which we define as the \textit{raw markout} and denote by $\mu_s$:
\begin{gather*}
    \mu_s = \frac{p_{s,b,h}-p_e}{p_{s,b,h}}.
\end{gather*}
Note that the markout is from the LP perspective and includes fees but excludes gas costs.

\subsection{LP Segmentation}\label{sec:lp-seg}

We now describe our methodology for segmenting liquidity providers into two degrees of activity level.

\paraheader{Active LP identification.}
We identify active liquidity providers by searching for liquidity positions that are opened and closed in a narrow time window.
Formally, we identify liquidity positions by combination of wallet address, liquidity pool, and price ranges (lower tick and upper tick). 
A liquidity position is \textit{active} if the total liquidity level burnt equals the total liquidity level minted in a given price range on a given pool, and all mint and burn transactions occur within $T$ seconds. 
\ifshort
We use a matching horizon of $T=20$ seconds on all chains; the choice of $T$, its
distinction from the markout horizon $h$, and robustness to $T=60$ seconds and to a
FIFO convention are discussed at the end of this section and in Appendix~\ref{app:matchrobust}.
\else
{\color{blue}Note that this mint-burn matching horizon ($T$) is distinct from the markout horizon ($h$). 
The matching horizon determines which liquidity positions are classified as active, whereas the markout horizon determines when the swap is valued relative to the external market.

On Ethereum, we classify liquidity as active when the matched addition and removal occur within the same block. 
On Arbitrum and Base, short-lived positions may span multiple rapidly produced blocks, so we use a common matching horizon of 20 seconds. 
This choice preserves a consistent economic definition of short duration liquidity across the two L2 environments without requiring the same number of blocks to elapse on each chain. 
Because Arbitrum and Base have different timing structures, the 20-second window corresponds to different numbers and types of block or preconfirmation updates. 
We group the chains as L2s for the principal analysis but retain chain-specific results throughout. 
As a robustness check we re-run the analysis with a 60-second matching horizon and with a FIFO matching convention and observe essentially no difference in the results (Appendix~\ref{app:matchrobust}).
}
\fi

While ``just-in-time'' (JIT) liquidity typically refers to \textit{same-block} mint-burn pairs that provide added depth for one or a few swaps, we relax this notion to allow for liquidity positions that last for longer than a single block (but are still relatively short); this accounts for the difference in transaction ordering and block construction on Layer 2 blockchains compared with Ethereum. 
This identification strategy catches several notable patterns of ``edge-case'' behavior, such as partial burns and multiple positions within the same functional price range. To handle these, we propose two separate methods of identifying active and passive LP behavior.

\paraheader{Passive LP P\&L: LIFO subtraction method.}
To identify the portion of order flow interacting with short-lived liquidity, we process additions and removals at the individual position level in on-chain order. The position identifier differs across protocol versions: the tuple for v2 is $(\texttt{Chain}, \texttt{Pool}, \texttt{LP})$, for v3 is $(\texttt{Chain}, \texttt{Pool}, \tau_\ell, \tau_u, \texttt{LP})$, and for v4 we use $(\texttt{Chain}, \texttt{Pool}, \tau_\ell, \tau_u, \texttt{salt})$. Within each position, we maintain a last-in-first-out (LIFO) stack of outstanding liquidity additions. 
Each removal is matched against the most recent unmatched addition(s). 
Partial matches are allowed: one addition may be closed through multiple removals, and one removal may consume liquidity from multiple earlier additions. 
This produces a collection of matched mint-burn allocation legs, each associated with a matched liquidity amount and an opening and closing time.

{\color{blue}LIFO is an identification convention rather than an accounting assumption about how LPs themselves record their positions. alternative conventions are discussed in \Cref{app:matchrobust}.}

Note that since this procedure is implemented at the address-position level, it does not capture active LP strategies that are split across multiple addresses, nor does it identify the ultimate controller of liquidity supplied through smart contracts or other intermediaries. Thus, the method should be interpreted as classifying short-lived liquidity at the observed on-chain address level rather than at the economic-agent level.

\ifshort
Each matched leg $a$ with liquidity $\ell_a$ \emph{covers} a swap $s$ if the swap falls strictly between the leg's mint and burn in on-chain order and (on v3 and v4) within its tick range; the active share of the swap attributable to the leg is $w_{s,a}^{\mathrm{act}}=L_{s,a}/L_s$, where $L_{s,a}$ encodes this coverage and $L_s$ is the pool liquidity at the swap. The passive component is the residual, $w_{s}^{\mathrm{pass}}=1-\sum_a w_{s,a}^{\mathrm{act}}$, so passive dollar volume and markout are $w_s^{\mathrm{pass}}q_{s,in}p_{in,h}$ and $w_s^{\mathrm{pass}}M_s$; equivalently, passive equals total minus active after aggregation. The full coverage definition and aggregation are given in Appendix~\ref{app:methoddetails}.
\else
% LIFO coverage/aggregation algebra. Inline under Methodology in the AFT/LIPIcs
% build; relocated to the appendix in the WINE build.
For a matched allocation leg \(a\), let \(\ell_a\) be the amount of liquidity matched between its mint and burn. We say that leg \(a\) covers swap \(s\) if the swap occurs strictly between the mint and burn in on-chain order and if the swap tick lies within the leg's tick range on v3 and v4. Formally,
\[
L_{s,a}
=
\ell_a\,
\mathbf{1}\{o(\mathrm{mint}_a)<o(s)<o(\mathrm{burn}_a)\}\,
\mathbf{1}\{\tau_{\ell,a}\leq \tau_s<\tau_{u,a}\}.
\]
Let \(L_s\) denote the in-range active liquidity recorded on the swap event, interpreted as the tick liquidity available after the execution.
We define the active share of swap $s$ attributable to that
matched leg as
\begin{gather*}
    w_{s,a}^{act}=\frac{L_{s,a}}{L_s}.
\end{gather*}
Using the dollar markout $M_s$ defined above, the active dollar volume and active dollar markout attributed to allocation leg $a$ from swap $s$ are
\begin{gather*}
    V^{\mathrm{act}}_{s,a}=w_{s,a}^{\mathrm{act}}\, q_{s,in} p_{in},
    \qquad
    M^{\mathrm{act}}_{s,a}=w_{s,a}^{\mathrm{act}}\,M_s.
\end{gather*}
Summing across all matched allocation legs and swaps yields the LIFO-based active volume and active markout.

The passive component is defined residually as the part of the swap not attributed to the matched short-lived liquidity. Thus, for swap $s$, the passive share is $1-\sum_a w_{s,a}^\mathrm{act}$, and the corresponding passive dollar volume and passive dollar markout are
\begin{gather*}
    V^{\mathrm{pass}}_{s}=w_{s}^{\mathrm{pass}}\, q_{s,in} p_{in,h},
    \qquad
    M^{\mathrm{pass}}_{s}=w_{s}^{\mathrm{pass}}\,M_s,
    \qquad
    w_{s}^{\mathrm{pass}} = 1 - \sum_a w_{s,a}^{\mathrm{act}}
\end{gather*}
Equivalently, after aggregating active quantities, passive volume and passive markout can be obtained as total volume and total markout minus their active counterparts.
\fi

We also implement a method for exact subtraction that finds active liquidity by consecutive mint-burn matches with exactly offsetting liquidity. The exact method produces passive markout estimates that are very close to the LIFO estimates throughout our sample, so we present LIFO as our main subtraction-based method, and leave the details and results of the exact-matching method to Appendix~\ref{app:exact}.

\paraheader{Passive LP P\&L: Infinitesimal method.}
Another method we use to evaluate the performance of passive LPs is to consider an \textit{infinitesimal LP}.\ifthenelse{\boolean{showcredits}}{\footnote{We thank Dan Robinson for suggesting the infinitesimal LP method.}}{}
The infinitesimal method approaches passive profitability without identifying active LP positions. 
Consider a hypothetical LP that is (i) full-range (or equivalently, always in-range) and (ii) contributes an infinitesimal amount of liquidity, so that it does not impact the price changes caused by on-chain swaps. 
{\color{blue}Given the pool prices immediately before and after a swap, the constant-product reserve formulas determine the hypothetical LP's token inflows and outflows per unit of liquidity, after accounting for fees.
This calculation does not require identifying other LPs' positions or classifying their liquidity as active or passive; their effect on execution is reflected in the observed price movement.}\footnote{We include fee revenue in the infinitesimal LP's markout, consistent with the fee-inclusive convention used for LIFO subtraction. For v3 and v4, we divide the curve-implied input amount by $1-f$, where $f$ is the pool's fee rate; v2 requires separate treatment, as described in Appendix~\ref{app:methoddetails}.}

% \textbf{BRIAN EDIT:}
% Specifically, our procedure is as follows.\footnote{Derivations for these formulas are corroborated by \citet{elsts2021liquidity}.} Taking the liquidity level $L$ to be 1, the reserves of token X and token Y in a constant product AMM as a function of the internal pool price $p$ (of token X relative to token Y) are given by
% \begin{gather*}
%     x(p) = p^{-1/2}, \quad y(p) = p^{1/2}.
% \end{gather*}
% Suppose that a user swapping tokens X in for tokens Y out moves the pool price from $p_0$ to $p_1$ where $p_1<p_0$. Then, the corresponding token amounts in and out are
% \begin{gather*}
%     q_{s,in} = \frac{|x(p_1)-x(p_0)|}{1-f}, \qquad q_{s,out} = |y(p_1)-y(p_0)|.
% \end{gather*}
% where $f$ is the fee tier of the pool. Conversely, if a user swaps tokens Y in for tokens X out, moving the pool price from $p_0$ to $p_1$ where $p_1>p_0$, then
% \begin{gather*}
%     q_{s,in} = \frac{|y(p_1)-y(p_0)|}{1-f}, \qquad q_{s,out} = |x(p_1)-x(p_0)|.
% \end{gather*}

% \textbf{AGATHE EDIT:}
\ifshort
We use these token amounts as $q_{s,in}$ and $q_{s,out}$ to compute the infinitesimal LP's raw and dollar markouts; the formulas are given in Appendix~\ref{app:methoddetails}.
\else
% Infinitesimal-LP reserve-change derivation. Inline under Methodology in the
% AFT/LIPIcs build; relocated to the appendix in the WINE build.
For each swap $s$, let $p_0$ and $p_1$ denote the pool price immediately before and after the swap. We compute the curve implied reserve change magnitudes
\begin{gather*}
    \Delta x_s = |x(p_1)-x(p_0)|,
    \qquad
    \Delta y_s = |y(p_1)-y(p_0)|.
\end{gather*}
We then assign these two reserve changes to input and output according to the observed swap direction. If the trader sells token X and receives token Y, then
\begin{gather*}
    q_{s,in} = \frac{\Delta x_s}{1-f},
    \qquad
    q_{s,out} = \Delta y_s.
\end{gather*}
If instead the trader sells token Y and receives token X, then
\begin{gather*}
    q_{s,in} = \frac{\Delta y_s}{1-f},
    \qquad
    q_{s,out} = \Delta x_s.
\end{gather*}
Here \(f\) is the pool fee tier. The factor \(1/(1-f)\) accounts for the fact that Uniswap fees are taken from the input amount before the post fee amount moves the invariant in v3 and v4. These \(q_{s,in}\) and \(q_{s,out}\) are then used to compute the raw and dollar markouts for the infinitesimal LP.

{\color{blue}For v3 and v4, we calculate the infinitesimal benchmark at unit liquidity, $L=1$ and fees are deducted upfront so only $(1-f)x$ is used in the swap accounting, when a swapper sells token X and buys token Y.
On the other hand, v2 requires a different treatment because fees are handled differently. 
In v2, fees are not taken out of the swap input; instead, the full input amount enters the pool and the fee is embedded in the reserves. 
This means that the pool receives the full $x$, not $(1-f)x$, and the fee is reflected in the post swap price change.}

These serve as the amounts that are used to compute the raw and dollar markouts for the infinitesimal LP. 
\fi
Note that the on-chain AMM price path is sufficient to infer the infinitesimal LP’s token exposure, but not to value that exposure. 
The markout calculation requires synchronized off-chain benchmark quotes and thus cannot be calculated using only RPC data. 
However, the key is that the method only requires the benchmark mid-price, and does not require the full reconstruction of CEX order book.

\paraheader{Comparing passive LP P\&L methods.}
\ifshort
\Cref{tab:method_comparison} summarizes the differences between the two passive-LP methods. The key contrast is that LIFO subtraction profiles both active and passive LPs by decomposing aggregate outcomes at the swap level and recovers dollar-scale P\&L, but it requires assumptions about which liquidity is active; the infinitesimal LP method measures only a marginal passive LP directly from the pool price path, needs no active-LP identification and only swap data, but yields a normalized benchmark and is more sensitive to sandwich-induced price reversals. A row-by-row discussion is provided in Appendix~\ref{app:methoddetails}.
\else
% Row-by-row discussion of the two passive-LP methods (Table~\ref{tab:method_comparison}).
% Inline under Methodology in the AFT/LIPIcs build; relocated to the appendix in WINE.
\cref{tab:method_comparison} summarizes the main differences and discusses the two methods of breaking down passive and active markouts. The first two rows ask whether or not a method can attribute P\&L to a given type of LP. Both methods can profile passive LPs, but only LIFO subtraction also profiles active LPs, since it explicitly decomposes aggregate LP outcomes into active and passive components. The infinitesimal LP method instead focuses only on the exposure of a marginal passive LP and therefore does not provide estimates for active LP performance.

The ``LP decomposition granularity'' row describes the level at which passive and active LP outcomes are separated. LIFO subtraction performs this decomposition at the swap level, assigning each trade between active and passive liquidity. The infinitesimal LP method also operates at the swap level, but only for passive exposure: it computes the markout of an infinitesimal passive LP along the realized price path rather than decomposing observed LP outcomes into active and passive components.

The confidence-interval rows describe whether each method can produce uncertainty estimates for markouts. Both methods can produce confidence intervals for passive markouts, but only LIFO subtraction can also produce confidence intervals for active markouts, because only LIFO subtraction estimates active LP performance. The ``Dollar scale of passive P\&L'' row asks whether the method recovers the aggregate dollar magnitude of passive LP gains and losses. LIFO subtraction does so because it starts from observed aggregate LP outcomes and subtracts the active component. The infinitesimal LP benchmark does not; it instead produces a normalized benchmark for the performance of a marginal passive LP.

The ``Assumptions on active LP behavior'' row captures whether the method requires a rule for identifying active liquidity provision. LIFO subtraction does require such assumptions, since it must decide which LP actions count as active in order to subtract them from the aggregate. The infinitesimal LP method does not require assumptions about active LP behavior because it does not attempt to identify active LPs. This is important when evaluating passive LP outcomes, since the former method attributes what remains after subtracting out active LP outcomes to passive LPs, while the latter method attempts to measure passive LP outcomes directly.

The ``Robust to sandwiches'' row concerns whether the method is naturally robust to mechanically induced price reversals such as sandwich attacks. LIFO subtraction is less sensitive to this issue because it measures realized LP outcomes through the active-passive decomposition. The infinitesimal LP method is more sensitive because it evaluates passive exposure along the full realized pool-price path, so temporary price distortions can generate anomalously large markouts unless sandwich-related swaps are filtered out.

The final row describes the data required to compute each method. LIFO subtraction requires mint, burn, and swap data, since it reconstructs LP position changes and allocates swap-level outcomes across active and passive liquidity. The infinitesimal LP method requires only swap data, or more precisely, the price path for a pool obtained from its trade history, since it computes the benchmark from the realized sequence of pool prices and swap quantities rather than from observed LP mint and burn behavior. In this regard, for computing passive LP outcomes, the infinitesimal method has a substantial practical advantage in terms of the data and complexity required to implement it relative to the LIFO subtraction method.
\fi

\begin{table}[thb]
\centering
% \footnotesize
\scriptsize
\setlength{\tabcolsep}{4pt}
\begin{tabular}{lp{3.6cm}p{3.8cm}}
\toprule
Attribute & LIFO subtraction & Infinitesimal LP \\
\midrule
Profiles passive LPs & Yes & Yes \\
Profiles active LPs & Yes & No \\
LP decomposition granularity & Swap-level & Swap-level passive only \\
Passive markout CI & Yes & Yes \\
Active markout CI & Yes & No \\
Dollar scale of passive P\&L & Yes & No; only normalized P\&L \\
Assumptions on active LP behavior & Yes & No \\
Robust to sandwiches & Yes & No \\
Data required to compute & Mint, burn, and swap history & Time series of pool spot prices \\
\bottomrule
\end{tabular}
\caption{Comparison of the two passive-LP methodologies.}
\label{tab:method_comparison}
\end{table}

\paraheader{Evaluating passive LP performance.}
{\color{blue}We use two time horizons in the empirical implementation. 
The mint-burn matching horizon $T$ determines which liquidity positions are classified as active, while the markout horizon $h$ determines when swap profitability is evaluated relative to the external market.}

On Ethereum mainnet, where block times are approximately 12 seconds, almost all the matching happens in the same block, since active LP behavior is primarily concentrated within a single block. 
On L2 chains, where block times are much shorter, {\color{blue}i.e., Arbitrum producing blocks at about 250 millisecond cadence and Base approximately every two seconds and additionally providing Flashblock preconfirmations every 200 milliseconds}, active LP positions often span multiple consecutive blocks; we therefore allow matching for all chains over the subsequent 20 seconds. 
{\color{blue}We run the analysis for 60 seconds matching horizon for robustness check and observe there is no difference in the results. Results can be found in Appendix~\ref{app:matchrobust}.}

{\color{blue}For the infinitesimal method, we reconstruct the complete sequence of valid pre and post swap pool prices and then exclude swaps whose absolute internal pool price movement exceeds 20\% or have been detected as sandwiches or victims of the sandwich attack. 
\ifshort
Both standard sandwiches (attacker frontrun/backrun around victim swaps) and same-transaction sandwiches are identified and removed---attacker legs and victim swaps alike---because the temporary mechanical price reversal can produce disproportionately large infinitesimal markouts; the identification rules are detailed in Appendix~\ref{app:methoddetails}.}
\else
We have two types of sandwich attacks. 
Standard sandwiches are identified as front running and back running swaps attributed to the same observed attacker around one or more victim transactions in the same pool. 
Same transaction sandwiches consist of a corresponding front run, victim, and back run sequence contained within a single transaction hash and identified using log order. 
We remove the attacker legs and the associated victim swaps because the temporary mechanical reversal in the pool price can produce disproportionately large infinitesimal markouts.
\fi

For all the methods, we report the aggregate passive markout normalized by aggregate passive volume, expressed in basis points:
\begin{gather*}
    M^{\mathrm{pass}}
    =
    \frac{\sum_s M_s^{\mathrm{pass}}}{\sum_s V_s^{\mathrm{pass}}},
\end{gather*}
where $M_s^{\mathrm{pass}}$ and $V_s^{\mathrm{pass}}$ denote passive dollar markout and passive dollar volume, respectively. 
For the subtraction methods, passive volume is defined as total volume minus active volume. 
Under LIFO this subtraction is done at the swap level before aggregation and under the infinitesimal method, $V_s^{\mathrm{pass}}$ is the simulated infinitesimal-LP volume for swap $s$. 
Thus, although the construction of passive quantities differs across methods, the reported normalized passive markout is the same aggregate volume-weighted object in all cases.

{\color{blue}The markout horizon determines how much subsequent price movement is incorporated into the benchmark.
We use a common horizon of 15 seconds across all chains to measure performance shortly after each swap.
The horizon is measured in elapsed time rather than blocks, so differences in block times do not mechanically change the valuation horizon.
This provides a consistent basis for comparing markouts across chains without assuming that price discovery proceeds identically on each chain.
Appendix~\ref{app:horizonrobust} examines horizons from 0 seconds to 6 hours and reports analytical and day-block bootstrap confidence intervals.
The plotted passive LIFO estimates are generally more stable between 15 seconds and one hour than at multi-hour horizons, although sensitivity varies across pools.}

\ifshort
Confidence intervals are reported for the LIFO and infinitesimal methods, which produce passive observations at the swap level: we take the volume-weighted mean of the per-swap normalized passive markout and report the normal-approximation interval $\bar m^{\mathrm{pass}} \pm 1.96\,\mathrm{SE}(\bar m^{\mathrm{pass}})$. The construction is detailed in Appendix~\ref{app:methoddetails}.
\else
% Confidence-interval construction for passive markouts.
% Inline under Methodology in the AFT/LIPIcs build; relocated to the appendix in WINE.
Confidence intervals are reported for the LIFO and infinitesimal methods, since these produce passive observations at the swap level. Specifically, for each swap $s$ we first compute the normalized passive markout
\begin{gather*}
    m_s^{\mathrm{pass}}=\frac{M_s^{\mathrm{pass}}}{V_s^{\mathrm{pass}}}\,,
\end{gather*}
and use passive volume $V_s^{\mathrm{pass}}$ as the weight.
% We then form the weighted mean and weighted standard error across swaps and report the normal-approximation interval
% \begin{gather*}
%     \bar m^{\mathrm{pass}} \pm 1.96\,\mathrm{SE}(\bar m^{\mathrm{pass}}).
% \end{gather*}
We then form the volume weighted mean and its linearized (ratio estimator) standard error,
\begin{gather*}
\bar m^{\mathrm{pass}}
= \frac{\sum_s V_s^{\mathrm{pass}}\, m_s^{\mathrm{pass}}}{\sum_s V_s^{\mathrm{pass}}}\,,
\qquad
\mathrm{SE}\!\left(\bar m^{\mathrm{pass}}\right)^2
= \frac{\sum_s \left(V_s^{\mathrm{pass}}\right)^2
        \left(m_s^{\mathrm{pass}} - \bar m^{\mathrm{pass}}\right)^2}
       {\left(\sum_s V_s^{\mathrm{pass}}\right)^2}\,,
\end{gather*}
and report the normal approximation interval $\bar m^{\mathrm{pass}} \pm 1.96\,\mathrm{SE}(\bar m^{\mathrm{pass}})$.

For the difference between the overall and passive markout, $D = \bar m^{\mathrm{all}} - \bar m^{\mathrm{pass}}$, the two ratios are built from the same swaps and are therefore correlated, so their variances do not add. 
We instead linearize the difference directly: defining the per-swap influence contribution
\begin{gather*}
d_s
= \frac{V_s\left(m_s - \bar m^{\mathrm{all}}\right)}{\sum_s V_s}
- \frac{V_s^{\mathrm{pass}}\left(m_s^{\mathrm{pass}} - \bar m^{\mathrm{pass}}\right)}{\sum_s V_s^{\mathrm{pass}}}\,,
\qquad
\mathrm{SE}(D)^2 = \sum_s d_s^2\, .
\end{gather*}
We report $D \pm 1.96\,\mathrm{SE}(D)$\,.
\fi

\section{Results}\label{sec:result}
\ifshort
We first summarize the composition of the sample by Uniswap versions, chains, pairs and fee tiers, documenting where trading volume, liquidity events, sandwich activity and active LIFO volume are concentrated. 
We then compare aggregate pool-level markouts with the passive LP markouts of the LIFO subtraction and infinitesimal methods.
\else
We present our results in two sections. 
In the first part, we summarize the composition of the sample by Uniswap versions, chains, pairs and fee tiers, documenting where trading volume, liquidity events, sandwich activity and active LIFO volume are concentrated. 
Second, we conduct an analysis of the LP profitability using the markout-based decomposition. 
We compare the aggregate pool-level markouts with the passive LP markouts of the LIFO subtraction and infinitesimal methods. 
Such structure separates descriptive patterns of market activity from the central economic question of how much aggregate pool profitability differs from the profitability of passive liquidity provision.
\fi

% Note that we calculate markouts relative to the benchmark price 15 seconds after the swap. This horizon captures the price adjustment in the short-term post-trade period, but leaves enough time for the trade information to be incorporated into the external reference market. Ethereum block times are on the order of 12 seconds, so the 15 seconds horizon is roughly equal to one Ethereum block plus a small propagation/finality buffer.

% \begin{itemize}
%     \item Descriptive Statistics and LP Activity Overview: How much of volume is active vs passive per pool?; Cross-chain comparison of active LP prevalence

%     \item The Passive LP Gap: Method A results: infinitesimal bp; Method B results: passive markout lifo/ passive vol lifo; Method C results: passive markout exact/ passive vol exact ; Cross-method comparison

%     \item P\&L and Volume Decomposition: Share of total volume: active vs passive; Share of total markout P\&L: active vs passive; Is the active share higher on L2s?

%     \item Drivers of the Passive LP Gap: Cross-pool and cross-chain variation in the gap; Candidate drivers: fee tier, volatility, block time, gas cost, asset type; Structured empirical analysis (regression or comparison framework)

% \end{itemize}

\subsection{Descriptive Statistics}\label{subsec:res_descriptive}
\ifshort
Trading volume is concentrated in WETH-stablecoin pools across all three protocol versions, with the largest v3 pools (USDC-WETH on Arbitrum and Ethereum at the 5 bps tier) exceeding \$45B each. Active LIFO volume is essentially zero in full-range v2, becomes pronounced in v3 and v4, and is highly concentrated in a small set of pools. Sandwich activity is concentrated almost exclusively in low-fee Ethereum pools (above 50\% of volume in the 1 bps USDC-WETH and USDT-WETH pools), so these pools should be interpreted separately. Full per-version volume-share tables and details for v2 and v4 are reported in Appendix~\ref{app:descriptive}.
\else
% Descriptive-statistics block. Rendered inline under Results in the AFT/LIPIcs
% build; relocated to the appendix in the WINE build. No top-level heading here
% (the caller supplies \subsection or \section as appropriate).
\Cref{tab:v2-volume-shares}--\ref{tab:v4-volume-shares} summarize trading activity, liquidity-event intensity, sandwich volume, and active LIFO volume for the v2, v3, and v4 Uniswap protocol versions. Trading volume is quite concentrated in WETH-stablecoin pools. V2’s reported volume comes mainly from the Ethereum USDT-WETH and USDC-WETH pools, which stood at around \$1.04B and \$968M respectively. The largest pools in v3 are USDC-WETH on Arbitrum and Ethereum at the 5 bps tier with \$61.13B and \$46.84B in volume, respectively, followed by Ethereum USDT-WETH at the 1 bps and 5 bps tiers. While v4 volumes are smaller than the mature v3 pools, they are still concentrated in Ethereum WETH-stablecoin markets, especially USDC-WETH and USDT-WETH (5 bps tier).
% In your main preamble, include:
% \usepackage{pgfplotstable}
% \usepackage{booktabs}
% \usepackage{makecell}
% \usepackage{float}

\begin{table}[H]
\centering
\scriptsize
\setlength{\tabcolsep}{5pt}
\renewcommand{\arraystretch}{1.08}

\begin{filecontents*}{sections/updated/v2_table_volume_shares.csv}
pool_label;num_trades;num_liquidity_events;total_volume;sandwich_volume_pct;active_lifo_volume_pct
\textbf{USDC-WETH};;;;;
Base 30 bps;2,463,854;2,464,144;515.67M;0.03;0.00
Arbitrum 30 bps;131,597;131,612;11.49M;0.22;0.00
Ethereum 30 bps;560,857;561,059;967.72M;1.47;0.00
\textbf{USDT-WETH};;;;;
Arbitrum 30 bps;97,124;97,744;10.66M;0.12;0.00
Ethereum 30 bps;525,709;525,854;1.04B;0.62;0.00
\end{filecontents*}
\pgfplotstabletypeset[
    col sep=semicolon,
    string type,
    columns={pool_label,num_trades,num_liquidity_events,total_volume,sandwich_volume_pct,active_lifo_volume_pct},
    columns/pool_label/.style={column name={\makecell[l]{v2 pool\\label}}, column type={l}},
    columns/num_trades/.style={column name={\makecell[c]{number of\\trades}}, column type={r}},
    columns/num_liquidity_events/.style={column name={\makecell[c]{number of\\liquidity events}}, column type={r}},
    columns/total_volume/.style={column name={\makecell[c]{total\\volume (\$)}}, column type={r}},
    columns/sandwich_volume_pct/.style={column name={\makecell[c]{sandwich\\volume (\%)}}, column type={r}},
    columns/active_lifo_volume_pct/.style={column name={\makecell[c]{active LIFO\\volume (\%)}}, column type={r}},
    every head row/.style={before row=\toprule, after row=\midrule},
    every row no 4/.style={before row=\midrule},
    every last row/.style={after row=\bottomrule},
]{sections/updated/v2_table_volume_shares.csv}

\caption{Uniswap v2 pool-level volume shares.}
\label{tab:v2-volume-shares}
\end{table}

Liquidity-event activity fluctuates dramatically with protocol version. In v2, active LIFO volume is basically zero across all pools. 
{\color{blue}Note that Uniswap v2 does not permit range specific liquidity placement, but LPs can mint and burn in full-range liquidity. Our statement that active liquidity is absent in the selected v2 pools is therefore an empirical conclusion under the LIFO definition rather than a general restriction imposed by the protocol. Specifically, within the matching horizons used in the analysis, the LIFO procedure identifies no mint and burn sequences satisfying the active liquidity criterion. The estimated active share is consequently zero, and the passive LIFO markout equals the aggregate pool markout.}

Liquidity management is much more pronounced in v3, especially in the Base and Arbitrum USDC-WETH pools with millions of liquidity events. v4 also has high event intensity, with liquidity-event counts often close to swap counts, reflecting the more granular event structure and presence of actively managed positions in some pools.

% In your main preamble, include:
% \usepackage{pgfplotstable}
% \usepackage{booktabs}
% \usepackage{makecell}

\begin{table}[!htbp]
\centering
\scriptsize
\setlength{\tabcolsep}{5pt}
\renewcommand{\arraystretch}{1.08}

\begin{filecontents*}{sections/updated/v3_table_volume_shares.csv}
pool_label;num_trades;num_liquidity_events;total_volume;sandwich_volume_pct;active_lifo_volume_pct
\textbf{USDC-WETH};;;;;
Base 5 bps;15,265,500;4,525,073;21.68B;0.41;3.56
Arbitrum 5 bps;12,369,562;2,325,108;61.13B;0.24;0.01
Ethereum 1 bps;3,540,909;44,004;22.21B;55.94;0.32
Ethereum 5 bps;1,860,687;152,851;46.84B;3.09;1.30
Ethereum 30 bps;176,519;44,000;6.19B;0.46;0.67
\textbf{USDT-WETH};;;;;
Base 5 bps;1,426,442;25,781;91.41M;0.14;0.00
Arbitrum 5 bps;6,462,674;317,030;8.20B;0.27;0.00
Ethereum 1 bps;3,696,392;62,839;30.24B;52.60;0.42
Ethereum 5 bps;1,200,301;62,141;16.98B;4.78;1.16
Ethereum 30 bps;276,561;123,397;12.35B;0.40;0.92
\textbf{USDC-cbBTC};;;;;
Base 5 bps;1,501,129;249,242;4.44B;0.16;0.00
\textbf{USDC-WBTC};;;;;
Arbitrum 5 bps;2,542,725;178,390;6.01B;0.10;0.00
Ethereum 5 bps;382,480;19,860;5.42B;4.88;1.78
Ethereum 30 bps;100,393;43,438;9.18B;0.24;1.51
\textbf{USDT-WBTC};;;;;
Arbitrum 5 bps;2,341,470;206,250;5.33B;0.05;0.00
Ethereum 5 bps;437,375;18,679;7.95B;4.46;1.14
Ethereum 30 bps;65,243;9,282;2.40B;0.32;0.79
\end{filecontents*}
\pgfplotstabletypeset[
    col sep=semicolon,
    string type,
    columns={pool_label,num_trades,num_liquidity_events,total_volume,sandwich_volume_pct,active_lifo_volume_pct},
    columns/pool_label/.style={column name={\makecell[l]{v3 pool\\label}}, column type={l}},
    columns/num_trades/.style={column name={\makecell[c]{number of\\trades}}, column type={r}},
    columns/num_liquidity_events/.style={column name={\makecell[c]{number of\\liquidity events}}, column type={r}},
    columns/total_volume/.style={column name={\makecell[c]{total\\volume (\$)}}, column type={r}},
    columns/sandwich_volume_pct/.style={column name={\makecell[c]{sandwich\\volume (\%)}}, column type={r}},
    columns/active_lifo_volume_pct/.style={column name={\makecell[c]{active LIFO\\volume (\%)}}, column type={r}},
    every head row/.style={before row=\toprule, after row=\midrule},
    every row no 6/.style={before row=\midrule},
    every row no 12/.style={before row=\midrule},
    every row no 14/.style={before row=\midrule},
    every row no 18/.style={before row=\midrule},
    every last row/.style={after row=\bottomrule},
]{sections/updated/v3_table_volume_shares.csv}

\caption{Uniswap v3 pool-level volume shares.}
\label{tab:v3-volume-shares}
\end{table}

% In your main preamble, include:
% \usepackage{pgfplotstable}
% \usepackage{booktabs}
% \usepackage{makecell}

\begin{table}[!htbp]
\centering
\scriptsize
\setlength{\tabcolsep}{5pt}
\renewcommand{\arraystretch}{1.08}

\begin{filecontents*}{sections/updated/v4_table_volume_shares.csv}
pool_label;num_trades;num_liquidity_events;total_volume;sandwich_volume_pct;active_lifo_volume_pct
\textbf{USDC-WETH};;;;;
Base 5 bps;2,303,175;2,305,263;529.85M;0.19;0.00
Arbitrum 5 bps;1,661,317;1,661,675;2.01B;0.04;0.00
Ethereum 1 bps;684,725;713,579;4.20B;47.65;0.17
Ethereum 5 bps;571,770;575,688;10.07B;1.61;0.13
Ethereum 30 bps;79,471;80,182;1.16B;0.39;0.11
\textbf{USDT-WETH};;;;;
Arbitrum 5 bps;1,140,885;1,141,370;496.08M;0.09;41.99
Ethereum 1 bps;167,133;185,815;322.33M;76.12;0.01
Ethereum 5 bps;568,893;573,825;8.19B;3.54;0.36
Ethereum 30 bps;41,258;41,376;288.42M;0.29;0.08
\textbf{USDC-cbBTC};;;;;
Base 5 bps;481,175;481,260;208.45M;0.02;84.08
\textbf{USDC-WBTC};;;;;
Arbitrum 5 bps;824,319;824,421;1.41B;0.04;0.00
Ethereum 5 bps;169,207;170,603;1.89B;4.66;0.23
Ethereum 30 bps;23,394;23,484;410.02M;0.86;0.07
\textbf{USDT-WBTC};;;;;
Arbitrum 5 bps;213,931;213,954;125.05M;0.02;69.42
Ethereum 5 bps;59,648;60,306;239.31M;5.84;0.10
Ethereum 30 bps;36,844;37,031;1.19B;0.96;0.13
\end{filecontents*}
\pgfplotstabletypeset[
    col sep=semicolon,
    string type,
    columns={pool_label,num_trades,num_liquidity_events,total_volume,sandwich_volume_pct,active_lifo_volume_pct},
    columns/pool_label/.style={column name={\makecell[l]{v4 pool\\label}}, column type={l}},
    columns/num_trades/.style={column name={\makecell[c]{number of\\trades}}, column type={r}},
    columns/num_liquidity_events/.style={column name={\makecell[c]{number of\\liquidity events}}, column type={r}},
    columns/total_volume/.style={column name={\makecell[c]{total\\volume (\$)}}, column type={r}},
    columns/sandwich_volume_pct/.style={column name={\makecell[c]{sandwich\\volume (\%)}}, column type={r}},
    columns/active_lifo_volume_pct/.style={column name={\makecell[c]{active LIFO\\volume (\%)}}, column type={r}},
    every head row/.style={before row=\toprule, after row=\midrule},
    every row no 6/.style={before row=\midrule},
    every row no 11/.style={before row=\midrule},
    every row no 13/.style={before row=\midrule},
    every row no 17/.style={before row=\midrule},
    every last row/.style={after row=\bottomrule},
]{sections/updated/v4_table_volume_shares.csv}

\caption{Uniswap v4 pool-level volume shares.}
\label{tab:v4-volume-shares}
\end{table}

The sandwich activity is concentrated almost exclusively on Ethereum pools with low fees. Sandwich volume shares above 50\% are evident in v3 Ethereum 1 bps USDC-WETH and USDT-WETH pools, while most L2 pools and higher fee Ethereum pools have sandwich shares less than 1\%. V4 follows a similar pattern with Ethereum 1 bps USDC-WETH and USDT-WETH having sandwich shares of 47.65\% and 76.12\%, respectively. The concentration implies that when interpreting markout based passive LP estimates, low fee Ethereum pools should be separated.

Active LIFO volume is also very concentrated. The biggest active share in v3 is Base USDC-WETH 5 bps with 3.56\% and there are a few Ethereum 5 bps and 30 bps pools with active share of 1-2\% range. Arbitrum v3 pools typically have active shares near zero. In v4, most pools on Ethereum have small active LIFO shares, but a few L2 pools have very large active shares, e.g. Arbitrum USDT-WETH, Base USDC-cbBTC, and Arbitrum USDT-WBTC. Outliers here suggest that active liquidity provision in v4 is pool specific and may be a function of specialized LP strategies or pool-level design features rather than a general active LP participation across markets.

\fi

\subsection{Markout-Based LP Profitability Decomposition}\label{subsec:res_lp_prof}
Total dollar markouts, total markouts per unit of total volume, and passive LP markouts under the LIFO and infinitesimal methods are presented in 
% \ifshort
% \Cref{tab:v2-markout-estimates}--\ref{tab:v4-markout-estimates} (Appendix~\ref{app:makrout24}).
% \else
\Cref{tab:v2-markout-estimates}--\ref{tab:v4-markout-estimates} 
% \fi
The clearest pattern is that v2 is a benchmark case: overall and passive LIFO markouts are nearly identical across all pools, consistent with the lack of concentrated active liquidity provision. 
The infinitesimal estimates are also close to the LIFO estimates for the larger v2 pools, indicating that when LPs are effectively homogeneous, aggregate pool profitability is informative for passive LP profitability.
% \ifshort
% \else
% In your main preamble, include:
% \usepackage{pgfplotstable}
% \usepackage{booktabs}
% \usepackage{makecell}
% \usepackage{float}

\begin{table}[H]
\centering
\scriptsize
\setlength{\tabcolsep}{4pt}
\renewcommand{\arraystretch}{1.08}

\begin{filecontents*}{sections/updated/v2_table_markout_estimates.csv}
pool_label;total_markout;overall_bps;passive_lifo_bps;infinitesimal_bps;passive_lifo_ci;infinitesimal_ci
\textbf{USDC-WETH};;;;;;
Base 30 bps;173,731;20.822100;20.822080;20.866623;[18.87,22.78];[19.01,22.73]
Arbitrum 30 bps;22,013;19.153800;19.153833;18.283203;[9.33,28.98];[9.15,27.42]
Ethereum 30 bps;901,955;9.320400;9.320412;8.6150225;[3.23,15.41];[3.5,13.73]
\textbf{USDT-WETH};;;;;;
Arbitrum 30 bps;22,895;21.486400;21.486375;17.455259;[-1.78,44.75];[9.76,25.15]
Ethereum 30 bps;973,349;9.329900;9.329889;7.6290297;[-7.18,25.85];[-7.93,23.19]
\end{filecontents*}
\pgfplotstabletypeset[
    col sep=semicolon,
    string type,
    columns={pool_label,total_markout,overall_bps,passive_lifo_bps,infinitesimal_bps,passive_lifo_ci,infinitesimal_ci},
    columns/pool_label/.style={column name={\makecell[l]{v2 pool\\label}}, column type={l}},
    columns/total_markout/.style={column name={\makecell[c]{total\\(\$)}}, column type={r}},
    columns/overall_bps/.style={column name={\makecell[c]{overall\\(bps)}}, column type={r}},
    columns/passive_lifo_bps/.style={column name={\makecell[c]{passive LIFO\\(bps)}}, column type={r}},
    columns/infinitesimal_bps/.style={column name={\makecell[c]{infinitesimal\\(bps)}}, column type={r}},
    columns/passive_lifo_ci/.style={column name={\makecell[c]{passive LIFO\\CI (bps)}}, column type={r}},
    columns/infinitesimal_ci/.style={column name={\makecell[c]{infinitesimal\\CI (bps)}}, column type={r}},
    every head row/.style={before row=\toprule, after row=\midrule},
    every row no 4/.style={before row=\midrule},
    every last row/.style={after row=\bottomrule},
]{sections/updated/v2_table_markout_estimates.csv}

\caption{Uniswap v2 pool-level total markout (\$), overall markout per total volume (bps), and passive markout per passive volume (bps) for the LIFO and infinitesimal methods, with confidence intervals for the LIFO and infinitesimal estimates.}
\label{tab:v2-markout-estimates}
\end{table}

% \fi

The results in v3 and v4 are more heterogeneous. Aggregate pool-level profitability can mask passive liquidity performance, as passive LIFO markouts are often different from overall markouts. 
For some of the Ethereum v3 pools the passive LIFO is much worse than the overall markout. 
For example USDC-WETH 5 bps goes from $-1.15$ bps overall to $-1.51$ bps for passive LIFO and USDT-WETH 5 bps goes from $-0.91$ bps to $-1.31$ bps. 
Similar patterns are seen in BTC-stable pools where aggregate markouts can be close to zero or positive while passive LIFO markouts are negative.

% \ifshort
% \else
% In your main preamble, include:
% \usepackage{pgfplotstable}
% \usepackage{booktabs}
% \usepackage{makecell}

\begin{table}[!htbp]
\centering
\scriptsize
\setlength{\tabcolsep}{4pt}
\renewcommand{\arraystretch}{1.08}

\begin{filecontents*}{sections/updated/v3_table_markout_estimates.csv}
pool_label;total_markout;overall_bps;passive_lifo_bps;infinitesimal_bps;passive_lifo_ci;infinitesimal_ci
\textbf{USDC-WETH};;;;;;
Base 5 bps;-1,476,525;-0.681176;-0.706874;-0.990307;[-0.81,-0.6];[-1.79,-0.19]
Arbitrum 5 bps;-4,477,895;-0.732521;-0.769710;-1.670802;[-6.08,4.54];[-1.95,-1.4]
Ethereum 1 bps;1,052,140;0.473692;-0.000559;-1.755563;[-18.38,18.38];[-2.52,-0.99]
Ethereum 5 bps;-5,388,028;-1.150404;-1.506019;-1.229423;[-2,-1.02];[-2.27,-0.19]
Ethereum 30 bps;-438,047;-0.707246;-0.866415;0.171106;[-2.63,0.9];[-5.31,5.66]
\textbf{USDT-WETH};;;;;;
Base 5 bps;-5,256;-0.574968;-0.574969;-1.011193;[-1.34,0.19];[-2.22,0.2]
Arbitrum 5 bps;-368,878;-0.449665;-0.449636;-1.350178;[-0.86,-0.04];[-1.74,-0.96]
Ethereum 1 bps;1,097,440;0.362955;-0.138684;-1.756175;[-6.87,6.59];[-2.69,-0.83]
Ethereum 5 bps;-1,546,285;-0.910758;-1.309165;-1.535256;[-4.42,1.8];[-2.61,-0.46]
Ethereum 30 bps;1,464,532;1.185598;1.001649;2.047672;[-0.71,2.71];[-5.5,9.6]
\textbf{USDC-cbBTC};;;;;;
Base 5 bps;-252,298;-0.568330;-0.568331;-0.625847;[-0.77,-0.37];[-1.32,0.06]
\textbf{USDC-WBTC};;;;;;
Arbitrum 5 bps;-354,545;-0.589463;-0.589305;-1.474014;[-12.73,11.55];[-2.51,-0.44]
Ethereum 5 bps;24,703;0.045536;-0.408402;-1.278491;[-5.49,4.67];[-2.76,0.21]
Ethereum 30 bps;481,238;0.524304;0.209764;9.906399;[-1.4,1.82];[-1.72,21.54]
\textbf{USDT-WBTC};;;;;;
Arbitrum 5 bps;-384,921;-0.722053;-0.722146;-1.279479;[-0.85,-0.6];[-1.71,-0.85]
Ethereum 5 bps;151,553;0.190694;-0.254329;-0.247134;[-1.76,1.29];[-2.69,2.19]
Ethereum 30 bps;649,333;2.705426;2.519784;3.164769;[0.51,4.53];[-2.29,8.62]
\end{filecontents*}
\pgfplotstabletypeset[
    col sep=semicolon,
    string type,
    columns={pool_label,total_markout,overall_bps,passive_lifo_bps,infinitesimal_bps,passive_lifo_ci,infinitesimal_ci},
    columns/pool_label/.style={column name={\makecell[l]{v3 pool\\label}}, column type={l}},
    columns/total_markout/.style={column name={\makecell[c]{total\\(\$)}}, column type={r}},
    columns/overall_bps/.style={column name={\makecell[c]{overall\\(bps)}}, column type={r}},
    columns/passive_lifo_bps/.style={column name={\makecell[c]{passive LIFO\\(bps)}}, column type={r}},
    columns/infinitesimal_bps/.style={column name={\makecell[c]{infinitesimal\\(bps)}}, column type={r}},
    columns/passive_lifo_ci/.style={column name={\makecell[c]{passive LIFO\\CI (bps)}}, column type={r}},
    columns/infinitesimal_ci/.style={column name={\makecell[c]{infinitesimal\\CI (bps)}}, column type={r}},
    every head row/.style={before row=\toprule, after row=\midrule},
    every row no 6/.style={before row=\midrule},
    every row no 12/.style={before row=\midrule},
    every row no 14/.style={before row=\midrule},
    every row no 18/.style={before row=\midrule},
    every last row/.style={after row=\bottomrule},
]{sections/updated/v3_table_markout_estimates.csv}

\caption{Uniswap v3 pool-level total markout (\$), overall markout per total volume (bps), and passive markout per passive volume (bps) for the LIFO and infinitesimal methods, with confidence intervals.}
\label{tab:v3-markout-estimates}
\end{table}

% In your main preamble, include:
% \usepackage{pgfplotstable}
% \usepackage{booktabs}
% \usepackage{makecell}

\begin{table}[!htbp]
\centering
\scriptsize
\setlength{\tabcolsep}{4pt}
\renewcommand{\arraystretch}{1.08}

\begin{filecontents*}{sections/updated/v4_table_markout_estimates.csv}
pool_label;total_markout;overall_bps;passive_lifo_bps;infinitesimal_bps;passive_lifo_ci;infinitesimal_ci
\textbf{USDC-WETH};;;;;;
Base 5 bps;-51,811;-0.977900;-0.977859;-0.868197;[-1.13,-0.82];[-1.14,-0.6]
Arbitrum 5 bps;-288,235;-1.431700;-1.431670;-1.875089;[-1.5,-1.37];[-2.16,-1.59]
Ethereum 1 bps;-17,974;-0.042800;-0.427445;-2.612735;[-15.61,14.76];[-5.42,0.2]
Ethereum 5 bps;-3,054,024;-3.032900;-3.070001;-2.826631;[-3.99,-2.15];[-3.97,-1.68]
Ethereum 30 bps;-101,573;-0.876000;-0.909249;-1.908843;[-7.63,5.81];[-10.23,6.41]
\textbf{USDT-WETH};;;;;;
Arbitrum 5 bps;-85,126;-1.716000;-1.972888;-1.918914;[-2.32,-1.62];[-2.51,-1.32]
Ethereum 1 bps;2,412;0.074800;-0.084154;-8.306552;[-48.2,48.03];[-11.77,-4.84]
Ethereum 5 bps;-1,322,337;-1.615100;-1.756679;-2.235749;[-2.29,-1.23];[-3.21,-1.26]
Ethereum 30 bps;-87,112;-3.020300;-3.044381;-7.365738;[-4.1,-1.98];[-11.24,-3.49]
\textbf{USDC-cbBTC};;;;;;
Base 5 bps;-19,013;-0.912100;-0.365854;-1.204660;[-0.83,0.1];[-3.71,1.3]
\textbf{USDC-WBTC};;;;;;
Arbitrum 5 bps;-182,076;-1.287200;-1.287180;-1.716737;[-1.51,-1.06];[-2.37,-1.06]
Ethereum 5 bps;-187,952;-0.993700;-1.278666;-0.922808;[-3.99,1.43];[-2.58,0.74]
Ethereum 30 bps;-82,145;-2.003400;-2.020871;-3.119593;[-6.02,1.98];[-17.11,10.87]
\textbf{USDT-WBTC};;;;;;
Arbitrum 5 bps;-30,895;-2.470600;-2.788936;-3.304469;[-3.16,-2.42];[-4.24,-2.37]
Ethereum 5 bps;-4,452;-0.186100;-1.147261;-6.655240;[-6.75,4.46];[-10.23,-3.08]
Ethereum 30 bps;15,042;0.126500;0.016537;-0.340322;[-2.38,2.42];[-34.75,34.07]
\end{filecontents*}
\pgfplotstabletypeset[
    col sep=semicolon,
    string type,
    columns={pool_label,total_markout,overall_bps,passive_lifo_bps,infinitesimal_bps,passive_lifo_ci,infinitesimal_ci},
    columns/pool_label/.style={column name={\makecell[l]{v4 pool\\label}}, column type={l}},
    columns/total_markout/.style={column name={\makecell[c]{total\\(\$)}}, column type={r}},
    columns/overall_bps/.style={column name={\makecell[c]{overall\\(bps)}}, column type={r}},
    columns/passive_lifo_bps/.style={column name={\makecell[c]{passive LIFO\\(bps)}}, column type={r}},
    columns/infinitesimal_bps/.style={column name={\makecell[c]{infinitesimal\\(bps)}}, column type={r}},
    columns/passive_lifo_ci/.style={column name={\makecell[c]{passive LIFO\\CI (bps)}}, column type={r}},
    columns/infinitesimal_ci/.style={column name={\makecell[c]{infinitesimal\\CI (bps)}}, column type={r}},
    every head row/.style={before row=\toprule, after row=\midrule},
    every row no 6/.style={before row=\midrule},
    every row no 11/.style={before row=\midrule},
    every row no 13/.style={before row=\midrule},
    every row no 17/.style={before row=\midrule},
    every last row/.style={after row=\bottomrule},
]{sections/updated/v4_table_markout_estimates.csv}

\caption{Uniswap v4 pool-level total markout (\$), overall markout per total volume (bps), and passive markout per passive volume (bps) for the LIFO and infinitesimal methods, with confidence intervals.}
\label{tab:v4-markout-estimates}
\end{table}

% \fi

The direction of the infinitesimal estimates is generally consistent with LIFO, but they tend to be more negative, especially in concentrated-liquidity pools. 
This is especially visible on Arbitrum v3 WETH-stablecoin pools and some v4 pools. 
This difference reflects the fact that the infinitesimal method is measuring a fully passive, always in range marginal LP whereas the LIFO subtraction is measuring the residual passive component after the identified active liquidity has been subtracted. 
Thus, the two methods give complementary views on passive LP exposure: LIFO measures the passive part of realized pool P\&L, and infinitesimal method offers a tighter passive benchmark relative to the AMM price trajectory.

In general, aggregate pool and passive LP profitability align in v2 but diverge in v3 and v4, motivating the plot-based analysis of how the gap depends on active liquidity share, chain, and fee tier.

\paraheader{Overall versus passive LP profitability.}
% Figures~\ref{fig:v3-active-lifo-gap} and \ref{fig:v4-active-lifo-gap} demonstrate that aggregate pool-level profitability can sometimes deviate from passive LP profitability. 
Figures in \ref{fig:gapv3v4} demonstrate that aggregate pool-level profitability can sometimes deviate from passive LP profitability.
The Ethereum 1 bps pools are excluded from these figures because their sandwich activity and markout behavior make them extreme outliers; omitting them improves visualization of the main cross-pool patterns. 
In concentrated-liquidity pools, passive LIFO markouts are on average worse than overall markouts, suggesting that aggregate pool measures may overstate the returns available to passive LPs. 
This is in contrast to the v2 results in \Cref{tab:v2-markout-estimates}, where overall and passive markouts are nearly identical, as expected in the absence of concentrated active liquidity.

The active-passive gap in v3 is most prominent in Ethereum pools, especially at the 5 bps tier, while Arbitrum pools generally have near-zero active LIFO volume and much smaller gaps. 
Active shares are compressed near zero, but range across several orders of magnitude, so in v4 we plot active LIFO volume on a log scale. 
There are a few outliers for v4 pools, particularly Arbitrum USDT-WETH, Base USDC-cbBTC, and Arbitrum USDT-WBTC, all of which have unusually high active LIFO shares. 
This is likely due to pool-specific liquidity management strategies or LPs with specialized behavior. 
In the main, the numbers confirm the key result that when liquidity can be actively repositioned, the performance of passive LPs can materially differ from the aggregate profitability of the pool.
%
% \begin{figure}[!htbp]
%     \centering
%     \gapplot{sections/updated/active_v3_plot.tex}
%     \caption{Uniswap v3 pools overall minus passive LIFO markout by active LIFO volume share.}
%     \label{fig:v3-active-lifo-gap}
% \end{figure}
% %
% \begin{figure}[!htbp]
%     \centering
%     \gapplot{sections/updated/active_v4_plot.tex}
%     \caption{Uniswap v4 pools overall minus passive LIFO markout by active LIFO volume share.}
%     \label{fig:v4-active-lifo-gap}
% \end{figure}

\begin{figure}[t]
\centering
\begin{minipage}{0.49\textwidth}\centering\scalebox{0.45}{\begin{tikzpicture}
\definecolor{pairblue}{RGB}{0,114,178}
\definecolor{pairgreen}{RGB}{0,158,115}
\definecolor{pairorange}{RGB}{230,159,0}
\definecolor{pairred}{RGB}{204,68,102}
\definecolor{pairpurple}{RGB}{136,34,204}

\begin{axis}[
width=13.5cm,
height=7.2cm,
grid=both,
major grid style={draw=gray!25},
minor grid style={draw=gray!10},
title style={font=\small\bfseries, align=center},
xlabel style={font=\scriptsize},
ylabel style={font=\scriptsize},
tick label style={font=\scriptsize},
tick align=outside,
legend style={
    at={(0.5,1.06)},
    anchor=south,
    legend columns=5,
    draw=none,
    fill=none,
    font=\tiny,
    column sep=0.08cm,
    /tikz/every even column/.append style={column sep=0.08cm}
},
ylabel={Overall $-$ Passive LIFO (bps)},
xlabel={Active LIFO Volume (\%)},
xmin=0,
xmax=3.6,
]

% -------------------------
% ETH-USDC
% -------------------------

% L1
\addplot+[
only marks,
forget plot,
mark=*,
mark size=2.6pt,
color=pairblue,
line width=0.8pt,
mark options={fill=pairblue, draw=pairblue, fill opacity=0.42, draw opacity=0.9, solid}
] coordinates {
(1.3,0.355615)
};

% L2
\addplot+[
only marks,
forget plot,
mark=*,
mark size=2.6pt,
color=pairblue,
line width=0.8pt,
mark options={fill=pairblue, draw=pairblue, fill opacity=0.42, draw opacity=0.9, dash pattern=on 1pt off 1pt}
] coordinates {
(3.56,0.025698)
(0.01,0.037189)
};

% L1 / 30 bps larger marker
\addplot+[
only marks,
forget plot,
mark=*,
mark size=4.0pt,
color=pairblue,
line width=0.8pt,
mark options={fill=pairblue, draw=pairblue, fill opacity=0.42, draw opacity=0.9, solid}
] coordinates {
(0.67,0.159169)
};

% -------------------------
% ETH-USDT
% -------------------------

% L1
\addplot+[
only marks,
forget plot,
mark=*,
mark size=2.6pt,
color=pairgreen,
line width=0.8pt,
mark options={fill=pairgreen, draw=pairgreen, fill opacity=0.42, draw opacity=0.9, solid}
] coordinates {
(1.16,0.398407)
};

% L2
\addplot+[
only marks,
forget plot,
mark=*,
mark size=2.6pt,
color=pairgreen,
line width=0.8pt,
mark options={fill=pairgreen, draw=pairgreen, fill opacity=0.42, draw opacity=0.9, dash pattern=on 1pt off 1pt}
] coordinates {
(0,1e-06)
(0,-2.9e-05)
};

% L1 / 30 bps larger marker
\addplot+[
only marks,
forget plot,
mark=*,
mark size=4.0pt,
color=pairgreen,
line width=0.8pt,
mark options={fill=pairgreen, draw=pairgreen, fill opacity=0.42, draw opacity=0.9, solid}
] coordinates {
(0.92,0.183949)
};

% -------------------------
% USDC-cbBTC
% -------------------------

% L2
\addplot+[
only marks,
forget plot,
mark=*,
mark size=2.6pt,
color=pairorange,
line width=0.8pt,
mark options={fill=pairorange, draw=pairorange, fill opacity=0.42, draw opacity=0.9, dash pattern=on 1pt off 1pt}
] coordinates {
(0,1e-06)
};

% -------------------------
% WBTC-USDC
% -------------------------

% L1
\addplot+[
only marks,
forget plot,
mark=*,
mark size=2.6pt,
color=pairred,
line width=0.8pt,
mark options={fill=pairred, draw=pairred, fill opacity=0.42, draw opacity=0.9, solid}
] coordinates {
(1.78,0.453938)
};

% L2
\addplot+[
only marks,
forget plot,
mark=*,
mark size=2.6pt,
color=pairred,
line width=0.8pt,
mark options={fill=pairred, draw=pairred, fill opacity=0.42, draw opacity=0.9, dash pattern=on 1pt off 1pt}
] coordinates {
(0,-0.000158)
};

% L1 / 30 bps larger marker
\addplot+[
only marks,
forget plot,
mark=*,
mark size=4.0pt,
color=pairred,
line width=0.8pt,
mark options={fill=pairred, draw=pairred, fill opacity=0.42, draw opacity=0.9, solid}
] coordinates {
(1.51,0.31454)
};

% -------------------------
% WBTC-USDT
% -------------------------

% L1
\addplot+[
only marks,
forget plot,
mark=*,
mark size=2.6pt,
color=pairpurple,
line width=0.8pt,
mark options={fill=pairpurple, draw=pairpurple, fill opacity=0.42, draw opacity=0.9, solid}
] coordinates {
(1.14,0.445023)
};

% L2
\addplot+[
only marks,
forget plot,
mark=*,
mark size=2.6pt,
color=pairpurple,
line width=0.8pt,
mark options={fill=pairpurple, draw=pairpurple, fill opacity=0.42, draw opacity=0.9, dash pattern=on 1pt off 1pt}
] coordinates {
(0,9.3e-05)
};

% L1 / 30 bps larger marker
\addplot+[
only marks,
forget plot,
mark=*,
mark size=4.0pt,
color=pairpurple,
line width=0.8pt,
mark options={fill=pairpurple, draw=pairpurple, fill opacity=0.42, draw opacity=0.9, solid}
] coordinates {
(0.79,0.185642)
};

% -------------------------
% Fitted line
% Fitted y = 0.1041906378 + 0.0775871054 x
% -------------------------
\addplot[
forget plot,
no markers,
black,
dashed,
dash pattern=on 4pt off 2pt,
line width=0.65pt,
domain=0:3.6,
samples=2
] {0.1041906378 + 0.0775871054*x};

% -------------------------
% Legend
% First row: pairs
% Second row: L1/L2 marker style and fitted line
% -------------------------

\addlegendimage{
only marks,
mark=*,
color=pairblue,
mark options={fill=pairblue, draw=pairblue, fill opacity=0.42, draw opacity=0.8}
}
\addlegendentry{\mbox{USDC-WETH}}

\addlegendimage{
only marks,
mark=*,
color=pairgreen,
mark options={fill=pairgreen, draw=pairgreen, fill opacity=0.42, draw opacity=0.8}
}
\addlegendentry{\mbox{USDT-WETH}}

\addlegendimage{
only marks,
mark=*,
color=pairorange,
mark options={fill=pairorange, draw=pairorange, fill opacity=0.42, draw opacity=0.8}
}
\addlegendentry{\mbox{USDC-cbBTC}}

\addlegendimage{
only marks,
mark=*,
color=pairred,
mark options={fill=pairred, draw=pairred, fill opacity=0.42, draw opacity=0.8}
}
\addlegendentry{\mbox{WBTC-USDC}}

\addlegendimage{
only marks,
mark=*,
color=pairpurple,
mark options={fill=pairpurple, draw=pairpurple, fill opacity=0.42, draw opacity=0.8}
}
\addlegendentry{\mbox{WBTC-USDT}}

\addlegendimage{
only marks,
mark=o,
mark size=3.2pt,
color=black,
mark options={fill=white, draw=black, solid}
}
\addlegendentry{L1}

\addlegendimage{
only marks,
mark=o,
mark size=3.2pt,
color=black,
mark options={fill=white, draw=black, dash pattern=on 1pt off 1pt}
}
\addlegendentry{L2}

\addlegendimage{
no markers,
black,
dashed,
dash pattern=on 4pt off 2pt,
line width=0.65pt
}
\addlegendentry{Fitted line}

\end{axis}
\end{tikzpicture}}\end{minipage}\hfill
\begin{minipage}{0.49\textwidth}\centering\scalebox{0.45}{\begin{tikzpicture}
\definecolor{pairblue}{RGB}{0,114,178}
\definecolor{pairgreen}{RGB}{0,158,115}
\definecolor{pairorange}{RGB}{230,159,0}
\definecolor{pairred}{RGB}{204,68,102}
\definecolor{pairpurple}{RGB}{136,34,204}

\begin{semilogxaxis}[
width=13.5cm,
height=7.2cm,
grid=both,
major grid style={draw=gray!25},
minor grid style={draw=gray!10},
title style={font=\small\bfseries, align=center},
xlabel style={font=\scriptsize},
ylabel style={font=\scriptsize},
tick label style={font=\scriptsize},
tick align=outside,
legend style={
    at={(0.5,1.06)},
    anchor=south,
    legend columns=4,
    draw=none,
    fill=none,
    font=\tiny,
    column sep=0.08cm,
    /tikz/every even column/.append style={column sep=0.08cm}
},
ylabel={Overall $-$ Passive LIFO (bps)},
xlabel={Active LIFO Volume (\%, log scale)},
xmin=0.00001,
xmax=100,
xtick={0.00001,0.0001,0.001,0.01,0.1,1,10,100},
xticklabels={$10^{-5}$,$10^{-4}$,$10^{-3}$,$10^{-2}$,$10^{-1}$,$1$,$10$,$10^2$},
]

% -------------------------
% ETH-USDC
% -------------------------

% L1
\addplot+[
only marks,
forget plot,
mark=square*,
mark size=2.6pt,
color=pairblue,
line width=0.8pt,
mark options={fill=pairblue, draw=pairblue, fill opacity=0.42, draw opacity=0.9, solid}
] coordinates {
(0.126840236,0.0371)
};

% L2
\addplot+[
only marks,
forget plot,
mark=square*,
mark size=2.6pt,
color=pairblue,
line width=0.8pt,
mark options={fill=pairblue, draw=pairblue, fill opacity=0.42, draw opacity=0.9, dash pattern=on 1pt off 1pt}
] coordinates {
(9.13e-05,0)
(1e-05,0)
};

% L1 / larger marker
\addplot+[
only marks,
forget plot,
mark=square*,
mark size=4.0pt,
color=pairblue,
line width=0.8pt,
mark options={fill=pairblue, draw=pairblue, fill opacity=0.42, draw opacity=0.9, solid}
] coordinates {
(0.111610447,0.0333)
};

% -------------------------
% ETH-USDT
% -------------------------

% L1
\addplot+[
only marks,
forget plot,
mark=square*,
mark size=2.6pt,
color=pairgreen,
line width=0.8pt,
mark options={fill=pairgreen, draw=pairgreen, fill opacity=0.42, draw opacity=0.9, solid}
] coordinates {
(0.363104221,0.1416)
};

% L2
\addplot+[
only marks,
forget plot,
mark=square*,
mark size=2.6pt,
color=pairgreen,
line width=0.8pt,
mark options={fill=pairgreen, draw=pairgreen, fill opacity=0.42, draw opacity=0.9, dash pattern=on 1pt off 1pt}
] coordinates {
(41.98641816,0.2569)
};

% L1 / larger marker
\addplot+[
only marks,
forget plot,
mark=square*,
mark size=4.0pt,
color=pairgreen,
line width=0.8pt,
mark options={fill=pairgreen, draw=pairgreen, fill opacity=0.42, draw opacity=0.9, solid}
] coordinates {
(0.075166248,0.024)
};

% -------------------------
% WBTC-USDC
% -------------------------

% L1
\addplot+[
only marks,
forget plot,
mark=square*,
mark size=2.6pt,
color=pairred,
line width=0.8pt,
mark options={fill=pairred, draw=pairred, fill opacity=0.42, draw opacity=0.9, solid}
] coordinates {
(0.234065768,0.2849)
};

% L2
\addplot+[
only marks,
forget plot,
mark=square*,
mark size=2.6pt,
color=pairred,
line width=0.8pt,
mark options={fill=pairred, draw=pairred, fill opacity=0.42, draw opacity=0.9, dash pattern=on 1pt off 1pt}
] coordinates {
(1e-05,0)
};

% L1 / larger marker
\addplot+[
only marks,
forget plot,
mark=square*,
mark size=4.0pt,
color=pairred,
line width=0.8pt,
mark options={fill=pairred, draw=pairred, fill opacity=0.42, draw opacity=0.9, solid}
] coordinates {
(0.065917404,0.0175)
};

% -------------------------
% WBTC-USDT
% -------------------------

% L1
\addplot+[
only marks,
forget plot,
mark=square*,
mark size=2.6pt,
color=pairpurple,
line width=0.8pt,
mark options={fill=pairpurple, draw=pairpurple, fill opacity=0.42, draw opacity=0.9, solid}
] coordinates {
(0.100620074,0.9612)
};

% L2
\addplot+[
only marks,
forget plot,
mark=square*,
mark size=2.6pt,
color=pairpurple,
line width=0.8pt,
mark options={fill=pairpurple, draw=pairpurple, fill opacity=0.42, draw opacity=0.9, dash pattern=on 1pt off 1pt}
] coordinates {
(69.41821706,0.3184)
};

% L1 / larger marker
\addplot+[
only marks,
forget plot,
mark=square*,
mark size=4.0pt,
color=pairpurple,
line width=0.8pt,
mark options={fill=pairpurple, draw=pairpurple, fill opacity=0.42, draw opacity=0.9, solid}
] coordinates {
(0.125882046,0.11)
};

% -------------------------
% Fitted line
% Fitted y = 0.2315449606 + 0.0466286420 log10(x)
% -------------------------
\addplot[
forget plot,
no markers,
black,
dashed,
dash pattern=on 4pt off 2pt,
line width=0.65pt,
domain=0.00001:100,
samples=2
] {0.2315449606 + 0.0466286420*ln(x)/ln(10)};

% -------------------------
% Legend
% First row: pairs
% Second row: L1/L2 marker style and fitted line
% -------------------------

\addlegendimage{
only marks,
mark=square*,
color=pairblue,
mark options={fill=pairblue, draw=pairblue, fill opacity=0.42, draw opacity=0.8}
}
\addlegendentry{\mbox{USDC-WETH}}

\addlegendimage{
only marks,
mark=square*,
color=pairgreen,
mark options={fill=pairgreen, draw=pairgreen, fill opacity=0.42, draw opacity=0.8}
}
\addlegendentry{\mbox{USDT-WETH}}

\addlegendimage{
only marks,
mark=square*,
color=pairred,
mark options={fill=pairred, draw=pairred, fill opacity=0.42, draw opacity=0.8}
}
\addlegendentry{\mbox{WBTC-USDC}}

\addlegendimage{
only marks,
mark=square*,
color=pairpurple,
mark options={fill=pairpurple, draw=pairpurple, fill opacity=0.42, draw opacity=0.8}
}
\addlegendentry{\mbox{WBTC-USDT}}

\addlegendimage{
only marks,
mark=square*,
mark size=3.2pt,
color=black,
mark options={fill=white, draw=black, solid}
}
\addlegendentry{L1}

\addlegendimage{
only marks,
mark=square*,
mark size=3.2pt,
color=black,
mark options={fill=white, draw=black, dash pattern=on 1pt off 1pt}
}
\addlegendentry{L2}

\addlegendimage{
no markers,
black,
dashed,
dash pattern=on 4pt off 2pt,
line width=0.65pt
}
\addlegendentry{Fitted line}

\end{semilogxaxis}
\end{tikzpicture}}\end{minipage}
\caption{Overall minus passive LIFO markout by active LIFO volume share, for Uniswap v3 (left) and v4 (right, log-scale volume share) pools.}
\label{fig:gapv3v4}
\end{figure}

\ifshort
\paraheader{Fee tiers, cross-chain differences, and method comparison.}
Beyond the headline gap, three further patterns emerge (figures in Appendix~\ref{app:additional}). 
First, high-fee pools are generally better for passive LPs: passive LIFO markouts are higher at the 30 bps tier than at 5 bps, consistent with fees cushioning adverse selection. 
Second, the gap is larger and more dispersed on Ethereum than on Arbitrum and Base, where it clusters near zero; we read this as descriptive, reflecting differences in block times, gas costs, and active-LP participation. 
Third, the LIFO and infinitesimal estimates agree closely---for almost all pools their difference is smaller than the infinitesimal confidence-interval width---so our conclusions are robust to the choice between them.
\else
% Secondary results: fee tiers, cross-chain differences, and the LIFO-vs-infinitesimal
% comparison. Rendered inline under Results in the AFT/LIPIcs build; relocated to the
% appendix in the WINE build. No top-level heading here (the caller supplies it).
\paraheader{Fee tiers and passive LP profitability.}
\Cref{fig:combined-fee-chain-passive-lifo} suggests high-fee pools are often, but not uniformly, better for passive LPs.
Passive LIFO markouts are typically higher in 30 bps pools than in 5 bps pools, consistent with fees providing a larger cushion against adverse selection.
This pattern is also related to active LP incentives.
Active liquidity provision is more attractive in low-fee pools when LPs can concentrate liquidity around expected trades and earn fees with limited exposure.
This behavior is more feasible on Ethereum because the public mempool allows LPs to observe incoming flow before block inclusion and provide short-lived liquidity around anticipated swaps.
By contrast, 30 bps pools tend to show less active LIFO volume and, in many cases, appear more favorable for passive liquidity provision.

\begin{figure}[!htbp]
    \centering
    \begin{tikzpicture}
\definecolor{pairblue}{RGB}{0,114,178}
\definecolor{pairgreen}{RGB}{0,158,115}
\definecolor{pairorange}{RGB}{230,159,0}
\definecolor{pairred}{RGB}{204,68,102}
\definecolor{pairpurple}{RGB}{136,34,204}
\begin{axis}[
width=13.5cm,
height=7.2cm,
major grid style={draw=gray!25},
minor grid style={draw=gray!10},
title style={font=\small\bfseries, align=center},
xlabel style={font=\scriptsize},
ylabel style={font=\scriptsize},
tick label style={font=\scriptsize},
tick align=outside,
legend style={at={(0.5,1.04)}, anchor=south, legend columns=5, draw=none, fill=none, font=\tiny, column sep=0.08cm, /tikz/every even column/.append style={column sep=0.08cm}},
xmajorgrids=false,
xminorgrids=false,
ymajorgrids=true,
yminorgrids=true,
ylabel={Passive LIFO (bps)},
xlabel={},
xmin=0.55,
xmax=3.45,
xtick={0.92,1.08,1.92,2.08,2.92,3.08},
xticklabels={5,30,5,30,5,30},
x tick label style={font=\scriptsize, yshift=0pt},
extra x ticks={0.92,1.08,1.92,2.08,2.92,3.08},
extra x tick labels={},
extra x tick style={grid=major, major grid style={draw=gray!25}, tick style={draw=none}},
clip=false,
]
\addplot+[only marks, forget plot, mark=triangle*, mark size=2.6pt, color=pairblue, line width=0.8pt, mark options={fill=pairblue, draw=pairblue, fill opacity=0.42, draw opacity=0.8}] coordinates {(1.08,19.15383325) (2.08,20.82207889) (3.08,9.320412415)};
\addplot+[only marks, forget plot, mark=*, mark size=2.6pt, color=pairblue, line width=0.8pt, mark options={fill=pairblue, draw=pairblue, fill opacity=0.42, draw opacity=0.8}] coordinates {(0.92,-0.76971) (1.92,-0.706874) (2.92,-1.506019) (3.08,-0.866415)};
\addplot+[only marks, forget plot, mark=square*, mark size=3.0pt, color=pairblue, line width=0.8pt, mark options={fill=pairblue, draw=pairblue, fill opacity=0.42, draw opacity=0.8}] coordinates {(0.92,-1.431669543) (1.92,-0.977858869) (2.92,-3.070001001) (3.08,-0.909249237)};
\addplot+[only marks, forget plot, mark=triangle*, mark size=2.6pt, color=pairgreen, line width=0.8pt, mark options={fill=pairgreen, draw=pairgreen, fill opacity=0.42, draw opacity=0.8}] coordinates {(1.08,21.48637526) (3.08,9.329889084)};
\addplot+[only marks, forget plot, mark=*, mark size=2.6pt, color=pairgreen, line width=0.8pt, mark options={fill=pairgreen, draw=pairgreen, fill opacity=0.42, draw opacity=0.8}] coordinates {(0.92,-0.449636) (1.92,-0.574969) (2.92,-1.309165) (3.08,1.001649)};
\addplot+[only marks, forget plot, mark=square*, mark size=3.0pt, color=pairgreen, line width=0.8pt, mark options={fill=pairgreen, draw=pairgreen, fill opacity=0.42, draw opacity=0.8}] coordinates {(0.92,-1.972888288) (2.92,-1.756678938) (3.08,-3.044381418)};
\addplot+[only marks, forget plot, mark=*, mark size=2.6pt, color=pairorange, line width=0.8pt, mark options={fill=pairorange, draw=pairorange, fill opacity=0.42, draw opacity=0.8}] coordinates {(1.92,-0.568331)};
\addplot+[only marks, forget plot, mark=square*, mark size=3.0pt, color=pairorange, line width=0.8pt, mark options={fill=pairorange, draw=pairorange, fill opacity=0.42, draw opacity=0.8}] coordinates {(1.92,-0.365854383)};
\addplot+[only marks, forget plot, mark=*, mark size=2.6pt, color=pairred, line width=0.8pt, mark options={fill=pairred, draw=pairred, fill opacity=0.42, draw opacity=0.8}] coordinates {(0.92,-0.589305) (2.92,-0.408402) (3.08,0.209764)};
\addplot+[only marks, forget plot, mark=square*, mark size=3.0pt, color=pairred, line width=0.8pt, mark options={fill=pairred, draw=pairred, fill opacity=0.42, draw opacity=0.8}] coordinates {(0.92,-1.287180197) (2.92,-1.278666208) (3.08,-2.020871007)};
\addplot+[only marks, forget plot, mark=*, mark size=2.6pt, color=pairpurple, line width=0.8pt, mark options={fill=pairpurple, draw=pairpurple, fill opacity=0.42, draw opacity=0.8}] coordinates {(0.92,-0.722146) (2.92,-0.254329) (3.08,2.519784)};
\addplot+[only marks, forget plot, mark=square*, mark size=3.0pt, color=pairpurple, line width=0.8pt, mark options={fill=pairpurple, draw=pairpurple, fill opacity=0.42, draw opacity=0.8}] coordinates {(0.92,-2.788935587) (2.92,-1.147260842) (3.08,0.01653679)};
\addplot+[forget plot, no markers, dashed, line width=0.55pt, color=black] coordinates {(0.92,-1.251433827) (1.08,20.32010426)};
\addplot+[forget plot, no markers, dashed, line width=0.55pt, color=black] coordinates {(1.92,-0.6387774504) (2.08,20.82207889)};
\addplot+[forget plot, no markers, dashed, line width=0.55pt, color=black] coordinates {(2.92,-1.341315249) (3.08,1.555711863)};
\addplot+[only marks, forget plot, mark=star, mark size=3.6pt, color=black, line width=1.0pt, mark options={fill=black, draw=black}] coordinates {(0.92,-1.251433827) (1.08,20.32010426) (1.92,-0.6387774504) (2.08,20.82207889) (2.92,-1.341315249) (3.08,1.555711863)};
\addlegendimage{only marks, mark=*, color=pairblue, mark options={fill=pairblue, draw=pairblue, fill opacity=0.42, draw opacity=0.8}}
\addlegendentry{\mbox{USDC-WETH}}
\addlegendimage{only marks, mark=*, color=pairgreen, mark options={fill=pairgreen, draw=pairgreen, fill opacity=0.42, draw opacity=0.8}}
\addlegendentry{\mbox{USDT-WETH}}
\addlegendimage{only marks, mark=*, color=pairorange, mark options={fill=pairorange, draw=pairorange, fill opacity=0.42, draw opacity=0.8}}
\addlegendentry{\mbox{USDC-cbBTC}}
\addlegendimage{only marks, mark=*, color=pairred, mark options={fill=pairred, draw=pairred, fill opacity=0.42, draw opacity=0.8}}
\addlegendentry{\mbox{WBTC-USDC}}
\addlegendimage{only marks, mark=*, color=pairpurple, mark options={fill=pairpurple, draw=pairpurple, fill opacity=0.42, draw opacity=0.8}}
\addlegendentry{\mbox{WBTC-USDT}}
\addlegendimage{only marks, mark=triangle*, color=black}
\addlegendentry{v2}
\addlegendimage{only marks, mark=*, color=black}
\addlegendentry{v3}
\addlegendimage{only marks, mark=square*, color=black, mark options={fill=black, draw=black}}
\addlegendentry{v4}
\addlegendimage{only marks, mark=star, mark size=3.6pt, color=black}
\addlegendentry{overall mean}

\node[anchor=north, font=\scriptsize] at (axis description cs:0.155,-0.12) {Arbitrum};
\node[anchor=north, font=\scriptsize] at (axis description cs:0.500,-0.12) {Base};
\node[anchor=north, font=\scriptsize] at (axis description cs:0.845,-0.12) {Ethereum};
\node[anchor=north, font=\scriptsize] at (axis description cs:0.500,-0.24) {Fee Tier (bps) within Chain};
\end{axis}
\end{tikzpicture}
%%%%%%%%%%%
    \caption{Passive LIFO markout across chains and fee tiers for Uniswap v2, v3, and v4 pools.}
    \label{fig:combined-fee-chain-passive-lifo}
\end{figure}

The figure also shows differences across chains. In many 5 bps pools on Arbitrum and Base, passive markouts tend to cluster around zero, while Ethereum pools show more dispersion across pairs and fee tiers. This suggests that fee tier interacts with market structure at the chain level: higher fees improve passive outcomes, but the size of the active-passive gap also depends on the ease of LPs actively repositioning liquidity around trades.

\paraheader{Cross-chain differences in the active-passive gap.}
\Cref{fig:combined-chain-lifo-gap} shows that the difference between overall and passive LP profitability is generally larger and more dispersed on Ethereum than on the L2s. Arbitrum and Base are mostly clustered around zero, while Ethereum pools more often display a positive gap, suggesting that passive LPs there tend to perform worse than aggregate pool-level measures imply.

\begin{figure}[!htbp]
    \centering
    \begin{tikzpicture}
\definecolor{pairblue}{RGB}{0,114,178}
\definecolor{pairgreen}{RGB}{0,158,115}
\definecolor{pairorange}{RGB}{230,159,0}
\definecolor{pairred}{RGB}{204,68,102}
\definecolor{pairpurple}{RGB}{136,34,204}
\begin{axis}[
width=13.5cm,
height=7.2cm,
grid=both,
major grid style={draw=gray!25},
minor grid style={draw=gray!10},
title style={font=\small\bfseries, align=center},
xlabel style={font=\scriptsize},
ylabel style={font=\scriptsize},
tick label style={font=\scriptsize},
tick align=outside,
legend style={at={(0.5,1.04)}, anchor=south, legend columns=5, draw=none, fill=none, font=\tiny, column sep=0.08cm, /tikz/every even column/.append style={column sep=0.08cm}},
ylabel={Overall $-$ Passive LIFO (bps)},
xlabel={Chain},
symbolic x coords={Arbitrum,Base,Ethereum},
xtick=data,
enlarge x limits=0.18,
x tick label style={anchor=north, yshift=-3pt, font=\scriptsize},
]
\addplot+[only marks, forget plot, mark=triangle*, mark size=2.6pt, color=pairblue, line width=0.8pt, mark options={fill=pairblue, draw=pairblue, fill opacity=0.42, draw opacity=0.8}] coordinates {(Arbitrum,0) (Base,0) (Ethereum,0)};
\addplot+[only marks, forget plot, mark=*, mark size=2.6pt, color=pairblue, line width=0.8pt, mark options={fill=pairblue, draw=pairblue, fill opacity=0.42, draw opacity=0.8}] coordinates {(Arbitrum,0.037189) (Base,0.025698) (Ethereum,0.355615) (Ethereum,0.159169)};
\addplot+[only marks, forget plot, mark=square*, mark size=3.0pt, color=pairblue, line width=0.8pt, mark options={fill=pairblue, draw=pairblue, fill opacity=0.42, draw opacity=0.8}] coordinates {(Arbitrum,0) (Base,0) (Ethereum,0.0371) (Ethereum,0.0333)};
\addplot+[only marks, forget plot, mark=triangle*, mark size=2.6pt, color=pairgreen, line width=0.8pt, mark options={fill=pairgreen, draw=pairgreen, fill opacity=0.42, draw opacity=0.8}] coordinates {(Arbitrum,0) (Ethereum,0)};
\addplot+[only marks, forget plot, mark=*, mark size=2.6pt, color=pairgreen, line width=0.8pt, mark options={fill=pairgreen, draw=pairgreen, fill opacity=0.42, draw opacity=0.8}] coordinates {(Arbitrum,-2.9e-05) (Base,1e-06) (Ethereum,0.398407) (Ethereum,0.183949)};
\addplot+[only marks, forget plot, mark=square*, mark size=3.0pt, color=pairgreen, line width=0.8pt, mark options={fill=pairgreen, draw=pairgreen, fill opacity=0.42, draw opacity=0.8}] coordinates {(Arbitrum,0.2569) (Ethereum,0.1416) (Ethereum,0.024)};
\addplot+[only marks, forget plot, mark=*, mark size=2.6pt, color=pairorange, line width=0.8pt, mark options={fill=pairorange, draw=pairorange, fill opacity=0.42, draw opacity=0.8}] coordinates {(Base,1e-06)};
\addplot+[only marks, forget plot, mark=square*, mark size=3.0pt, color=pairorange, line width=0.8pt, mark options={fill=pairorange, draw=pairorange, fill opacity=0.42, draw opacity=0.8}] coordinates {(Base,-0.5463)};
\addplot+[only marks, forget plot, mark=*, mark size=2.6pt, color=pairred, line width=0.8pt, mark options={fill=pairred, draw=pairred, fill opacity=0.42, draw opacity=0.8}] coordinates {(Arbitrum,-0.000158) (Ethereum,0.453938) (Ethereum,0.31454)};
\addplot+[only marks, forget plot, mark=square*, mark size=3.0pt, color=pairred, line width=0.8pt, mark options={fill=pairred, draw=pairred, fill opacity=0.42, draw opacity=0.8}] coordinates {(Arbitrum,0) (Ethereum,0.2849) (Ethereum,0.0175)};
\addplot+[only marks, forget plot, mark=*, mark size=2.6pt, color=pairpurple, line width=0.8pt, mark options={fill=pairpurple, draw=pairpurple, fill opacity=0.42, draw opacity=0.8}] coordinates {(Arbitrum,9.3e-05) (Ethereum,0.445023) (Ethereum,0.185642)};
\addplot+[only marks, forget plot, mark=square*, mark size=3.0pt, color=pairpurple, line width=0.8pt, mark options={fill=pairpurple, draw=pairpurple, fill opacity=0.42, draw opacity=0.8}] coordinates {(Arbitrum,0.3184) (Ethereum,0.9612) (Ethereum,0.11)};

% Volume-weighted mean across all included pools within each chain.
\addplot+[forget plot, mark=star, mark size=3.6pt, color=black, line width=0.9pt, dashed, mark options={fill=black, draw=black}] coordinates {(Arbitrum,0.02879277192) (Base,0.01613893326) (Ethereum,0.2899732533)};
\addlegendimage{only marks, mark=*, color=pairblue, mark options={fill=pairblue, draw=pairblue, fill opacity=0.42, draw opacity=0.8}}
\addlegendentry{\mbox{USDC-WETH}}
\addlegendimage{only marks, mark=*, color=pairgreen, mark options={fill=pairgreen, draw=pairgreen, fill opacity=0.42, draw opacity=0.8}}
\addlegendentry{\mbox{USDT-WETH}}
\addlegendimage{only marks, mark=*, color=pairorange, mark options={fill=pairorange, draw=pairorange, fill opacity=0.42, draw opacity=0.8}}
\addlegendentry{\mbox{USDC-cbBTC}}
\addlegendimage{only marks, mark=*, color=pairred, mark options={fill=pairred, draw=pairred, fill opacity=0.42, draw opacity=0.8}}
\addlegendentry{\mbox{WBTC-USDC}}
\addlegendimage{only marks, mark=*, color=pairpurple, mark options={fill=pairpurple, draw=pairpurple, fill opacity=0.42, draw opacity=0.8}}
\addlegendentry{\mbox{WBTC-USDT}}
\addlegendimage{only marks, mark=triangle*, color=black}
\addlegendentry{v2}
\addlegendimage{only marks, mark=*, color=black}
\addlegendentry{v3}
\addlegendimage{only marks, mark=square*, color=black, mark options={fill=black, draw=black}}
\addlegendentry{v4}
\addlegendimage{mark=star, mark size=3.6pt, color=black, dashed, line width=0.9pt, mark options={fill=black, draw=black}}
\addlegendentry{VW mean}
\end{axis}
\end{tikzpicture}
    \caption{Overall minus passive LIFO markout across chains for Uniswap v2, v3, and v4 pools.}
    \label{fig:combined-chain-lifo-gap}
\end{figure}

We see this pattern as descriptive rather than causal. Ethereum, Arbitrum and Base are different not only in terms of block times, but also in terms of gas costs, liquidity depth, user and flow composition, MEV infrastructure and the set of active LPs participating on each chain. All these factors may affect the active-passive gap. Mechanistically, there are two forces pulling in opposite directions on L2s: shorter block times could reduce the value of timing liquidity around incoming flow, narrowing the gap, and lower gas costs could make frequent liquidity updates cheaper, potentially widening the gap. The small gaps observed on Arbitrum and Base suggest that the first force may dominate in our sample, or that active LP participation is less developed in these pools.  A causal separation of these channels would require a matched cross-chain analysis controlling for volatility, volume, liquidity depth, gas costs and flow composition.

\paraheader{Comparison of LIFO and infinitesimal estimates.}
\Cref{fig:combined-ci-range-lifo-inf} plots the absolute difference between their estimates against the infinitesimal confidence interval width and compares the passive LIFO and infinitesimal methods.
Points below the 45-degree line indicate cases where the difference between methods is smaller than the uncertainty range of the infinitesimal estimate.
Almost all pools fall below this line, showing that the two approaches provide broadly similar passive LP profitability estimates despite very different constructions.
The exceptions are pools where the passive estimate is more sensitive to the choice of methodology.

\begin{figure}[!htbp]
    \centering
    \begin{tikzpicture}
\definecolor{pairblue}{RGB}{0,114,178}
\definecolor{pairgreen}{RGB}{0,158,115}
\definecolor{pairorange}{RGB}{230,159,0}
\definecolor{pairred}{RGB}{204,68,102}
\definecolor{pairpurple}{RGB}{136,34,204}

\begin{loglogaxis}[
width=13.5cm,
height=7.2cm,
xmajorgrids=true,
xminorgrids=true,
ymajorgrids=true,
yminorgrids=true,
major grid style={draw=gray!25},
minor grid style={draw=gray!10},
title style={font=\small\bfseries, align=center},
xlabel style={font=\scriptsize},
ylabel style={font=\scriptsize},
tick label style={font=\scriptsize},
tick align=outside,
legend style={at={(0.5,1.08)}, anchor=south, legend columns=5, draw=none, fill=none, font=\tiny, column sep=0.08cm, /tikz/every even column/.append style={column sep=0.08cm}},
ylabel={|passive LIFO $-$ infinitesimal| (bps, log scale)},
xlabel={Infinitesimal CI Range (bps, log scale)},
xmin=0.5,
xmax=100,
ymin=0.005,
ymax=100,
minor tick num=8,
xtick={0.5,1,2,5,10,20,50,100},
xticklabels={0.5,1,2,5,10,20,50,100},
]
\addplot+[only marks, forget plot, mark=triangle*, mark size=4.0pt, color=pairblue, line width=0.8pt, mark options={fill=pairblue, draw=pairblue, fill opacity=0.42, draw opacity=0.8, solid}] coordinates {(10.235191051,0.320412415)};
\addplot+[only marks, forget plot, mark=triangle*, mark size=4.0pt, color=pairblue, line width=0.8pt, mark options={fill=pairblue, draw=pairblue, fill opacity=0.42, draw opacity=0.8, dash pattern=on 1pt off 1pt}] coordinates {(3.723135660,0.044543460) (18.268190172,0.870630750)};
\addplot+[only marks, forget plot, mark=*, mark size=2.6pt, color=pairblue, line width=0.8pt, mark options={fill=pairblue, draw=pairblue, fill opacity=0.42, draw opacity=0.8, solid}] coordinates {(2.087603813,0.276595793)};
\addplot+[only marks, forget plot, mark=*, mark size=4.0pt, color=pairblue, line width=0.8pt, mark options={fill=pairblue, draw=pairblue, fill opacity=0.42, draw opacity=0.8, solid}] coordinates {(10.96784081,1.037521026)};
\addplot+[only marks, forget plot, mark=*, mark size=2.6pt, color=pairblue, line width=0.8pt, mark options={fill=pairblue, draw=pairblue, fill opacity=0.42, draw opacity=0.8, dash pattern=on 1pt off 1pt}] coordinates {(1.595592732,0.2834333) (0.550290463,0.901092413)};
\addplot+[only marks, forget plot, mark=square*, mark size=2.6pt, color=pairblue, line width=0.8pt, mark options={fill=pairblue, draw=pairblue, fill opacity=0.42, draw opacity=0.8, solid}] coordinates {(2.290031478,0.243370158)};
\addplot+[only marks, forget plot, mark=square*, mark size=4.0pt, color=pairblue, line width=0.8pt, mark options={fill=pairblue, draw=pairblue, fill opacity=0.42, draw opacity=0.8, solid}] coordinates {(16.64228006,0.999593702)};
\addplot+[only marks, forget plot, mark=square*, mark size=2.6pt, color=pairblue, line width=0.8pt, mark options={fill=pairblue, draw=pairblue, fill opacity=0.42, draw opacity=0.8, dash pattern=on 1pt off 1pt}] coordinates {(0.537556262,0.109661764) (0.560240946,0.443419364)};
\addplot+[only marks, forget plot, mark=triangle*, mark size=4.0pt, color=pairgreen, line width=0.8pt, mark options={fill=pairgreen, draw=pairgreen, fill opacity=0.42, draw opacity=0.8, solid}] coordinates {(31.128023538,1.329889084)};
\addplot+[only marks, forget plot, mark=triangle*, mark size=4.0pt, color=pairgreen, line width=0.8pt, mark options={fill=pairgreen, draw=pairgreen, fill opacity=0.42, draw opacity=0.8, dash pattern=on 1pt off 1pt}] coordinates {(15.388740056,4.031116580)};
\addplot+[only marks, forget plot, mark=*, mark size=2.6pt, color=pairgreen, line width=0.8pt, mark options={fill=pairgreen, draw=pairgreen, fill opacity=0.42, draw opacity=0.8, solid}] coordinates {(2.142031248,0.22609105)};
\addplot+[only marks, forget plot, mark=*, mark size=4.0pt, color=pairgreen, line width=0.8pt, mark options={fill=pairgreen, draw=pairgreen, fill opacity=0.42, draw opacity=0.8, solid}] coordinates {(15.09606331,1.046023139)};
\addplot+[only marks, forget plot, mark=*, mark size=2.6pt, color=pairgreen, line width=0.8pt, mark options={fill=pairgreen, draw=pairgreen, fill opacity=0.42, draw opacity=0.8, dash pattern=on 1pt off 1pt}] coordinates {(2.416778547,0.436223854) (0.772240033,0.900541971)};
\addplot+[only marks, forget plot, mark=square*, mark size=2.6pt, color=pairgreen, line width=0.8pt, mark options={fill=pairgreen, draw=pairgreen, fill opacity=0.42, draw opacity=0.8, solid}] coordinates {(1.95020931,0.4790698)};
\addplot+[only marks, forget plot, mark=square*, mark size=4.0pt, color=pairgreen, line width=0.8pt, mark options={fill=pairgreen, draw=pairgreen, fill opacity=0.42, draw opacity=0.8, solid}] coordinates {(7.742995806,4.321356848)};
\addplot+[only marks, forget plot, mark=square*, mark size=2.6pt, color=pairgreen, line width=0.8pt, mark options={fill=pairgreen, draw=pairgreen, fill opacity=0.42, draw opacity=0.8, dash pattern=on 1pt off 1pt}] coordinates {(1.190733102,0.053974063)};
\addplot+[only marks, forget plot, mark=*, mark size=2.6pt, color=pairorange, line width=0.8pt, mark options={fill=pairorange, draw=pairorange, fill opacity=0.42, draw opacity=0.8, dash pattern=on 1pt off 1pt}] coordinates {(1.379603113,0.057516356)};
\addplot+[only marks, forget plot, mark=square*, mark size=2.6pt, color=pairorange, line width=0.8pt, mark options={fill=pairorange, draw=pairorange, fill opacity=0.42, draw opacity=0.8, dash pattern=on 1pt off 1pt}] coordinates {(5.013885476,0.838805514)};
\addplot+[only marks, forget plot, mark=*, mark size=2.6pt, color=pairred, line width=0.8pt, mark options={fill=pairred, draw=pairred, fill opacity=0.42, draw opacity=0.8, solid}] coordinates {(2.970337147,0.870088628)};
\addplot+[only marks, forget plot, mark=*, mark size=4.0pt, color=pairred, line width=0.8pt, mark options={fill=pairred, draw=pairred, fill opacity=0.42, draw opacity=0.8, solid}] coordinates {(23.26112581,9.696634825)};
\addplot+[only marks, forget plot, mark=*, mark size=2.6pt, color=pairred, line width=0.8pt, mark options={fill=pairred, draw=pairred, fill opacity=0.42, draw opacity=0.8, dash pattern=on 1pt off 1pt}] coordinates {(2.065013574,0.884709478)};
\addplot+[only marks, forget plot, mark=square*, mark size=2.6pt, color=pairred, line width=0.8pt, mark options={fill=pairred, draw=pairred, fill opacity=0.42, draw opacity=0.8, solid}] coordinates {(3.322851324,0.3558586)};
\addplot+[only marks, forget plot, mark=square*, mark size=4.0pt, color=pairred, line width=0.8pt, mark options={fill=pairred, draw=pairred, fill opacity=0.42, draw opacity=0.8, solid}] coordinates {(27.97851943,1.098721718)};
\addplot+[only marks, forget plot, mark=square*, mark size=2.6pt, color=pairred, line width=0.8pt, mark options={fill=pairred, draw=pairred, fill opacity=0.42, draw opacity=0.8, dash pattern=on 1pt off 1pt}] coordinates {(1.303687466,0.429556978)};
\addplot+[only marks, forget plot, mark=*, mark size=2.6pt, color=pairpurple, line width=0.8pt, mark options={fill=pairpurple, draw=pairpurple, fill opacity=0.42, draw opacity=0.8, solid}] coordinates {(4.879413032,0.007194778)};
\addplot+[only marks, forget plot, mark=*, mark size=4.0pt, color=pairpurple, line width=0.8pt, mark options={fill=pairpurple, draw=pairpurple, fill opacity=0.42, draw opacity=0.8, solid}] coordinates {(10.90928374,0.644984915)};
\addplot+[only marks, forget plot, mark=*, mark size=2.6pt, color=pairpurple, line width=0.8pt, mark options={fill=pairpurple, draw=pairpurple, fill opacity=0.42, draw opacity=0.8, dash pattern=on 1pt off 1pt}] coordinates {(0.85349707,0.557332585)};
\addplot+[only marks, forget plot, mark=square*, mark size=2.6pt, color=pairpurple, line width=0.8pt, mark options={fill=pairpurple, draw=pairpurple, fill opacity=0.42, draw opacity=0.8, solid}] coordinates {(7.159430784,5.507979217)};
\addplot+[only marks, forget plot, mark=square*, mark size=4.0pt, color=pairpurple, line width=0.8pt, mark options={fill=pairpurple, draw=pairpurple, fill opacity=0.42, draw opacity=0.8, solid}] coordinates {(68.81283344,0.356858371)};
\addplot+[only marks, forget plot, mark=square*, mark size=2.6pt, color=pairpurple, line width=0.8pt, mark options={fill=pairpurple, draw=pairpurple, fill opacity=0.42, draw opacity=0.8, dash pattern=on 1pt off 1pt}] coordinates {(1.86807785,0.515533644)};
\addplot[forget plot, domain=0.5:100, samples=2, dashed, line width=0.65pt, color=black, mark=none] {x};
\addlegendimage{only marks, mark=*, color=pairblue, mark options={fill=pairblue, draw=pairblue, fill opacity=0.42, draw opacity=0.8}}
\addlegendentry{\mbox{USDC-WETH}}
\addlegendimage{only marks, mark=*, color=pairgreen, mark options={fill=pairgreen, draw=pairgreen, fill opacity=0.42, draw opacity=0.8}}
\addlegendentry{\mbox{USDT-WETH}}
\addlegendimage{only marks, mark=*, color=pairorange, mark options={fill=pairorange, draw=pairorange, fill opacity=0.42, draw opacity=0.8}}
\addlegendentry{\mbox{USDC-cbBTC}}
\addlegendimage{only marks, mark=*, color=pairred, mark options={fill=pairred, draw=pairred, fill opacity=0.42, draw opacity=0.8}}
\addlegendentry{\mbox{WBTC-USDC}}
\addlegendimage{only marks, mark=*, color=pairpurple, mark options={fill=pairpurple, draw=pairpurple, fill opacity=0.42, draw opacity=0.8}}
\addlegendentry{\mbox{WBTC-USDT}}
\addlegendimage{only marks, mark=triangle*, color=black}
\addlegendentry{v2}
\addlegendimage{only marks, mark=*, color=black}
\addlegendentry{v3}
\addlegendimage{only marks, mark=square*, color=black, mark options={fill=black, draw=black}}
\addlegendentry{v4}
\addlegendimage{empty legend}
\addlegendentry{}
\addlegendimage{empty legend}
\addlegendentry{}
\addlegendimage{dashed, line width=0.65pt, color=black}
\addlegendentry{$y=x$}
\addlegendimage{only marks, mark=*, color=black, mark options={fill=white, draw=black, solid}}
\addlegendentry{L1}
\addlegendimage{only marks, mark=*, color=black, mark options={fill=white, draw=black, dash pattern=on 1pt off 1pt}}
\addlegendentry{L2}
\end{loglogaxis}
\end{tikzpicture}
    \caption{Difference between passive LIFO and infinitesimal estimates by infinitesimal confidence interval range.}
    \label{fig:combined-ci-range-lifo-inf}
\end{figure}

\fi

\section{Discussion and Conclusion}\label{sec:conclusion}
\paraheader{Summary of findings.}
Using v2 as a homogeneous-LP benchmark, we verify that aggregate pool and passive LP markouts coincide when liquidity is full-range and active provision is effectively absent.
In v3 and v4 concentrated-liquidity pools a systematic active--passive gap opens up: passive LPs systematically underperform pool-aggregate markouts, with the largest gaps on Ethereum and at the 5\,bps tier where concentrated liquidity can be targeted most efficiently around incoming flow.
The LIFO and infinitesimal estimators agree directionally and lie within each other's confidence intervals across the great majority of pools, providing cross-method evidence that the gap is a feature of the data rather than an artifact of either identification strategy.

\ifshort
\paraheader{Practical use of the methodology.}
The two estimators differ sharply in cost: the infinitesimal benchmark uses only the on-chain price path and can be computed block-by-block for real-time monitoring, while the LIFO decomposition requires a stateful matching pass and is best used retrospectively. 
This supports rolling-window pool monitoring, ex-ante pool selection, and---since the infinitesimal benchmark needs no behavioral assumptions---a standardized market-quality indicator. Appendix~\ref{app:practical} develops these use cases.
\else
\paraheader{Practical use of the methodology.}
% Practical use of the methodology. Inline under Discussion in the AFT/LIPIcs
% build; relocated to the appendix in the WINE build.
Our two estimators differ sharply in cost, which determines how they map onto practitioner workflows.
The infinitesimal-LP benchmark depends only on the internal pool price path and per-swap input/output amounts---quantities emitted on-chain as standard pool events.
It can be computed block-by-block from any Ethereum or L2 RPC with negligible overhead, and is naturally suited to real-time monitoring.
The LIFO and exact-matching estimators, by contrast, require the full per-pool mint--burn--swap event log together with a stateful matching pass over LP positions.
They are linear in the number of events per pool and run in minutes per pool-month on a single machine, but are best treated as retrospective rather than real-time tools.

A natural application for passive LPs is a rolling-window pool monitor: practitioners compute the infinitesimal passive markout over a trailing day or week, and treat the lower bound of the associated confidence band as a decision threshold for exiting a position.
The same procedure can be inverted to support pool selection ex ante, ranking candidate pools, fee tiers, and chains by realized passive markout per dollar of TVL over a recent reference window.
For LPs considering whether to invest in active management, the realized gap between active and passive markout from the LIFO decomposition gives an estimate of the rent available to strategic provision.

The active--passive gap is also a natural target for protocol and pool designers.
Proposed design interventions---JIT-mitigation hooks, sequencing changes, faster block times---can be evaluated by re-estimating the gap before and after the change.
Because the infinitesimal benchmark requires no behavioral assumptions about LP heterogeneity, it is well-suited as a standardized market-quality indicator that pool operators and aggregators could publish alongside conventional TVL and volume statistics.
\fi

\paraheader{Limitations.}
\ifshort
Four caveats bound the interpretation of our estimates.
Our LIFO classification operates at the on-chain address level (\Cref{sec:methodology}), so multi-address strategies and pooled vaults are misclassified and the LIFO active share is a lower bound on strategic provision.
The infinitesimal benchmark corresponds to a full-range position; finite-width passive LPs face more adverse selection per unit of liquidity, making the estimate a floor on their markout cost.
Our cross-chain comparison is descriptive---Ethereum, Arbitrum, and Base differ in gas, MEV infrastructure, block times, and LP composition, channels we do not separate (\Cref{sec:result})---and the sample is restricted to selected high-volume pools above a \$10M floor, excluding the noisier long tail (\Cref{sec:data}).
% Finally, results reflect a single window (February--November 2025): we report per-estimator confidence intervals but not formal per-pool tests of the gap, and read the directional agreement between the two estimators as cross-method evidence rather than a hypothesis test or a longitudinal claim.
Finally, results reflect a single window (February--November 2025), and we read the directional agreement between the two estimators as cross-method evidence rather than a formal hypothesis test.
\else
Several caveats in our identification strategy are flagged in the relevant sections and worth restating together.
Our LIFO classification operates at the on-chain address level (\Cref{sec:methodology}); multi-address active strategies and pooled vault contracts are misclassified, so the LIFO active share is best read as a lower bound on strategic provision at the economic-agent level.
Our cross-chain comparison is descriptive: Ethereum, Arbitrum, and Base differ in gas costs, mempool visibility, MEV infrastructure, and LP composition as well as block times, and we do not attempt to separate these channels (\Cref{sec:result}).
Sample selection is restricted to high-volume WETH/USDC/USDT/WBTC pools above a \$10M volume floor (\Cref{sec:data}); the long tail of smaller pools, where active LP participation is thinner and noise dominates, is outside scope.
Three further limitations are not addressed in the body.
First, the infinitesimal benchmark assumes an always-in-range marginal LP and therefore corresponds to a full-range passive position; realistic v3/v4 passive LPs choose finite tick widths and face proportionally more adverse selection per unit of liquidity, so the infinitesimal estimate is best interpreted as a floor on the markout cost of typical concentrated passive positions.
Second, although we report confidence intervals for the passive markout level under each estimator, we do not yet report formal per-pool tests of whether the gap (overall minus passive) is statistically different from zero; computing pool-level gap standard errors and significance tests is a planned extension, and in the meantime the directional consistency between the LIFO and infinitesimal estimates should be read as cross-method agreement rather than as a hypothesis test.
Third, results reflect a single sample window (February--November 2025; see~\Cref{sec:data}): LP composition, JIT prevalence, and active-passive incentives all evolve over time, and the cross-sectional snapshot here is not a longitudinal claim about how the gap is changing.
\fi

\paraheader{Conclusion.}
We have decomposed pool-level markouts into active and passive components using two complementary estimators built from very different identification strategies, and shown that the active--passive gap is robust across both methods, widens in concentrated-liquidity settings, and is most pronounced where active LPs can most efficiently target incoming flow.
Several natural extensions remain: re-aggregating LIFO positions by smart-contract deployer to recover an economic-agent-level decomposition, matched cross-chain comparisons that control for volatility, liquidity depth, and gas costs, and longitudinal analysis tracking how the gap evolves as L2 sequencing, JIT competition, and v4 hook usage mature.
More broadly, the results suggest that aggregate pool statistics---TVL, volume, and pool-level markout---are an incomplete picture of the returns to passive capital in concentrated-liquidity AMMs, and that LP-side decompositions of the kind we propose are a useful complement to existing market-quality metrics.

\iffalse
\subsection{Discussion}
\begin{itemize}
    \item Economic interpretation of the passive LP gap: who bears the cost?
    \item Implications for LP strategy: when does passive provision make sense?
    \item Implications for DEX mechanism design: fee structures, JIT mitigation
    \item Flashblocks / Unichain results: what do faster block times do to the gap?
    \item Limitations: markout as a proxy, pool selection, time period
\end{itemize}

\subsection{Conclusion}
\begin{itemize}
    \item Summary of main findings
    \item Contribution to market microstructure and DeFi literature
    \item Future work: fee-adjusted P\&L, broader pool coverage, dynamic LP strategies
\end{itemize}
\fi

%%% Local Variables:
%%% mode: LaTeX
%%% TeX-master: "../main"
%%% End:

\bibliography{active-passive-gap-bibliography}

\appendix
% fc-version appendices, rendered only when \shorttrue (see main.tex).

% Paragraphs in blue are NEW relative to the main build.

\section{Extended Related Work: Empirical Measurement of LP Profitability}\label{app:related-empirical}

Empirical studies of decentralized exchanges have progressed from simple return calculations to more detailed analyses of LP performance and flow composition. 
Early empirical work such as \citet{heimbach2022risks} evaluates overall LP returns across Uniswap v3 pools, documenting wide cross-sectional dispersion in realized profitability. \citet{lehar2025decentralized} use the full Uniswap transaction history to compare automated market making to limit-order-book equilibria, characterizing when AMM liquidity dominates and showing the absence of long-lived arbitrage. 
\citet{hasbrouck2025economic} develop a structural economic model of concentrated liquidity provision
in which LP supply is pinned down by a no-arbitrage risk-adjusted return condition, providing a benchmark against which realized LP performance can be evaluated. Most directly related to our setting, \citet{fritsch2024measuring} measure arbitrage losses against fee income across the largest Uniswap pools and find that arbitrage losses exceed fees in many pools, with v2 positions often more profitable than their v3 counterparts in the passive limit.

A parallel literature analyzes the composition of on-chain order flow. \citet{qin2022quantifying} provide a large-scale quantification of blockchain extractable value across DEX arbitrage, sandwich attacks, and liquidations, establishing the empirical scale of value transferred from passive participants to searchers. 
Markout-based methodologies have also been adopted in both academic and practitioner analyses to identify informed or arbitrage-driven flow in Uniswap markets; for example, \citet{zhu2024drives} use markout as an explanatory variable for liquidity supply and document its predictive power for future market depth. These approaches use post-trade price movements to estimate the extent to which LPs trade against price-moving order flow.

However, most existing empirical work analyzes LP performance at the pool level, implicitly treating liquidity providers as homogeneous. 
In practice, concentrated liquidity enables a wide range of strategies with different exposure to arbitrage and fee income.

Our paper addresses this gap by decomposing pool-level P\&L into active and passive liquidity provision using markout-based metrics across multiple chains. 
This allows us to quantify how arbitrage and trading flow are distributed across different types of LPs and to measure the resulting profitability gap between passive and strategically managed liquidity.

\section{Uniswap Protocol Mechanics}\label{app:mechanics}

Uniswap v2 implements a constant-product automated market maker in which token reserves $x$ and $y$ satisfy $xy = k$. Liquidity in v2 is supplied across the full price range, so all liquidity in the pool participates in a swap at the prevailing price. LPs may add or remove liquidity over time, but they cannot choose a narrower price range over which their capital is active.

Uniswap v3 introduced concentrated liquidity, extending the constant-product automated market maker framework. While the underlying invariant remains based on the constant-product relationship ($xy = k$), liquidity providers can allocate capital within user-specified price ranges rather than across the entire price curve.

Prices in Uniswap are discretized into ticks, where the price at tick $i$ is defined as $P(i) = 1.0001^i$. Each liquidity position specifies a price interval $[P_a, P_b]$ over which the provided liquidity is active. When the pool price $P$ lies within this interval, the position participates in trades and earns a pro-rata share of trading fees. 
If the price moves outside the specified range, the position becomes fully composed of a single asset and no longer earns fees until the price re-enters the range.

Each LP position is represented as a non-fungible position defined by its price bounds and liquidity amount. This design allows liquidity providers to adopt heterogeneous strategies ranging from wide passive ranges to highly concentrated liquidity that closely tracks the market price.

Uniswap v4 retains the concentrated-liquidity framework of v3 while introducing a singleton architecture and programmable hooks that allow additional pool-specific logic. 
The core distinction relevant for our analysis remains the same: liquidity can be concentrated over specific price ranges and actively repositioned over time. The version-specific event data used to reconstruct swaps and liquidity changes are described in Section~\ref{sec:data}.

\section{Methodology Details}\label{app:methoddetails}
\paraheader{LIFO subtraction: coverage and aggregation.} For a matched allocation leg $a$, let $\ell_a$ be the amount of liquidity matched between its mint and burn. 
We say that leg $a$ covers swap $s$ if the swap occurs strictly between the mint and burn in on-chain order and if the swap tick lies within the leg's tick range on v3 and v4. 
Formally,
\begin{gather*}
L_{s,a} = \ell_a\, \mathds{1}\{o(\mathrm{mint}_a) < o(s) < o(\mathrm{burn}_a)\}\,
\mathds{1}\{\tau_{\ell,a} \le \tau_s < \tau_{u,a}\}.
\end{gather*}
Let $L_s$ denote the in-range active liquidity recorded on the swap event, interpreted as the tick liquidity available after the execution. 
We define the active share of swap $s$ attributable to that matched leg as
\begin{gather*}
w^{\mathrm{act}}_{s,a} = \frac{L_{s,a}}{L_s}.
\end{gather*}
Using the dollar markout $M_s$ defined above, the active dollar volume and active dollar markout attributed to allocation leg $a$ from swap $s$ are
\begin{gather*}
V^{\mathrm{act}}_{s,a} = w^{\mathrm{act}}_{s,a}\, q_{s,\mathrm{in}}\, p_{s,\mathrm{in},h}, \qquad
M^{\mathrm{act}}_{s,a} = w^{\mathrm{act}}_{s,a}\, M_s.
\end{gather*}
Summing across all matched allocation legs and swaps yields the LIFO-based active volume and active markout.

The passive component is defined residually as the part of the swap not attributed to the matched short-lived liquidity. 
Thus, for swap $s$, the passive share is $1 - \sum_a w^{\mathrm{act}}_{s,a}$, and the corresponding passive dollar volume and passive dollar markout are
\begin{gather*}
V^{\mathrm{pass}}_s = w^{\mathrm{pass}}_s\, q_{s,\mathrm{in}}\, p_{\mathrm{in},h}, \qquad
M^{\mathrm{pass}}_s = w^{\mathrm{pass}}_s\, M_s, \qquad
w^{\mathrm{pass}}_s = 1 - \sum_a w^{\mathrm{act}}_{s,a}.
\end{gather*}
Equivalently, after aggregating active quantities, passive volume and passive markout can be obtained as total volume and total markout minus their active counterparts.

\paraheader{Infinitesimal method: reserve-change derivation.} For each swap $s$, let $p_0$ and $p_1$ denote
the pool price immediately before and after the swap. We compute the curve implied reserve change magnitudes
\begin{gather*}
\Delta x_s = |x(p_1) - x(p_0)|, \qquad \Delta y_s = |y(p_1) - y(p_0)|.
\end{gather*}
We then assign these two reserve changes to input and output according to the observed swap direction. 
If the trader sells token X and receives token Y, then
\begin{gather*}
q_{s,\mathrm{in}} = \frac{\Delta x_s}{1-f}, \qquad q_{s,\mathrm{out}} = \Delta y_s.
\end{gather*}
If instead the trader sells token Y and receives token X, then
\begin{gather*}
q_{s,\mathrm{in}} = \frac{\Delta y_s}{1-f}, \qquad q_{s,\mathrm{out}} = \Delta x_s.
\end{gather*}
Here $f$ is the pool fee tier. The factor $1/(1-f)$ accounts for the fact that Uniswap fees are taken from the input amount before the post fee amount moves the invariant in v3 and v4. 
These $q_{s,\mathrm{in}}$ and $q_{s,\mathrm{out}}$ are then used to compute the raw and dollar markouts for the infinitesimal LP.

For v3 and v4, we calculate the infinitesimal benchmark at unit liquidity, $L = 1$ and fees are deducted upfront so only $(1-f)x$ is used in the swap accounting, when a swapper sells token X and buys token Y. 
On the other hand, v2 requires a different treatment because fees are handled differently. 
In v2, fees are not taken out of the swap input; instead, the full input amount enters the pool and the fee is embedded in the reserves. This means that the pool receives the full $x$, not $(1-f)x$, and the fee is reflected in the post swap price change.

% \begin{table}[ht]
% \centering
% \small
% \renewcommand{\arraystretch}{1.08}
% \begin{tabular}{lll}
% \toprule
% Attribute & LIFO subtraction & Infinitesimal LP \\
% \midrule
% Profiles passive LPs & Yes & Yes \\
% Profiles active LPs & Yes & No \\
% LP decomposition granularity & Swap-level & Swap-level passive only \\
% Passive markout CI & Yes & Yes \\
% Active markout CI & Yes & No \\
% Dollar scale of passive P\&L & Yes & No; only normalized P\&L \\
% Assumptions on active LP behavior & Yes & No \\
% Robust to sandwiches & Yes & No \\
% Data required to compute & Mint, burn, and swap history & Time series of pool spot prices \\
% \bottomrule
% \end{tabular}
% \caption{Comparison of the two passive-LP methodologies.}
% \label{tab:methodcomparison}
% \end{table}

\ifshort
\paraheader{Identifying Sandwiches.}
We have two types of sandwich attacks. 
Standard sandwiches are identified as front running and back running swaps attributed to the same observed attacker around one or more victim transactions in the same pool. 
Same transaction sandwiches consist of a corresponding front run, victim, and back run sequence contained within a single transaction hash and identified using log order. 
We remove the attacker legs and the associated victim swaps because the temporary mechanical reversal in the pool price can produce disproportionately large infinitesimal markouts.
\else
\fi

\paraheader{Comparing passive LP P\&L methods (row by row).}
Table~\ref{tab:method_comparison} summarizes the main differences and discusses the two methods of breaking down passive and active markouts. The first two rows ask whether or not a method can attribute P\&L to a given type of LP. 
Both methods can profile passive LPs, but only LIFO subtraction also profiles active LPs, since it explicitly decomposes aggregate LP outcomes into active and passive components. The infinitesimal LP method instead focuses only on the exposure of a marginal passive LP and therefore does not provide estimates for active LP performance.

The ``LP decomposition granularity'' row describes the level at which passive and active LP outcomes are separated. LIFO subtraction performs this decomposition at the swap level, assigning each trade between active and passive liquidity. The infinitesimal LP method also operates at the swap level, but only for passive exposure: it computes the markout of an infinitesimal passive LP along the realized price path rather than decomposing observed LP outcomes into active and passive components.

The confidence-interval rows describe whether each method can produce uncertainty estimates for markouts. Both methods can produce confidence intervals for passive markouts, but only LIFO subtraction can also produce confidence intervals for active markouts, because only LIFO subtraction estimates active LP performance. 
The ``Dollar scale of passive P\&L'' row asks whether the method recovers the aggregate dollar magnitude of passive LP gains and losses. 
LIFO subtraction does so because it starts from observed aggregate LP outcomes and subtracts the active component. 
The infinitesimal LP benchmark does not; it instead produces a normalized benchmark for the performance of a marginal passive LP.

The ``Assumptions on active LP behavior'' row captures whether the method requires a rule for identifying active liquidity provision. LIFO subtraction does require such assumptions, since it must decide which LP actions count as active in order to subtract them from the aggregate. 
The infinitesimal LP method does not require assumptions about active LP behavior because it does not attempt to identify active LPs. 
This is important when evaluating passive LP outcomes, since the former method attributes what remains after subtracting out active LP outcomes to passive LPs, while the latter method attempts to measure passive LP outcomes directly.

The ``Robust to sandwiches'' row concerns whether the method is naturally robust to mechanically induced price reversals such as sandwich attacks. 
LIFO subtraction is less sensitive to this issue because it measures realized LP outcomes through the active-passive decomposition. 
The infinitesimal LP method is more sensitive because it evaluates passive exposure along the full realized pool-price path, so temporary price distortions can generate anomalously large markouts unless sandwich-related swaps are filtered out.

The final row describes the data required to compute each method. 
LIFO subtraction requires mint, burn, and swap data, since it reconstructs LP position changes and allocates swap-level outcomes across active and passive liquidity. 
The infinitesimal LP method requires only swap data, or more precisely, the price path for a pool obtained from its trade history, since it computes the benchmark from the realized sequence of pool prices and swap quantities rather than from observed LP mint and burn behavior. 
In this regard, for computing passive LP outcomes, the infinitesimal method has a substantial practical advantage in terms of the data and complexity required to implement it relative to the LIFO subtraction method.

\paraheader{Confidence intervals for passive markouts.} Confidence intervals are reported for the LIFO and
infinitesimal methods, since these produce passive observations at the swap level. Specifically,
for each swap $s$ we first compute the normalized passive markout
\begin{gather*}
m^{\mathrm{pass}}_s = \frac{M^{\mathrm{pass}}_s}{V^{\mathrm{pass}}_s},
\end{gather*}
and use passive volume $V^{\mathrm{pass}}_s$ as the weight. We then form the volume weighted mean
and its linearized (ratio estimator) standard error,
\begin{gather*}
\bar m^{\mathrm{pass}} = \frac{\sum_s V^{\mathrm{pass}}_s\, m^{\mathrm{pass}}_s}{\sum_s V^{\mathrm{pass}}_s},
\qquad
\mathrm{SE}(\bar m^{\mathrm{pass}})^2
= \frac{\sum_s (V^{\mathrm{pass}}_s)^2\, (m^{\mathrm{pass}}_s - \bar m^{\mathrm{pass}})^2}
{\left(\sum_s V^{\mathrm{pass}}_s\right)^2},
\end{gather*}
and report the normal approximation interval $\bar m^{\mathrm{pass}} \pm 1.96\,\mathrm{SE}(\bar m^{\mathrm{pass}})$.

For the difference between the overall and passive markout, $D = \bar m^{\mathrm{all}} - \bar m^{\mathrm{pass}}$, the two ratios are built from the same swaps and are therefore correlated, so their variances do not add. 
We instead linearize the difference directly: defining the per-swap influence contribution
\begin{gather*}
d_s = \frac{V_s\left(m_s - \bar m^{\mathrm{all}}\right)}{\sum_s V_s}
- \frac{V^{\mathrm{pass}}_s\left(m^{\mathrm{pass}}_s - \bar m^{\mathrm{pass}}\right)}{\sum_s V^{\mathrm{pass}}_s},
\qquad
\mathrm{SE}(D)^2 = \sum_s d_s^2.
\end{gather*}
We report $D \pm 1.96\,\mathrm{SE}(D)$.

\section{Passive LP P\&L: Exact Subtraction Method}\label{app:exact}

We also consider an exact-matching approach defined on the same tuple (Chain, Pool, $\tau_\ell$, $\tau_u$, LP) as in the LIFO procedure, consisting of chain, pool, lower tick, upper tick, and LP address. 
In this method, we retain only mint-burn matches that are consecutive in on-chain order within the tuple and exactly matched in liquidity. 
Exact matching may occur either one-to-one or across multiple consecutive burns: a single mint may be matched with multiple consecutive burns, provided that the sum of the burned liquidity equals the minted liquidity exactly, with no residual unmatched amount on either side.

For each such exact mint-burn match, we gather all associated collect events occurring between the mint and the final matched burn, so that any fees withdrawn during the life of the position are included in the realized position outcome. 
Let $m_0, m_1$ denote the token amounts deposited at mint, and let $c_0, c_1$ denote the total token amounts collected over all collect events associated with that exact matched position, including intermediate and terminal collects. 
The active dollar markout is then computed as
\begin{gather*}
M^{\mathrm{act}} = (c_0 - m_0)p_0 + (c_1 - m_1)p_1,
\end{gather*}
where $p_0$ and $p_1$ are reference prices of token 0 and token 1 in a common numeraire.

This construction yields active markout and the corresponding active volume for exactly matched positions, i.e., $c - m$ of either token. 
We then aggregate these active quantities across positions and define the passive component residually as total minus active. Hence, under exact subtraction, passive markout and passive volume are identified only at the aggregate level:
\begin{gather*}
\text{Passive aggregate} = \text{Total aggregate} - \text{Exact-match active aggregate}.
\end{gather*}

Table~\ref{tab:v3_active_lifo_exact} compares the LIFO and exact subtraction methods. 
The estimates of active volume shares and passive markouts across pools are very similar for both approaches. 
The differences are small relative to the level of the markouts in most cases. 
This implies that the subtraction-based decomposition is not sensitive to the particular mint-burn matching convention and justifies the use of LIFO as the main specification in the main analysis.

% In your main preamble, include:
% \usepackage{pgfplotstable}
% \usepackage{booktabs}
% \usepackage{makecell}
% \usepackage{float}

\begin{table}[!htbp]
\centering
\scriptsize
\setlength{\tabcolsep}{5pt}
\renewcommand{\arraystretch}{1.08}

\begin{filecontents*}{sections/updated/v3_active_volume_passive_table.csv}
pool_label;active_lifo_volume_pct;active_exact_volume_pct;passive_lifo_bps;passive_exact_bps
\textbf{USDC-WETH};;;;
Base 5 bps;3.82;3.56;-0.706874;-0.707497
Arbitrum 5 bps;0.03;0.01;-0.769710;-0.769485
Ethereum 1 bps;0.33;0.32;-0.000559;0.002104
Ethereum 5 bps;1.36;1.30;-1.506019;-1.521428
Ethereum 30 bps;0.68;0.67;-0.866415;-0.863497
\textbf{USDT-WETH};;;;
Base 5 bps;0.00;0.00;-0.574969;-0.574969
Arbitrum 5 bps;0.00;0.00;-0.449636;-0.550026
Ethereum 1 bps;0.44;0.42;-0.138684;-0.137273
Ethereum 5 bps;1.23;1.16;-1.309165;-1.300011
Ethereum 30 bps;0.93;0.92;1.001649;1.001035
\textbf{USDC-cbBTC};;;;
Base 5 bps;0.00;0.00;-0.568331;-0.568547
\textbf{USDC-WBTC};;;;
Arbitrum 5 bps;0.00;0.00;-0.589305;-0.589479
Ethereum 1 bps;0.06;0.03;-2.962705;-2.960789
Ethereum 5 bps;1.90;1.78;-0.408402;-0.395788
Ethereum 30 bps;1.54;1.51;0.209764;0.221894
\textbf{USDT-WBTC};;;;
Arbitrum 5 bps;0.00;0.00;-0.722146;-0.722148
Ethereum 1 bps;1.05;0.97;-4.239112;-4.263411
Ethereum 5 bps;1.25;1.14;-0.254329;-0.231116
Ethereum 30 bps;0.84;0.79;2.519784;2.535369
\end{filecontents*}
\pgfplotstabletypeset[
    col sep=semicolon,
    string type,
    columns={pool_label,active_lifo_volume_pct,active_exact_volume_pct,passive_lifo_bps,passive_exact_bps},
    columns/pool_label/.style={column name={\makecell[l]{v3 pool\\label}}, column type={l}},
    columns/active_lifo_volume_pct/.style={column name={\makecell[c]{active LIFO\\volume (\%)}}, column type={r}},
    columns/active_exact_volume_pct/.style={column name={\makecell[c]{active exact\\volume (\%)}}, column type={r}},
    columns/passive_lifo_bps/.style={column name={\makecell[c]{passive LIFO\\(bps)}}, column type={r}},
    columns/passive_exact_bps/.style={column name={\makecell[c]{passive exact\\(bps)}}, column type={r}},
    every head row/.style={before row=\toprule, after row=\midrule},
    every row no 6/.style={before row=\midrule},
    every row no 12/.style={before row=\midrule},
    every row no 14/.style={before row=\midrule},
    every row no 19/.style={before row=\midrule},
    every last row/.style={after row=\bottomrule},
]{sections/updated/v3_active_volume_passive_table.csv}

\caption{Uniswap v3 active LIFO and exact volume shares with passive LIFO and exact markouts.}
\label{tab:v3_active_lifo_exact}
\end{table}

\section{Descriptive Statistics}\label{app:descriptive}

Table~\ref{tab:v2-volume-shares}--\ref{tab:v4-volume-shares} summarize trading activity, liquidity-event intensity, sandwich volume, and active LIFO volume for the v2, v3, and v4 Uniswap protocol versions. Trading
volume is quite concentrated in WETH-stablecoin pools. V2's reported volume comes mainly from the Ethereum USDT-WETH and USDC-WETH pools, which stood at around \$1.04B and \$968M respectively. 
The largest pools in v3 are USDC-WETH on Arbitrum and Ethereum at the 5 bps tier with \$61.13B and \$46.84B in volume, respectively, followed by Ethereum USDT-WETH at the 1 bps and Base USDC-WETH 5 bps. While v4 volumes are smaller than the mature v3 pools, they are still concentrated in Ethereum WETH-stablecoin markets, especially USDC-WETH and USDT-WETH (5 bps tier).

% In your main preamble, include:
% \usepackage{pgfplotstable}
% \usepackage{booktabs}
% \usepackage{makecell}
% \usepackage{float}

\begin{table}[H]
\centering
\scriptsize
\setlength{\tabcolsep}{5pt}
\renewcommand{\arraystretch}{1.08}

\begin{filecontents*}{sections/updated/v2_table_volume_shares.csv}
pool_label;num_trades;num_liquidity_events;total_volume;sandwich_volume_pct;active_lifo_volume_pct
\textbf{USDC-WETH};;;;;
Base 30 bps;2,463,854;2,464,144;515.67M;0.03;0.00
Arbitrum 30 bps;131,597;131,612;11.49M;0.22;0.00
Ethereum 30 bps;560,857;561,059;967.72M;1.47;0.00
\textbf{USDT-WETH};;;;;
Arbitrum 30 bps;97,124;97,744;10.66M;0.12;0.00
Ethereum 30 bps;525,709;525,854;1.04B;0.62;0.00
\end{filecontents*}
\pgfplotstabletypeset[
    col sep=semicolon,
    string type,
    columns={pool_label,num_trades,num_liquidity_events,total_volume,sandwich_volume_pct,active_lifo_volume_pct},
    columns/pool_label/.style={column name={\makecell[l]{v2 pool\\label}}, column type={l}},
    columns/num_trades/.style={column name={\makecell[c]{number of\\trades}}, column type={r}},
    columns/num_liquidity_events/.style={column name={\makecell[c]{number of\\liquidity events}}, column type={r}},
    columns/total_volume/.style={column name={\makecell[c]{total\\volume (\$)}}, column type={r}},
    columns/sandwich_volume_pct/.style={column name={\makecell[c]{sandwich\\volume (\%)}}, column type={r}},
    columns/active_lifo_volume_pct/.style={column name={\makecell[c]{active LIFO\\volume (\%)}}, column type={r}},
    every head row/.style={before row=\toprule, after row=\midrule},
    every row no 4/.style={before row=\midrule},
    every last row/.style={after row=\bottomrule},
]{sections/updated/v2_table_volume_shares.csv}

\caption{Uniswap v2 pool-level volume shares.}
\label{tab:v2-volume-shares}
\end{table}

Liquidity-event activity fluctuates dramatically with protocol version. In v2, active LIFO volume is basically zero across all pools. Note that Uniswap v2 does not permit range specific liquidity placement, but LPs can mint and burn in full-range liquidity. 
Our statement that active liquidity is absent in the selected v2 pools is therefore an empirical conclusion under the LIFO definition rather than a general restriction imposed by the protocol. Specifically, within the matching horizons used in the analysis, the LIFO procedure identifies no mint and burn sequences satisfying the active liquidity criterion. The estimated active share is consequently zero, and the passive LIFO markout equals the aggregate pool markout.

Liquidity management is much more pronounced in v3, especially in the Base and Arbitrum USDC-WETH pools with millions of liquidity events. 
V4 also has high event intensity, with liquidity-event counts often close to swap counts, reflecting the more granular event structure and presence of actively managed positions in some pools.

The sandwich activity is concentrated almost exclusively on Ethereum pools with low fees. 
Sandwich volume shares above 50\% are evident in v3 Ethereum 1 bps USDC-WETH and USDT-WETH pools, while most L2 pools and higher fee Ethereum pools have sandwich shares less than 1\%. 
V4 follows a similar pattern with Ethereum 1 bps USDC-WETH and USDT-WETH having sandwich shares of 47.65\% and 76.12\%, respectively. 
The concentration implies that when interpreting markout based passive LP estimates, low fee Ethereum pools should be separated.

Active LIFO volume is also very concentrated. 
The biggest active share in v3 is Base USDC-WETH 5 bps with 3.56\% and there are a few Ethereum 5 bps and 30 bps pools with active share of 1-2\% range. Arbitrum v3 pools typically have active shares near zero. 
In v4, most pools on Ethereum have small active LIFO shares, but a few L2 pools have very large active shares, e.g.\ Arbitrum USDT-WETH, Base USDC-cbBTC, and Arbitrum WBTC-USDT. Outliers here suggest that active liquidity provision in v4 is pool specific and may be a function of specialized LP strategies or pool-level design features rather than a general active LP participation across markets.

\section{Additional Results: Fee Tiers, Cross-Chain, and Method Comparison}\label{app:additional}

\paraheader{Fee tiers and passive LP profitability.}

Figure~\ref{fig:feetier} suggests high-fee pools are often, but not uniformly, better for passive LPs. Passive LIFO markouts are typically higher in 30 bps pools than in 5 bps pools, consistent with fees providing a larger cushion against adverse selection. 
This pattern is also related to active LP incentives. Active liquidity provision is more attractive in low-fee pools when LPs can concentrate liquidity around expected trades and earn fees with limited exposure. 
This behavior is more feasible on Ethereum because the public mempool allows LPs to observe incoming flow before block inclusion and provide short-lived liquidity around anticipated swaps. 
By contrast, 30 bps pools tend to show less active LIFO volume and, in many cases, appear more favorable for passive liquidity provision.

The figure also shows differences across chains. 
In many 5 bps pools on Arbitrum and Base, passive markouts tend to cluster around zero, while Ethereum pools show more dispersion across pairs and fee tiers. 
This suggests that fee tier interacts with market structure at the chain level: higher fees improve passive outcomes, but the size of the active-passive gap also depends on the ease of LPs actively repositioning liquidity around trades.

\begin{figure}[!htbp]
\centering
\scalebox{0.58}{\begin{tikzpicture}
\definecolor{pairblue}{RGB}{0,114,178}
\definecolor{pairgreen}{RGB}{0,158,115}
\definecolor{pairorange}{RGB}{230,159,0}
\definecolor{pairred}{RGB}{204,68,102}
\definecolor{pairpurple}{RGB}{136,34,204}
\begin{axis}[
width=13.5cm,
height=7.2cm,
major grid style={draw=gray!25},
minor grid style={draw=gray!10},
title style={font=\small\bfseries, align=center},
xlabel style={font=\scriptsize},
ylabel style={font=\scriptsize},
tick label style={font=\scriptsize},
tick align=outside,
legend style={at={(0.5,1.04)}, anchor=south, legend columns=5, draw=none, fill=none, font=\tiny, column sep=0.08cm, /tikz/every even column/.append style={column sep=0.08cm}},
xmajorgrids=false,
xminorgrids=false,
ymajorgrids=true,
yminorgrids=true,
ylabel={Passive LIFO (bps)},
xlabel={},
xmin=0.55,
xmax=3.45,
xtick={0.92,1.08,1.92,2.08,2.92,3.08},
xticklabels={5,30,5,30,5,30},
x tick label style={font=\scriptsize, yshift=0pt},
extra x ticks={0.92,1.08,1.92,2.08,2.92,3.08},
extra x tick labels={},
extra x tick style={grid=major, major grid style={draw=gray!25}, tick style={draw=none}},
clip=false,
]
\addplot+[only marks, forget plot, mark=triangle*, mark size=2.6pt, color=pairblue, line width=0.8pt, mark options={fill=pairblue, draw=pairblue, fill opacity=0.42, draw opacity=0.8}] coordinates {(1.08,19.15383325) (2.08,20.82207889) (3.08,9.320412415)};
\addplot+[only marks, forget plot, mark=*, mark size=2.6pt, color=pairblue, line width=0.8pt, mark options={fill=pairblue, draw=pairblue, fill opacity=0.42, draw opacity=0.8}] coordinates {(0.92,-0.76971) (1.92,-0.706874) (2.92,-1.506019) (3.08,-0.866415)};
\addplot+[only marks, forget plot, mark=square*, mark size=3.0pt, color=pairblue, line width=0.8pt, mark options={fill=pairblue, draw=pairblue, fill opacity=0.42, draw opacity=0.8}] coordinates {(0.92,-1.431669543) (1.92,-0.977858869) (2.92,-3.070001001) (3.08,-0.909249237)};
\addplot+[only marks, forget plot, mark=triangle*, mark size=2.6pt, color=pairgreen, line width=0.8pt, mark options={fill=pairgreen, draw=pairgreen, fill opacity=0.42, draw opacity=0.8}] coordinates {(1.08,21.48637526) (3.08,9.329889084)};
\addplot+[only marks, forget plot, mark=*, mark size=2.6pt, color=pairgreen, line width=0.8pt, mark options={fill=pairgreen, draw=pairgreen, fill opacity=0.42, draw opacity=0.8}] coordinates {(0.92,-0.449636) (1.92,-0.574969) (2.92,-1.309165) (3.08,1.001649)};
\addplot+[only marks, forget plot, mark=square*, mark size=3.0pt, color=pairgreen, line width=0.8pt, mark options={fill=pairgreen, draw=pairgreen, fill opacity=0.42, draw opacity=0.8}] coordinates {(0.92,-1.972888288) (2.92,-1.756678938) (3.08,-3.044381418)};
\addplot+[only marks, forget plot, mark=*, mark size=2.6pt, color=pairorange, line width=0.8pt, mark options={fill=pairorange, draw=pairorange, fill opacity=0.42, draw opacity=0.8}] coordinates {(1.92,-0.568331)};
\addplot+[only marks, forget plot, mark=square*, mark size=3.0pt, color=pairorange, line width=0.8pt, mark options={fill=pairorange, draw=pairorange, fill opacity=0.42, draw opacity=0.8}] coordinates {(1.92,-0.365854383)};
\addplot+[only marks, forget plot, mark=*, mark size=2.6pt, color=pairred, line width=0.8pt, mark options={fill=pairred, draw=pairred, fill opacity=0.42, draw opacity=0.8}] coordinates {(0.92,-0.589305) (2.92,-0.408402) (3.08,0.209764)};
\addplot+[only marks, forget plot, mark=square*, mark size=3.0pt, color=pairred, line width=0.8pt, mark options={fill=pairred, draw=pairred, fill opacity=0.42, draw opacity=0.8}] coordinates {(0.92,-1.287180197) (2.92,-1.278666208) (3.08,-2.020871007)};
\addplot+[only marks, forget plot, mark=*, mark size=2.6pt, color=pairpurple, line width=0.8pt, mark options={fill=pairpurple, draw=pairpurple, fill opacity=0.42, draw opacity=0.8}] coordinates {(0.92,-0.722146) (2.92,-0.254329) (3.08,2.519784)};
\addplot+[only marks, forget plot, mark=square*, mark size=3.0pt, color=pairpurple, line width=0.8pt, mark options={fill=pairpurple, draw=pairpurple, fill opacity=0.42, draw opacity=0.8}] coordinates {(0.92,-2.788935587) (2.92,-1.147260842) (3.08,0.01653679)};
\addplot+[forget plot, no markers, dashed, line width=0.55pt, color=black] coordinates {(0.92,-1.251433827) (1.08,20.32010426)};
\addplot+[forget plot, no markers, dashed, line width=0.55pt, color=black] coordinates {(1.92,-0.6387774504) (2.08,20.82207889)};
\addplot+[forget plot, no markers, dashed, line width=0.55pt, color=black] coordinates {(2.92,-1.341315249) (3.08,1.555711863)};
\addplot+[only marks, forget plot, mark=star, mark size=3.6pt, color=black, line width=1.0pt, mark options={fill=black, draw=black}] coordinates {(0.92,-1.251433827) (1.08,20.32010426) (1.92,-0.6387774504) (2.08,20.82207889) (2.92,-1.341315249) (3.08,1.555711863)};
\addlegendimage{only marks, mark=*, color=pairblue, mark options={fill=pairblue, draw=pairblue, fill opacity=0.42, draw opacity=0.8}}
\addlegendentry{\mbox{USDC-WETH}}
\addlegendimage{only marks, mark=*, color=pairgreen, mark options={fill=pairgreen, draw=pairgreen, fill opacity=0.42, draw opacity=0.8}}
\addlegendentry{\mbox{USDT-WETH}}
\addlegendimage{only marks, mark=*, color=pairorange, mark options={fill=pairorange, draw=pairorange, fill opacity=0.42, draw opacity=0.8}}
\addlegendentry{\mbox{USDC-cbBTC}}
\addlegendimage{only marks, mark=*, color=pairred, mark options={fill=pairred, draw=pairred, fill opacity=0.42, draw opacity=0.8}}
\addlegendentry{\mbox{WBTC-USDC}}
\addlegendimage{only marks, mark=*, color=pairpurple, mark options={fill=pairpurple, draw=pairpurple, fill opacity=0.42, draw opacity=0.8}}
\addlegendentry{\mbox{WBTC-USDT}}
\addlegendimage{only marks, mark=triangle*, color=black}
\addlegendentry{v2}
\addlegendimage{only marks, mark=*, color=black}
\addlegendentry{v3}
\addlegendimage{only marks, mark=square*, color=black, mark options={fill=black, draw=black}}
\addlegendentry{v4}
\addlegendimage{only marks, mark=star, mark size=3.6pt, color=black}
\addlegendentry{overall mean}

\node[anchor=north, font=\scriptsize] at (axis description cs:0.155,-0.12) {Arbitrum};
\node[anchor=north, font=\scriptsize] at (axis description cs:0.500,-0.12) {Base};
\node[anchor=north, font=\scriptsize] at (axis description cs:0.845,-0.12) {Ethereum};
\node[anchor=north, font=\scriptsize] at (axis description cs:0.500,-0.24) {Fee Tier (bps) within Chain};
\end{axis}
\end{tikzpicture}
%%%%%%%%%%%
}
\caption{Passive LIFO markout (bps) by fee tier within each chain, for Uniswap v2, v3, and v4 pools. 
Each marker is a pool, colored by pair, with marker shape denoting the protocol version; the horizontal marks show the mean across pools at each chain--fee-tier combination.}
\label{fig:feetier}
\end{figure}

\paraheader{Cross-chain differences in the active-passive gap.}

Figure~\ref{fig:chain} shows that the difference between overall and passive LP profitability is generally larger and more dispersed on Ethereum than on the L2s. 
Arbitrum and Base are mostly clustered around zero, while Ethereum pools more often display a positive gap, suggesting that passive LPs there tend to perform worse than aggregate pool-level measures imply.

\begin{figure}[!htbp]
\centering
\scalebox{0.58}{\begin{tikzpicture}
\definecolor{pairblue}{RGB}{0,114,178}
\definecolor{pairgreen}{RGB}{0,158,115}
\definecolor{pairorange}{RGB}{230,159,0}
\definecolor{pairred}{RGB}{204,68,102}
\definecolor{pairpurple}{RGB}{136,34,204}
\begin{axis}[
width=13.5cm,
height=7.2cm,
grid=both,
major grid style={draw=gray!25},
minor grid style={draw=gray!10},
title style={font=\small\bfseries, align=center},
xlabel style={font=\scriptsize},
ylabel style={font=\scriptsize},
tick label style={font=\scriptsize},
tick align=outside,
legend style={at={(0.5,1.04)}, anchor=south, legend columns=5, draw=none, fill=none, font=\tiny, column sep=0.08cm, /tikz/every even column/.append style={column sep=0.08cm}},
ylabel={Overall $-$ Passive LIFO (bps)},
xlabel={Chain},
symbolic x coords={Arbitrum,Base,Ethereum},
xtick=data,
enlarge x limits=0.18,
x tick label style={anchor=north, yshift=-3pt, font=\scriptsize},
]
\addplot+[only marks, forget plot, mark=triangle*, mark size=2.6pt, color=pairblue, line width=0.8pt, mark options={fill=pairblue, draw=pairblue, fill opacity=0.42, draw opacity=0.8}] coordinates {(Arbitrum,0) (Base,0) (Ethereum,0)};
\addplot+[only marks, forget plot, mark=*, mark size=2.6pt, color=pairblue, line width=0.8pt, mark options={fill=pairblue, draw=pairblue, fill opacity=0.42, draw opacity=0.8}] coordinates {(Arbitrum,0.037189) (Base,0.025698) (Ethereum,0.355615) (Ethereum,0.159169)};
\addplot+[only marks, forget plot, mark=square*, mark size=3.0pt, color=pairblue, line width=0.8pt, mark options={fill=pairblue, draw=pairblue, fill opacity=0.42, draw opacity=0.8}] coordinates {(Arbitrum,0) (Base,0) (Ethereum,0.0371) (Ethereum,0.0333)};
\addplot+[only marks, forget plot, mark=triangle*, mark size=2.6pt, color=pairgreen, line width=0.8pt, mark options={fill=pairgreen, draw=pairgreen, fill opacity=0.42, draw opacity=0.8}] coordinates {(Arbitrum,0) (Ethereum,0)};
\addplot+[only marks, forget plot, mark=*, mark size=2.6pt, color=pairgreen, line width=0.8pt, mark options={fill=pairgreen, draw=pairgreen, fill opacity=0.42, draw opacity=0.8}] coordinates {(Arbitrum,-2.9e-05) (Base,1e-06) (Ethereum,0.398407) (Ethereum,0.183949)};
\addplot+[only marks, forget plot, mark=square*, mark size=3.0pt, color=pairgreen, line width=0.8pt, mark options={fill=pairgreen, draw=pairgreen, fill opacity=0.42, draw opacity=0.8}] coordinates {(Arbitrum,0.2569) (Ethereum,0.1416) (Ethereum,0.024)};
\addplot+[only marks, forget plot, mark=*, mark size=2.6pt, color=pairorange, line width=0.8pt, mark options={fill=pairorange, draw=pairorange, fill opacity=0.42, draw opacity=0.8}] coordinates {(Base,1e-06)};
\addplot+[only marks, forget plot, mark=square*, mark size=3.0pt, color=pairorange, line width=0.8pt, mark options={fill=pairorange, draw=pairorange, fill opacity=0.42, draw opacity=0.8}] coordinates {(Base,-0.5463)};
\addplot+[only marks, forget plot, mark=*, mark size=2.6pt, color=pairred, line width=0.8pt, mark options={fill=pairred, draw=pairred, fill opacity=0.42, draw opacity=0.8}] coordinates {(Arbitrum,-0.000158) (Ethereum,0.453938) (Ethereum,0.31454)};
\addplot+[only marks, forget plot, mark=square*, mark size=3.0pt, color=pairred, line width=0.8pt, mark options={fill=pairred, draw=pairred, fill opacity=0.42, draw opacity=0.8}] coordinates {(Arbitrum,0) (Ethereum,0.2849) (Ethereum,0.0175)};
\addplot+[only marks, forget plot, mark=*, mark size=2.6pt, color=pairpurple, line width=0.8pt, mark options={fill=pairpurple, draw=pairpurple, fill opacity=0.42, draw opacity=0.8}] coordinates {(Arbitrum,9.3e-05) (Ethereum,0.445023) (Ethereum,0.185642)};
\addplot+[only marks, forget plot, mark=square*, mark size=3.0pt, color=pairpurple, line width=0.8pt, mark options={fill=pairpurple, draw=pairpurple, fill opacity=0.42, draw opacity=0.8}] coordinates {(Arbitrum,0.3184) (Ethereum,0.9612) (Ethereum,0.11)};

% Volume-weighted mean across all included pools within each chain.
\addplot+[forget plot, mark=star, mark size=3.6pt, color=black, line width=0.9pt, dashed, mark options={fill=black, draw=black}] coordinates {(Arbitrum,0.02879277192) (Base,0.01613893326) (Ethereum,0.2899732533)};
\addlegendimage{only marks, mark=*, color=pairblue, mark options={fill=pairblue, draw=pairblue, fill opacity=0.42, draw opacity=0.8}}
\addlegendentry{\mbox{USDC-WETH}}
\addlegendimage{only marks, mark=*, color=pairgreen, mark options={fill=pairgreen, draw=pairgreen, fill opacity=0.42, draw opacity=0.8}}
\addlegendentry{\mbox{USDT-WETH}}
\addlegendimage{only marks, mark=*, color=pairorange, mark options={fill=pairorange, draw=pairorange, fill opacity=0.42, draw opacity=0.8}}
\addlegendentry{\mbox{USDC-cbBTC}}
\addlegendimage{only marks, mark=*, color=pairred, mark options={fill=pairred, draw=pairred, fill opacity=0.42, draw opacity=0.8}}
\addlegendentry{\mbox{WBTC-USDC}}
\addlegendimage{only marks, mark=*, color=pairpurple, mark options={fill=pairpurple, draw=pairpurple, fill opacity=0.42, draw opacity=0.8}}
\addlegendentry{\mbox{WBTC-USDT}}
\addlegendimage{only marks, mark=triangle*, color=black}
\addlegendentry{v2}
\addlegendimage{only marks, mark=*, color=black}
\addlegendentry{v3}
\addlegendimage{only marks, mark=square*, color=black, mark options={fill=black, draw=black}}
\addlegendentry{v4}
\addlegendimage{mark=star, mark size=3.6pt, color=black, dashed, line width=0.9pt, mark options={fill=black, draw=black}}
\addlegendentry{VW mean}
\end{axis}
\end{tikzpicture}}
\caption{Overall minus passive LIFO markout (bps) by chain, for Uniswap v2, v3, and v4 pools. 
Each marker is a pool, colored by pair, with marker shape denoting the protocol version; the VW-mean marker shows, for each chain, the volume-weighted mean across all included pools, protocol versions, and fee tiers, weighted by total pool dollar volume. Positive values indicate that passive LPs underperform the pool aggregate.}
\label{fig:chain}
\end{figure}

We see this pattern as descriptive rather than causal. Ethereum, Arbitrum and Base are different not only in terms of block times, but also in terms of gas costs, liquidity depth, user and flow composition, MEV infrastructure and the set of active LPs participating on each chain. 
All these factors may affect the active-passive gap. Mechanistically, there are two forces pulling in opposite directions on L2s: shorter block times could reduce the value of timing liquidity around incoming flow, narrowing the gap, and lower gas costs could make frequent liquidity updates cheaper, potentially widening the gap. The small gaps observed on Arbitrum and Base suggest that the first force may dominate in our sample, or that active LP participation is less developed in these pools. 
A causal separation of these channels would require a matched cross-chain analysis controlling for volatility, volume, liquidity depth, gas costs and flow composition.

\paraheader{Comparison of LIFO and infinitesimal estimates.}

Figure~\ref{fig:cirange} plots the absolute difference between their estimates against the infinitesimal confidence interval width and compares the passive LIFO and infinitesimal methods. 
Points below the 45-degree line indicate cases where the difference between methods is smaller than the uncertainty range of the infinitesimal estimate. 
Almost all pools fall below this line, showing that the two approaches provide broadly similar passive LP profitability estimates despite very different constructions. 
The exceptions are pools where the passive estimate is more sensitive to the choice of methodology.

\begin{figure}[!htbp]
\centering
\scalebox{0.58}{\begin{tikzpicture}
\definecolor{pairblue}{RGB}{0,114,178}
\definecolor{pairgreen}{RGB}{0,158,115}
\definecolor{pairorange}{RGB}{230,159,0}
\definecolor{pairred}{RGB}{204,68,102}
\definecolor{pairpurple}{RGB}{136,34,204}

\begin{loglogaxis}[
width=13.5cm,
height=7.2cm,
xmajorgrids=true,
xminorgrids=true,
ymajorgrids=true,
yminorgrids=true,
major grid style={draw=gray!25},
minor grid style={draw=gray!10},
title style={font=\small\bfseries, align=center},
xlabel style={font=\scriptsize},
ylabel style={font=\scriptsize},
tick label style={font=\scriptsize},
tick align=outside,
legend style={at={(0.5,1.08)}, anchor=south, legend columns=5, draw=none, fill=none, font=\tiny, column sep=0.08cm, /tikz/every even column/.append style={column sep=0.08cm}},
ylabel={|passive LIFO $-$ infinitesimal| (bps, log scale)},
xlabel={Infinitesimal CI Range (bps, log scale)},
xmin=0.5,
xmax=100,
ymin=0.005,
ymax=100,
minor tick num=8,
xtick={0.5,1,2,5,10,20,50,100},
xticklabels={0.5,1,2,5,10,20,50,100},
]
\addplot+[only marks, forget plot, mark=triangle*, mark size=4.0pt, color=pairblue, line width=0.8pt, mark options={fill=pairblue, draw=pairblue, fill opacity=0.42, draw opacity=0.8, solid}] coordinates {(10.235191051,0.320412415)};
\addplot+[only marks, forget plot, mark=triangle*, mark size=4.0pt, color=pairblue, line width=0.8pt, mark options={fill=pairblue, draw=pairblue, fill opacity=0.42, draw opacity=0.8, dash pattern=on 1pt off 1pt}] coordinates {(3.723135660,0.044543460) (18.268190172,0.870630750)};
\addplot+[only marks, forget plot, mark=*, mark size=2.6pt, color=pairblue, line width=0.8pt, mark options={fill=pairblue, draw=pairblue, fill opacity=0.42, draw opacity=0.8, solid}] coordinates {(2.087603813,0.276595793)};
\addplot+[only marks, forget plot, mark=*, mark size=4.0pt, color=pairblue, line width=0.8pt, mark options={fill=pairblue, draw=pairblue, fill opacity=0.42, draw opacity=0.8, solid}] coordinates {(10.96784081,1.037521026)};
\addplot+[only marks, forget plot, mark=*, mark size=2.6pt, color=pairblue, line width=0.8pt, mark options={fill=pairblue, draw=pairblue, fill opacity=0.42, draw opacity=0.8, dash pattern=on 1pt off 1pt}] coordinates {(1.595592732,0.2834333) (0.550290463,0.901092413)};
\addplot+[only marks, forget plot, mark=square*, mark size=2.6pt, color=pairblue, line width=0.8pt, mark options={fill=pairblue, draw=pairblue, fill opacity=0.42, draw opacity=0.8, solid}] coordinates {(2.290031478,0.243370158)};
\addplot+[only marks, forget plot, mark=square*, mark size=4.0pt, color=pairblue, line width=0.8pt, mark options={fill=pairblue, draw=pairblue, fill opacity=0.42, draw opacity=0.8, solid}] coordinates {(16.64228006,0.999593702)};
\addplot+[only marks, forget plot, mark=square*, mark size=2.6pt, color=pairblue, line width=0.8pt, mark options={fill=pairblue, draw=pairblue, fill opacity=0.42, draw opacity=0.8, dash pattern=on 1pt off 1pt}] coordinates {(0.537556262,0.109661764) (0.560240946,0.443419364)};
\addplot+[only marks, forget plot, mark=triangle*, mark size=4.0pt, color=pairgreen, line width=0.8pt, mark options={fill=pairgreen, draw=pairgreen, fill opacity=0.42, draw opacity=0.8, solid}] coordinates {(31.128023538,1.329889084)};
\addplot+[only marks, forget plot, mark=triangle*, mark size=4.0pt, color=pairgreen, line width=0.8pt, mark options={fill=pairgreen, draw=pairgreen, fill opacity=0.42, draw opacity=0.8, dash pattern=on 1pt off 1pt}] coordinates {(15.388740056,4.031116580)};
\addplot+[only marks, forget plot, mark=*, mark size=2.6pt, color=pairgreen, line width=0.8pt, mark options={fill=pairgreen, draw=pairgreen, fill opacity=0.42, draw opacity=0.8, solid}] coordinates {(2.142031248,0.22609105)};
\addplot+[only marks, forget plot, mark=*, mark size=4.0pt, color=pairgreen, line width=0.8pt, mark options={fill=pairgreen, draw=pairgreen, fill opacity=0.42, draw opacity=0.8, solid}] coordinates {(15.09606331,1.046023139)};
\addplot+[only marks, forget plot, mark=*, mark size=2.6pt, color=pairgreen, line width=0.8pt, mark options={fill=pairgreen, draw=pairgreen, fill opacity=0.42, draw opacity=0.8, dash pattern=on 1pt off 1pt}] coordinates {(2.416778547,0.436223854) (0.772240033,0.900541971)};
\addplot+[only marks, forget plot, mark=square*, mark size=2.6pt, color=pairgreen, line width=0.8pt, mark options={fill=pairgreen, draw=pairgreen, fill opacity=0.42, draw opacity=0.8, solid}] coordinates {(1.95020931,0.4790698)};
\addplot+[only marks, forget plot, mark=square*, mark size=4.0pt, color=pairgreen, line width=0.8pt, mark options={fill=pairgreen, draw=pairgreen, fill opacity=0.42, draw opacity=0.8, solid}] coordinates {(7.742995806,4.321356848)};
\addplot+[only marks, forget plot, mark=square*, mark size=2.6pt, color=pairgreen, line width=0.8pt, mark options={fill=pairgreen, draw=pairgreen, fill opacity=0.42, draw opacity=0.8, dash pattern=on 1pt off 1pt}] coordinates {(1.190733102,0.053974063)};
\addplot+[only marks, forget plot, mark=*, mark size=2.6pt, color=pairorange, line width=0.8pt, mark options={fill=pairorange, draw=pairorange, fill opacity=0.42, draw opacity=0.8, dash pattern=on 1pt off 1pt}] coordinates {(1.379603113,0.057516356)};
\addplot+[only marks, forget plot, mark=square*, mark size=2.6pt, color=pairorange, line width=0.8pt, mark options={fill=pairorange, draw=pairorange, fill opacity=0.42, draw opacity=0.8, dash pattern=on 1pt off 1pt}] coordinates {(5.013885476,0.838805514)};
\addplot+[only marks, forget plot, mark=*, mark size=2.6pt, color=pairred, line width=0.8pt, mark options={fill=pairred, draw=pairred, fill opacity=0.42, draw opacity=0.8, solid}] coordinates {(2.970337147,0.870088628)};
\addplot+[only marks, forget plot, mark=*, mark size=4.0pt, color=pairred, line width=0.8pt, mark options={fill=pairred, draw=pairred, fill opacity=0.42, draw opacity=0.8, solid}] coordinates {(23.26112581,9.696634825)};
\addplot+[only marks, forget plot, mark=*, mark size=2.6pt, color=pairred, line width=0.8pt, mark options={fill=pairred, draw=pairred, fill opacity=0.42, draw opacity=0.8, dash pattern=on 1pt off 1pt}] coordinates {(2.065013574,0.884709478)};
\addplot+[only marks, forget plot, mark=square*, mark size=2.6pt, color=pairred, line width=0.8pt, mark options={fill=pairred, draw=pairred, fill opacity=0.42, draw opacity=0.8, solid}] coordinates {(3.322851324,0.3558586)};
\addplot+[only marks, forget plot, mark=square*, mark size=4.0pt, color=pairred, line width=0.8pt, mark options={fill=pairred, draw=pairred, fill opacity=0.42, draw opacity=0.8, solid}] coordinates {(27.97851943,1.098721718)};
\addplot+[only marks, forget plot, mark=square*, mark size=2.6pt, color=pairred, line width=0.8pt, mark options={fill=pairred, draw=pairred, fill opacity=0.42, draw opacity=0.8, dash pattern=on 1pt off 1pt}] coordinates {(1.303687466,0.429556978)};
\addplot+[only marks, forget plot, mark=*, mark size=2.6pt, color=pairpurple, line width=0.8pt, mark options={fill=pairpurple, draw=pairpurple, fill opacity=0.42, draw opacity=0.8, solid}] coordinates {(4.879413032,0.007194778)};
\addplot+[only marks, forget plot, mark=*, mark size=4.0pt, color=pairpurple, line width=0.8pt, mark options={fill=pairpurple, draw=pairpurple, fill opacity=0.42, draw opacity=0.8, solid}] coordinates {(10.90928374,0.644984915)};
\addplot+[only marks, forget plot, mark=*, mark size=2.6pt, color=pairpurple, line width=0.8pt, mark options={fill=pairpurple, draw=pairpurple, fill opacity=0.42, draw opacity=0.8, dash pattern=on 1pt off 1pt}] coordinates {(0.85349707,0.557332585)};
\addplot+[only marks, forget plot, mark=square*, mark size=2.6pt, color=pairpurple, line width=0.8pt, mark options={fill=pairpurple, draw=pairpurple, fill opacity=0.42, draw opacity=0.8, solid}] coordinates {(7.159430784,5.507979217)};
\addplot+[only marks, forget plot, mark=square*, mark size=4.0pt, color=pairpurple, line width=0.8pt, mark options={fill=pairpurple, draw=pairpurple, fill opacity=0.42, draw opacity=0.8, solid}] coordinates {(68.81283344,0.356858371)};
\addplot+[only marks, forget plot, mark=square*, mark size=2.6pt, color=pairpurple, line width=0.8pt, mark options={fill=pairpurple, draw=pairpurple, fill opacity=0.42, draw opacity=0.8, dash pattern=on 1pt off 1pt}] coordinates {(1.86807785,0.515533644)};
\addplot[forget plot, domain=0.5:100, samples=2, dashed, line width=0.65pt, color=black, mark=none] {x};
\addlegendimage{only marks, mark=*, color=pairblue, mark options={fill=pairblue, draw=pairblue, fill opacity=0.42, draw opacity=0.8}}
\addlegendentry{\mbox{USDC-WETH}}
\addlegendimage{only marks, mark=*, color=pairgreen, mark options={fill=pairgreen, draw=pairgreen, fill opacity=0.42, draw opacity=0.8}}
\addlegendentry{\mbox{USDT-WETH}}
\addlegendimage{only marks, mark=*, color=pairorange, mark options={fill=pairorange, draw=pairorange, fill opacity=0.42, draw opacity=0.8}}
\addlegendentry{\mbox{USDC-cbBTC}}
\addlegendimage{only marks, mark=*, color=pairred, mark options={fill=pairred, draw=pairred, fill opacity=0.42, draw opacity=0.8}}
\addlegendentry{\mbox{WBTC-USDC}}
\addlegendimage{only marks, mark=*, color=pairpurple, mark options={fill=pairpurple, draw=pairpurple, fill opacity=0.42, draw opacity=0.8}}
\addlegendentry{\mbox{WBTC-USDT}}
\addlegendimage{only marks, mark=triangle*, color=black}
\addlegendentry{v2}
\addlegendimage{only marks, mark=*, color=black}
\addlegendentry{v3}
\addlegendimage{only marks, mark=square*, color=black, mark options={fill=black, draw=black}}
\addlegendentry{v4}
\addlegendimage{empty legend}
\addlegendentry{}
\addlegendimage{empty legend}
\addlegendentry{}
\addlegendimage{dashed, line width=0.65pt, color=black}
\addlegendentry{$y=x$}
\addlegendimage{only marks, mark=*, color=black, mark options={fill=white, draw=black, solid}}
\addlegendentry{L1}
\addlegendimage{only marks, mark=*, color=black, mark options={fill=white, draw=black, dash pattern=on 1pt off 1pt}}
\addlegendentry{L2}
\end{loglogaxis}
\end{tikzpicture}}
\caption{Absolute difference between the passive LIFO and infinitesimal estimates (bps, log scale) against the width of the infinitesimal confidence interval (bps, log scale), per pool. 
Points below the 45-degree line are pools where the disagreement between the two methods is smaller than the statistical uncertainty of the infinitesimal estimate.}
\label{fig:cirange}
\end{figure}

\section{Robustness: Matching Convention and Matching Horizon}\label{app:matchrobust}

{\color{blue}
This appendix examines the sensitivity of the passive markout estimates to the two main implementation choices of the mint-burn matching step: the matching convention (LIFO versus FIFO) and the matching horizon ($T = 20$ versus $60$ seconds). Robustness to the markout horizon $h$ is examined in Appendix~\ref{app:horizonrobust}.

LIFO is an identification convention rather than an accounting assumption about how LPs themselves record their positions. 
When several unmatched liquidity additions coexist within the same position, assigning a subsequent removal to the most recent addition is particularly suitable for identifying short-lived liquidity episodes. 
One can do the matching using other methods, e.g., FIFO, pro-rata or average cost allocation. The last two would distribute the removal across multiple outstanding lots and consequently provide a \textit{different definition} of which liquidity was active over a short horizon.}

\paraheader{FIFO matching convention.}

{\color{blue}
We re-implement the subtraction estimator replacing the LIFO stack with a first-in-first-out (FIFO) queue: each removal is matched against the \emph{oldest} outstanding addition within the same position tuple rather than the most recent. All other elements of the procedure are identical, including the partial-match bookkeeping (one addition may be closed through multiple removals and one removal may consume liquidity from multiple earlier additions), the coverage rule, and the liquidity-share weights. 
When a position never has more than one outstanding addition at a time---the typical short-lived liquidity pattern---FIFO and LIFO produce identical matches by construction; the two conventions can differ only when several unmatched additions coexist. 
Because FIFO assigns removals to older additions, measured position durations lengthen mechanically in precisely those multi-lot cases, so we run FIFO with a 60-second matching horizon to avoid clipping the episodes where the conventions differ.
}

\paraheader{Matching horizon.}

{\color{blue}
We additionally re-run the baseline LIFO estimator with the matching horizon widened from 20 to 60 seconds on all chains.
}

{\color{blue}
Table~\ref{tab:robustness-matching} reports the passive markout and its 95\% confidence interval for the v3 pools under the three variants: LIFO with $T=20$s (baseline), LIFO with $T=60$s, and FIFO with $T=60$s.\footnote{The robustness runs use a data vintage extending through December 1st, 2025,
which accounts for small level differences relative to Table~\ref{tab:v3-markout-estimates}. 
The Ethereum 1 bps pools are omitted from the robustness runs.} 
The estimates are essentially unchanged across all pools. FIFO and LIFO at the same horizon agree to four or more decimal places in every pool, confirming that the multi-lot reallocation the convention governs is immaterial at the pool level. 
Widening the matching horizon from 20 to 60 seconds moves the estimate visibly only in the pool with the largest active share (Base USDC-WETH 5 bps, from $-0.709$ to $-0.740$ bps), a shift well within the confidence interval; in several pools the 20- and 60-second estimates are numerically identical, indicating that the short-lived liquidity identified by the procedure closes within 20 seconds. We conclude that the subtraction decomposition is effectively agnostic to both the matching convention and the matching horizon.
}

% In your main preamble, include:
% \usepackage{pgfplotstable}
% \usepackage{booktabs}
% \usepackage{makecell}
% \usepackage{float}

\begin{table}[H]
\centering
\scriptsize
\setlength{\tabcolsep}{4pt}
\renewcommand{\arraystretch}{1.08}

\begin{filecontents*}{sections/updated/robustness_matching_table.csv}
pool_label;lifo20_bps;lifo20_ci;lifo60_bps;lifo60_ci;fifo60_bps;fifo60_ci
\textbf{USDC-WETH};;;;;;
Base 5 bps;-0.709;[-0.81,-0.61];-0.740;[-0.84,-0.64];-0.740;[-0.84,-0.64]
Arbitrum 5 bps;-0.770;[-6.08,4.54];-0.769;[-6.08,4.54];-0.769;[-6.08,4.54]
Ethereum 5 bps;-1.505;[-2.00,-1.01];-1.505;[-2.00,-1.01];-1.505;[-2.00,-1.01]
Ethereum 30 bps;-0.866;[-2.63,0.90];-0.866;[-2.63,0.90];-0.866;[-2.63,0.90]
\textbf{USDT-WETH};;;;;;
Base 5 bps;-0.575;[-1.34,0.19];-0.575;[-1.34,0.19];-0.575;[-1.34,0.19]
Arbitrum 5 bps;-0.450;[-0.86,-0.04];-0.449;[-0.86,-0.04];-0.449;[-0.86,-0.04]
Ethereum 5 bps;-1.309;[-4.42,1.80];-1.309;[-4.42,1.80];-1.309;[-4.42,1.80]
Ethereum 30 bps;1.002;[-0.71,2.71];1.002;[-0.71,2.71];1.002;[-0.71,2.71]
\textbf{USDC-cbBTC};;;;;;
Base 5 bps;-0.568;[-0.77,-0.37];-0.568;[-0.77,-0.37];-0.568;[-0.77,-0.37]
\textbf{WBTC-USDC};;;;;;
Arbitrum 5 bps;-0.589;[-12.73,11.55];-0.589;[-12.73,11.55];-0.589;[-12.73,11.55]
Ethereum 5 bps;-0.407;[-5.49,4.67];-0.407;[-5.49,4.67];-0.407;[-5.49,4.67]
Ethereum 30 bps;0.210;[-1.40,1.82];0.210;[-1.40,1.82];0.210;[-1.40,1.82]
\textbf{WBTC-USDT};;;;;;
Arbitrum 5 bps;-0.722;[-0.85,-0.60];-0.722;[-0.85,-0.60];-0.722;[-0.85,-0.60]
Ethereum 5 bps;-0.230;[-1.75,1.29];-0.230;[-1.75,1.29];-0.230;[-1.75,1.29]
Ethereum 30 bps;2.520;[0.51,4.53];2.520;[0.51,4.53];2.520;[0.51,4.53]
\end{filecontents*}
\pgfplotstabletypeset[
    col sep=semicolon,
    string type,
    columns={pool_label,lifo20_bps,lifo20_ci,lifo60_bps,lifo60_ci,fifo60_bps,fifo60_ci},
    columns/pool_label/.style={column name={\makecell[l]{v3 pool\\label}}, column type={l}},
    columns/lifo20_bps/.style={column name={\makecell[c]{passive LIFO\\$T{=}20$s\\(bps)}}, column type={r}},
    columns/lifo20_ci/.style={column name={\makecell[c]{passive LIFO\\$T{=}20$s\\CI (bps)}}, column type={r}},
    columns/lifo60_bps/.style={column name={\makecell[c]{passive LIFO\\$T{=}60$s\\(bps)}}, column type={r}},
    columns/lifo60_ci/.style={column name={\makecell[c]{passive LIFO\\$T{=}60$s\\CI (bps)}}, column type={r}},
    columns/fifo60_bps/.style={column name={\makecell[c]{passive FIFO\\$T{=}60$s\\(bps)}}, column type={r}},
    columns/fifo60_ci/.style={column name={\makecell[c]{passive FIFO\\$T{=}60$s\\CI (bps)}}, column type={r}},
    every head row/.style={before row=\toprule, after row=\midrule},
    every row no 5/.style={before row=\midrule},
    every row no 10/.style={before row=\midrule},
    every row no 12/.style={before row=\midrule},
    every row no 16/.style={before row=\midrule},
    every last row/.style={after row=\bottomrule},
]{sections/updated/robustness_matching_table.csv}

\caption{Robustness of the passive markout (15-second markout horizon) to the mint-burn matching
convention and matching horizon, for Uniswap v3 pools. Each variant reports the volume-weighted
passive markout and its 95\% confidence interval. The robustness runs use a data vintage extending
through December 1st, 2025, which accounts for small level differences relative to
Table~\ref{tab:v3-markout-estimates}; the Ethereum 1 bps pools are omitted from the robustness runs.}
\label{tab:robustness-matching}
\end{table}

\section{Robustness: Markout Horizon and Interval Estimates}\label{app:horizonrobust}

\begin{figure}[p]
\centering
\includegraphics[width=1.1\textwidth]{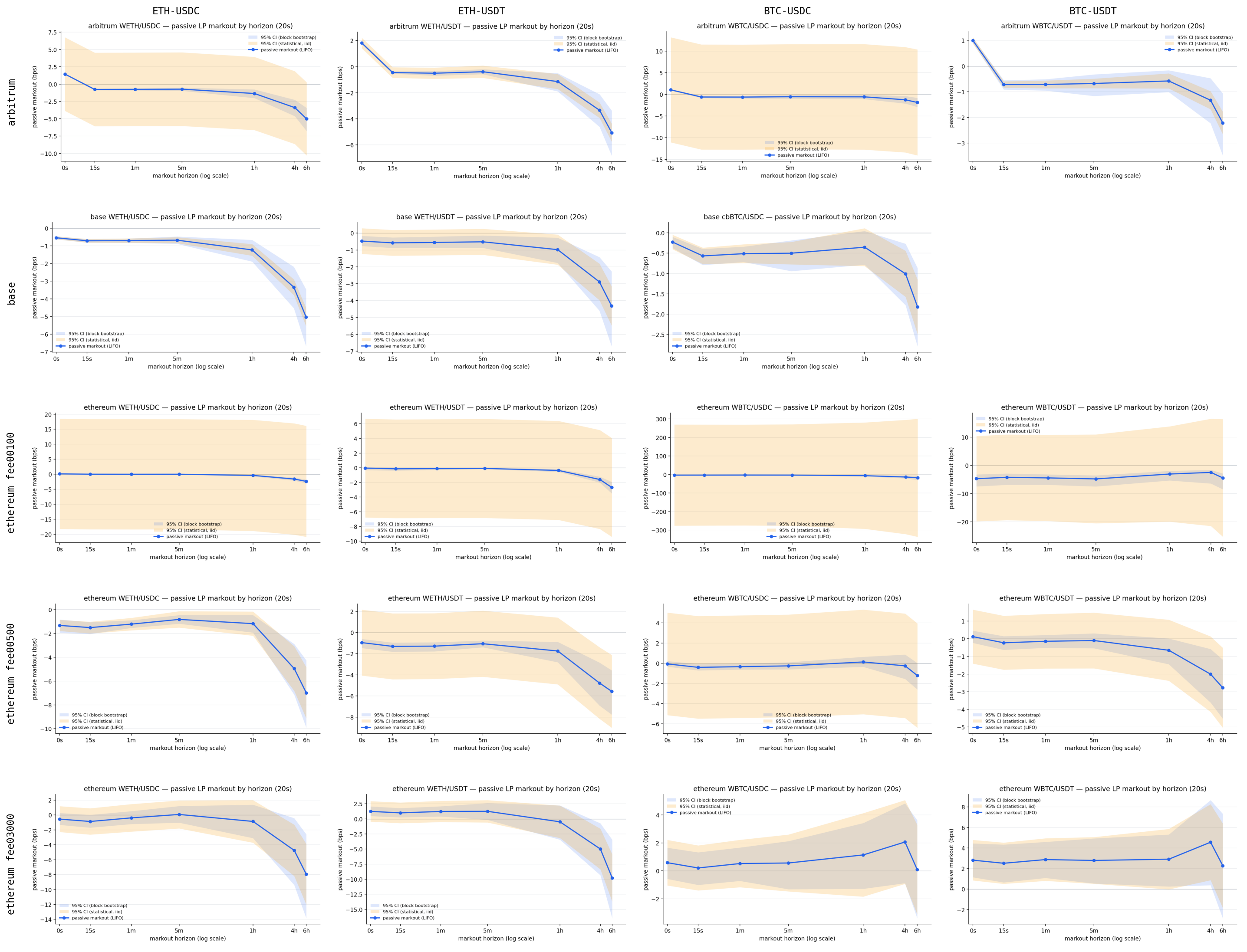}
\caption{Passive LIFO markout across markout horizons for Uniswap v3 pools (rows: chain and fee
tier; columns: pair), under the baseline 20-second matching horizon. Each panel shows the passive
markout point estimates with 95\% analytical (statistical, iid) and day-block bootstrap
confidence bands; the horizontal axis is log-scaled in the markout horizon.}
\label{fig:horizongrid20}
\end{figure}

{\color{blue}
We next vary the markout horizon $h$ over $\{0\mathrm{s}, 15\mathrm{s}, 1\mathrm{m}, 5\mathrm{m}, 1\mathrm{h}, 4\mathrm{h}, 6\mathrm{h}\}$, recomputing benchmark prices with the same construction as the baseline (last admissible Binance quote at or before $t+h$) and re-estimating the passive LIFO markout of every v3 pool at each horizon. 
Figure~\ref{fig:horizongrid20} reports the resulting term structures under the baseline 20-second matching horizon, with two interval estimates per pool: the analytical (statistical) confidence interval of Appendix~\ref{app:methoddetails}, which treats swaps as independent, and a block-bootstrap interval described below.
}

\paraheader{Term structure of the passive markout.}

{\color{blue}
The passive markout profile is flat from 0 seconds out to roughly one hour: estimates at the 0-second, 15-second, 1-minute, 5-minute, and 1-hour horizons are very close to the 15-second baseline in every pool. 
Beyond one hour, the estimates drift and the intervals widen sharply---the variance of benchmark price moves grows approximately linearly with $h$, so 4- and 6-hour markouts are increasingly dominated by market drift over the sample window rather than by swap-level adverse selection, and the decline at those horizons is synchronized across chains and pairs. 
The 4- and 6-hour points should therefore be read as noise-dominated; the precisely measured segment of the term structure, and the basis for our conclusions, is the 0-second to one-hour range, within which the 15-second baseline sits. 
We also recompute the \emph{total} pool markout at the same horizons: the total and passive profiles are nearly identical at every horizon (even in the pool with the largest active LIFO share, Base USDC-WETH 5 bps, the total-passive gap at the 15-second horizon is only about 0.03 bps), so the shape of the term structure is a property of the pool's overall order flow rather than of passive LP behavior, and the active-passive decomposition is not sensitive to the choice of markout horizon.
}

\paraheader{Block-bootstrap interval estimates.}

{\color{blue}
As an alternative to the analytical intervals, which assume independence across swaps, we compute a cluster bootstrap over calendar days: days are resampled with replacement, the volume-weighted passive markout is recomputed on each resample (5{,}000 replicates), and the 2.5th and 97.5th percentiles are reported. 
Because whole days are resampled, within-day dependence---including the overlap of markout windows across nearby swaps, which becomes substantial at multi-hour horizons---is preserved inside each block. 
The point estimates coincide with the analytical ones by construction. 
At short horizons the two interval estimates agree closely in the liquid pools, supporting the analytical confidence intervals reported in the main text. 
The bootstrap intervals are consistently narrower than the analytical ones, often dramatically so in pools where volume is concentrated in a small number of very large swaps (most visibly the Ethereum 1 bps pools in Figure~\ref{fig:horizongrid20}, where the analytical intervals span tens to hundreds of basis points while the bootstrap intervals remain tight around the estimate). 
The mechanism is that the analytical variance weights each swap by its squared volume and treats swaps as independent draws, so a handful of dominant swaps inflate the variance, whereas offsetting flow within a day makes day-level aggregates far less volatile than the swap-level formula implies. 
Our use of the analytical intervals in the main text is therefore conservative.
}

\paraheader{Interaction with the matching choices.}

\begin{figure}[p]
\centering
\includegraphics[width=1.1\textwidth]{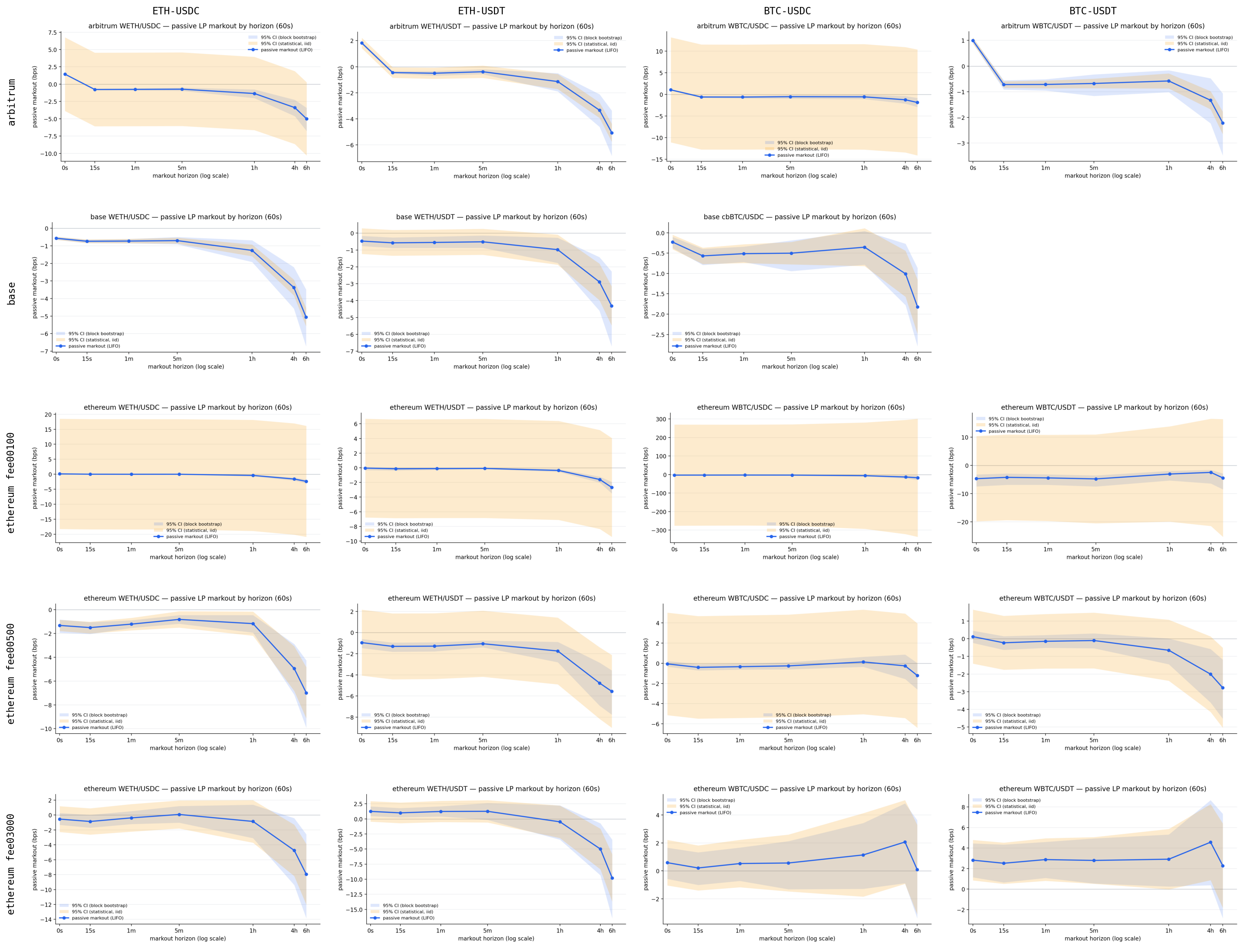}
\caption{Passive LIFO markout across markout horizons for Uniswap v3 pools, as in
Figure~\ref{fig:horizongrid20}, under a 60-second matching horizon. The panels are visually
indistinguishable from the 20-second case.}
\label{fig:horizongrid60}
\end{figure}

{\color{blue}
Figure~\ref{fig:horizongrid60} repeats the entire horizon analysis with the matching horizon widened to 60 seconds; the panels are visually indistinguishable from Figure~\ref{fig:horizongrid20}, consistent with the pool-level results in Table~\ref{tab:robustness-matching}. Repeating the analysis under the FIFO matching convention of Appendix~\ref{app:matchrobust} likewise leaves the term structures essentially unchanged. 
The
horizon, matching-window, and matching-convention robustness checks therefore reinforce one another: the passive markout estimates are stable along all three implementation dimensions.
}

\section{Practical Use of the Methodology}\label{app:practical}

Our two estimators differ sharply in cost, which determines how they map onto practitioner workflows. 
The infinitesimal-LP benchmark depends only on the internal pool price path and per-swap input/output amounts---quantities emitted on-chain as standard pool events. 
It can be computed block-by-block from any Ethereum or L2 RPC with negligible overhead, and is naturally suited to real-time monitoring. 
The LIFO and exact-matching estimators, by contrast, require the full per-pool mint--burn--swap event log together with a stateful matching pass over LP positions. 
They are linear in the number of events per pool and run in minutes per pool-month on a single machine, but are best treated as retrospective rather than real-time tools.

A natural application for passive LPs is a rolling-window pool monitor: practitioners compute the infinitesimal passive markout over a trailing day or week, and treat the lower bound of the associated confidence band as a decision threshold for exiting a position. 
The same procedure can be inverted to support pool selection ex ante, ranking candidate pools, fee tiers, and chains by realized passive markout per dollar of TVL over a recent reference window. 
For LPs considering whether to invest in active management, the realized gap between active and passive markout from the LIFO decomposition gives an estimate of the rent available to strategic provision.

The active--passive gap is also a natural target for protocol and pool designers. 
Proposed design interventions---JIT-mitigation hooks, sequencing changes, faster block times---can be evaluated by re-estimating the gap before and after the change. 
Because the infinitesimal benchmark requires no behavioral assumptions about LP heterogeneity, it is well-suited as a standardized market-quality indicator that pool operators and aggregators could publish alongside conventional TVL and volume statistics.

\end{document}